\RequirePackage{fix-cm}
\documentclass[smallextended]{svjour3}
\usepackage{times} 
\usepackage[T1]{fontenc}

\smartqed

\usepackage{graphicx}
\usepackage[latin9]{inputenc}
\usepackage{geometry}
\usepackage{color}
\usepackage[figuresright]{rotating}
\usepackage{amsmath,amstext,amssymb,bbm}

\usepackage{esint}
\usepackage[title]{appendix}
\usepackage[square,numbers,compress]{natbib}
\usepackage[unicode=true,
bookmarks=false,
breaklinks=false,pdfborder={0 0 1},colorlinks=true,urlcolor  = blue]
{hyperref}
\hypersetup{citecolor=blue,pdfencoding=auto}
\usepackage{breakurl}
\usepackage{tikz}
\usepackage[abs]{overpic}
\usepackage{xcolor}
\usepackage{comment}
\usepackage{authblk}
\usepackage{parskip}
\usepackage[title]{appendix}
\usepackage{enumerate}
\usepackage{graphicx,subfig}
\usepackage[american]{babel}

\def\makeheadbox{}

\usepackage{caption}
\newcommand{\allpar}{\theta}  
\newcommand{\data}{\mathcal{D}}  
\newcommand{\dist}{\rho}  
\newcommand{\thre}{\epsilon} 
\newcommand{\mh}{\alpha} 
\newcommand{\prior}{\pi} 
\newcommand{\trans}{q} 
\newcommand{\e}{\mathrm{e}}
\renewcommand{\leq}{\leqslant}
\renewcommand{\geq}{\geqslant}

\addto\captionsamerican{\renewcommand{\figurename}{\textbf{Fig.}}}

\journalname{Journal of Mathematical Biology}

\begin{document}

\title{Normal and pathological dynamics of platelets in humans}

\author{Gabriel P. Langlois \and
	 Morgan Craig \and
	 Antony R. Humphries \and
	 Michael C. Mackey \and
	 Joseph M. Mahaffy \and
	 Jacques  B\'{e}lair\and
	 Thibault Moulin \and
	 Sean R. Sinclair \and
	 Liangliang Wang}

\institute{G. P. Langlois \at
	Division of Applied Mathematics, Brown University, \\
	182 George St., Providence, RI 02912, USA  \\
	\email{Gabriel\_provencher\_langlois@brown.edu}
	\and
	M. Craig \at
	Program for Evolutionary Dynamics, Harvard University, \\  One Brattle Square  Cambridge MA 02138, USA \\
	\email{morganlainecraig@fas.harvard.edu}
	\and
	A. R. Humphries \at
	Department of Mathematics and Statistics, McGill University, Montreal, \\
	QC H3A 0B9, Canada \\
	\email{tony.humphries@mcgill.ca}
	\and
	M. C. Mackey \at
	Department of Mathematics, Physics, and Physiology, McGill University, \\ Montreal, QC H3G 1Y6, Canada \\
	\email{michael.mackey@mcgill.ca}
	\and
	J. M. Mahaffy \at
	Department of Mathematical Sciences, San Diego State University, San Diego, \\
	CA 92182-7720, USA \\
	\email{jmahaffy@mail.sdsu.edu}
	\and
	J. B\'{e}lair \at
	D{\'e}partement de math{\'e}matiques et de statistique, Universit{\'e} de Montr{\'e}al, \\
	Montr{\'e}al, QC H3C 3J7 Canada \\
	\email{belair@crm.umontreal.ca}
	\and
	T. Moulin \at
	Laboratoire Chrono-Environnement, Universit{\'e} de Bourgogne Franche-Comt{\'e}, \\
	16 route de Gray 25030, Besançon, France \\
	\email{thibault.moulin@univ-fcomte.fr}
	\and
	S. R. Sinclair \at
	Department of Mathematics and Statistics, McGill University, Montreal, \\
	QC H3A 0B9, Canada \\
	\email{sean.sinclair@mail.mcgill.ca}
	\and
	L. Wang \at
	Department of Statistics and Actuarial Science, Simon Fraser University, \\
	Burnaby, BC V5A 1S6, Canada \\
	\email{liangliang\_wang@sfu.ca}
}

\date{Received: date / Accepted: date}

\titlerunning{}        
\authorrunning{G.P. Langlois et al.} 

\maketitle

\begin{abstract}
We develop a mathematical model of platelet, megakaryocyte,
and thrombopoietin dynamics in humans. We show that there is a single stationary solution that can undergo a Hopf bifurcation, and use this information to investigate both normal and pathological platelet production, specifically cyclic thrombocytopenia. Carefully estimating model parameters from laboratory and clinical data, we then argue that a subset of parameters are involved in the genesis of cyclic thrombocytopenia based on clinical information. We provide model fits to the existing data for both platelet counts and thrombopoietin levels by changing four parameters that have physiological correlates. Our results indicate that the primary change in cyclic thrombocytopenia is an   interference with, or destruction of, the thrombopoietin receptor with secondary changes in other processes, including immune-mediated destruction of platelets and megakaryocyte deficiency and failure in platelet production. This study contributes to the understanding of the origin of cyclic thrombocytopenia as well as extending the modeling of thrombopoiesis.

\end{abstract}

\section{Introduction}

Mammalian blood contains three major types of cells that are essential in the maintenance of life: the red blood cells whose intracellular hemoglobin carries oxygen to tissues, the white blood cells responsible for all immune responses, and the platelets which maintain the integrity of clotting mechanisms. This tricellular system is known as the hematopoietic system.

The maintenance of hematological integrity in humans, as in all other mammals, is essential for normal physiological function, and under most circumstances is
wonderfully maintained by several intricate control mechanisms that are only partially understood. This control usually regulates the circulating levels of leukocytes (white blood cells), erythrocytes (red blood cells), and thrombocytes (platelets) within relatively narrow limits for a given individual notwithstanding the relatively wide variation within populations. For example, human platelet levels remain relatively stable in the range $150-450\times10^{9}\mbox{ platelets/L}$ with an average of about $290\times10^{9}\mbox{ platelets/L}$ of blood \cite{giles1981platelet}.

The hematopoietic cells are estimated to constitute $90\%$ of all cells in a human \citep{sender2016b}, and the lifetime production of these cells in humans is rather surprising as the average human produces the equivalent of their body weight in hematopoietic cells every decade of life \cite{mackey2001cell}.  Perhaps the most astonishing thing about this enormous production is that it usually proceeds without a flaw.

Disturbances to this tightly controlled regulation, however, can be quite harmful and often manifests itself as dynamic pathologies. Among these are a spectrum of periodic hematological diseases documented in the clinical literature that have provided rich fodder for those interested in mathematically modeling the regulation of hematopoiesis \citep{foley-mackey-2009}. Many of these periodic hematological diseases appear to be what are known as dynamic diseases \cite{glass1988clocks}.
Perhaps one of the best known and most studied of these periodic hematological diseases is cyclic neutropenia \citep{haurie:98}, a condition where
the neutrophils, erythrocyte precursors, and platelets all
oscillate at the same period in a given patient. A great deal is
known about the pathogenesis of cyclic neutropenia \citep{Colijn2007neutropenia}, and it is now generally believed that the disorder is linked to an abnormally high level of apoptosis in neutrophil precursors. This, in turn, leads to an elevated efflux of hematopoietic stem cells into the neutrophil lineage causing a destabilization of stem cell dynamics and an ensuing oscillation that is propagated into all of the hematopoietic lines.

Though the control of neutrophil production as well as the regulation
of erythropoiesis have been the subject of a number of modeling studies,
there have been fewer treating the regulation of
platelet production.  
One of the earliest
was that of \citet{wichmann79}, which was followed by an exposition
of their complete hematopoiesis model \citep{wichmann1985mathematical}. \citet{scholz2010}
used the same modeling framework to try to understand the response
to chemotherapy. Motivated by observed
oscillations in the platelet counts of healthy humans, \citet{vonschulthess1986} devised a conceptually different model for thrombopoiesis, which was followed
by \citet{belair1987model}. Building on this, \citet{santillan2000regulation} and \citet{apostu2008understanding} further refined the model to understand the origins of cyclic thrombocytopenia (CT).

In this work, we use recent laboratory and clinical data to develop a more physiologically realistic model for the regulation of mammalian platelet production concentrating on humans, which takes into account both the megakaryocytes and platelets and the effects of thrombopoietin on their dynamics. Sect.~\ref{sec:physiol} first reviews the relevant physiology of normal thrombogenesis and then briefly discusses cyclic thrombocytopenia. Next, in Sect.~\ref{sec:math_model} we derive the model for the dynamics of megakaryocytes, platelets, and thrombopoietin in humans. We present in Sect.~\ref{sec:analysis} several mathematical results that were derived for the model, including the existence of a unique positive equilibrium, the linearization and stability analysis of the model equations about the equilibrium, and a parameter sensitivity analysis for the model. In Sect.~\ref{sec:ctp}, we use data on cyclic thrombocytopenia patients as a benchmark against which to test the model. Starting with the parameters for a healthy subject, we change these parameters to those for a CT patient, showing a parameter set where a Hopf bifurcation occurs. We conclude with a brief discussion of our results and comparison with previous work in Sect.~\ref{sec:discussion}.

We have relegated more technical details to a series of appendices.  Appendix~\ref{app:parameter} describes our estimation of model parameters, Appendix~\ref{app:EU} contains a proof of the existence and uniqueness of the steady state solution of the model equations, while Appendix~\ref{app:linearization} contains a linearization of the full nonlinear model and a derivation of the characteristic equation for the stability analysis of the linearized model. Appendix~\ref{app:para-sense} gives the full results of our study of the sensitivity to parameter changes of the model when all parameters are held at the levels estimated for a healthy individual. Appendix~\ref{app:hopf_ct} continues the stability analysis of three CT patients from Sect.~\ref{subsec:hopf_ct}. Appendix~\ref{app:ABC} details the numerical techniques that we have used to fit the model to the cyclic thrombocytopenia data that we have available, while Appendix~\ref{app:numerics} gives the details of the numerical code we have used to solve the model equations.

\section{Physiological background}\label{sec:physiol}

\subsection{Normal thrombopoiesis}

Platelets are the hematological cells responsible for clotting and do so by adhering to the sites of damaged tissue to produce a hemostatic plug, which forms the surface on which coagulation factors are activated for clot formation. The mean platelet volume follows a log-normal distribution with respect to platelet count, with an average mean platelet volume of $8.6$ fL for an average platelet count of $290\times10^{9}\mbox{ platelets/L of blood}$
\citep{nakeff1970platelet,giles1981platelet}. About one-third of the total mass of
platelets is sequestered in an exchangeable splenic pool \citep{aster1966pooling}. Platelets have a life span of about 8 to 10 days, which is determined by an internal apoptotic regulating pathway, and are destroyed by the reticuloendothelial system \citep{mason2007programmed}.

Platelets are derived from megakaryocytes, large polyploid cells
found in the bone marrow. Megakaryocytes in turn are produced by the hematopoietic
stem cells, also found in the bone marrow, which are responsible
for generating all blood cells in the body. Among others, hematopoietic stem cells give rise to early bi-lineage progenitors that eventually undergo erythrocyte (red blood cell) or megakaryocyte differentiation. The differentiation process eventually produces the colony-forming unit-megakaryocyte
(CFU-Meg), a precursor cell committed to megakaryocyte
differentiation. These cells undergo mitosis (cell division) \citep{nakeff1977colony},
which stops some time after the CFU-Meg matures into a megakaryoblast,
an early maturation stage of megakaryocytes.

After cell division ceases,
megakaryocytes begin endomitosis -- a process in
which DNA replicates through nuclear division without cell division while the cytoplasm remains intact \citep{William}. DNA can replicate $2$ to $7$
times during endomitosis, resulting in cells with DNA content between $8$ and $128$ times
the normal diploid content of DNA in a single, highly lobated nucleus.
Megakaryocytes are generally classified by their ploidy, which is
the number of chromosomes that they have. A megakaryocyte ploidy of
$2$N refers to a megakaryocyte that has not undergone endomitosis. The
modal megakaryocyte ploidy in humans is $16$N \citep{jackson1984two,kuter1989analysis},
which corresponds to a megakaryocyte with $8$ times the normal diploid
content of DNA. In general the higher the ploidy number, the larger the megakaryocyte, with the diameter of megakaryocytes ranging
from $20$ to $60$ $\mbox{\ensuremath{\mu}m}$ depending on the ploidy \cite{tomer1996measurements}.

As DNA replicates, the cytoplasm of the megakaryocyte expands
and develops a demarcation membrane that eventually becomes the external
membrane of each platelet. Megakaryocytes are released from the bone
marrow and travel to the lungs, where the megakaryocytes shed
platelets \citep{kaufman1965circulating,pedersen1978occurrence,trowbridge1982evidence}.
On average, one megakaryocyte sheds between $1000$ and $3000$ platelets \citep{harker1969thrombokinetics}.
It is estimated that it takes $5$ to $7$ days for a megakaryocyte to begin endomitosis,
grow into a mature megakaryocyte, and shed platelets \citep{William}.

The principal hormone that regulates megakaryocyte and platelet development
is thrombopoietin (TPO). TPO is produced principally by the liver,
and to a smaller extent, in the kidney and bone marrow \citep{nomura1997cellular,qian1998primary}.
Its crystal structure is that of an anti-parallel four-helix bundle
fold with two different binding sites for the TPO receptor \citep{feese2004structure}.
It is released into blood as a $95$ kDa glycoprotein \citep{kuter2009thrombopoietin}, and acts as the ligand for the c-Mpl receptor, present
on the surface of CFU-Meg, megakaryocytes,
and platelets \citep{debili1995mpl,li1999interaction}. On binding
to thrombopoietin, the receptor dimerizes and initiates a number of
signal transduction events that eventually stimulate differentiation and mitosis
of CFU-Meg, increase the rate of endomitosis of megakaryocytes, and
reduce the rate of apoptosis of CFU-Meg and megakaryocytes \citep{kaushansky1995thrombopoietin,majka2000stromal,zauli1997vitro}.
The thrombopoietin is then internalized, degraded, and removed
from circulation. This internalization process is the major mechanism
of TPO removal from the blood by platelets and megakaryocytes \citep{li1999interaction}. 

Although TPO supports the survival of CFU-Meg and megakaryocytes, it is not essential. Elimination of the thrombopoietin gene or its receptor in mice reduces
megakaryocyte and platelet levels to approximately $10$\% of normal
\citep{de1996physiological}. The residual platelets and megakaryocytes
are normal and functional, and the other blood cells are also at their
normal levels. The same observation has also been made in humans (Kaushansky,
private communication).

\subsection{Cyclical thrombocytopenia} \label{subsec:ctp_physio}

Cyclic thrombocytopenia is a hematological disorder that causes the platelet count of an affected individual to undergo large periodic
fluctuations over time. In these individuals, platelet counts oscillate
from very low ($1\times10^{9}\mbox{ platelets/L}$) to normal or very
high levels ($2000\times10^{9}\mbox{ platelets/L}$) \citep{swinburne}. At the nadir, patients are at risk
of bruising and excessive bleeding, whereas at very high levels there is an increased risk of clot formation. Little is known about the pathogenesis
of the disease, which has been reviewed in \citep{apostu2008understanding,cohen,go2005idiopathic,swinburne}.  It is well established that in premenopausal women with CT there is often a relation between blood hormonal and platelet levels, but it is unclear whether this is causal.  In other cases, clinical findings suggest at least three possible origins:
immune-mediated platelet destruction (autoimmune CT), megakaryocyte
deficiency and cyclic failure in platelet production (amegakaryocytic
CT), and possible immune interference with or destruction of the TPO receptor \cite{go2005idiopathic}. Autoimmune CT is thought to be an unusual form of immune thrombocytopenia
purpura (a disease in which the platelet count is abnormally low).
The hematological profile of most affected patients reveals high
levels of antiplatelet antibodies, shorter platelet lifespans at the
platelet nadir, and normal to high levels of marrow megakaryocytes.
Amegakaryocytic CT is postulated to be a variant of acquired amegakaryocytic
thrombocytopenic purpura and is mainly characterized by the absence
of megakaryocytes in the thrombocytopenia phase and increased megakaryocyte
number during thrombocytosis.

A curious feature of the disease is that the fluctuations appear only
to be present in the platelet cell line and not in the white or red
blood cell lines. \citet{swinburne} and \citet{apostu2008understanding}
searched the English literature and found and analyzed well-documented
cases of cyclic thrombocytopenia. In no case was there a report
of fluctuations in the red or white blood cells. In other existing
cyclic hematological disorders like periodic chronic myelogenous leukemia \citep{colijn2005mathematicalI},
fluctuations appear in all major blood cell lines and are all at the
same period in a given subject. These diseases are believed to evolve
from the hematopoietic stem cell compartment in the bone marrow. In cyclic thrombocytopenia, fluctuations are observed in the platelet line only, and therefore a
destabilization of a peripheral control mechanism could play an important
role in the genesis of the disorder. This hypothesis was the starting
point of the investigation and mathematical modeling of \citet{santillan2000regulation}
and \citet{apostu2008understanding}.

\section{Mathematical model of thrombopoiesis \label{sec:math_model}}

\begin{figure}[!ht]
	\centering
	\scalebox{1.2}{
		\begin{overpic}[width=0.5\textwidth]{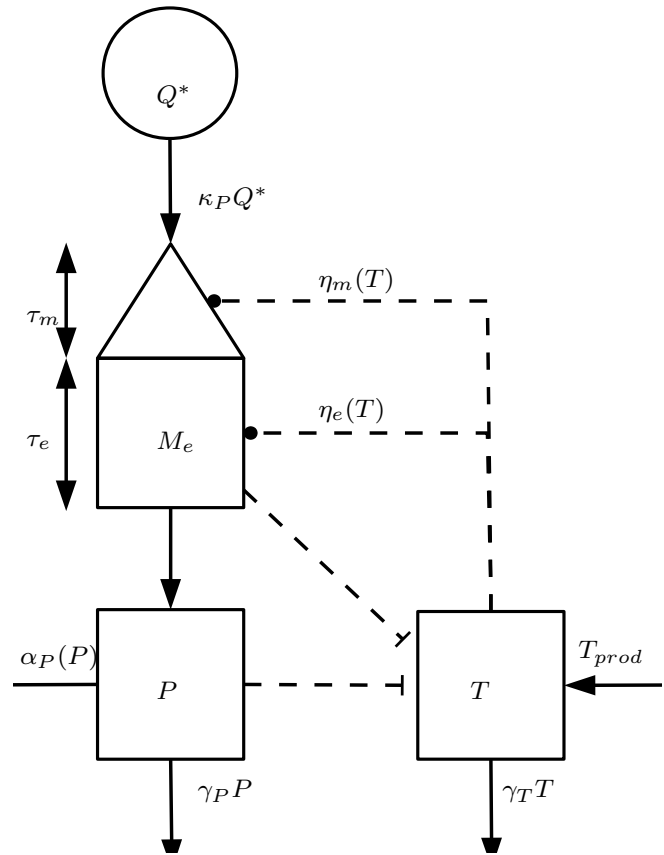}
			\put(45,138){$ \mbox{\footnotesize $M_e$} $}
			\put(45,247){$ \mbox{\footnotesize $Q^*$} $}
			\put(58,215){$ \mbox{\footnotesize $\kappa_PQ^{*}$} $}

			\put(04,139){$ \mbox{\footnotesize $\tau_e$} $}
			\put(04,178){$ \mbox{\footnotesize $\tau_m$} $}
			
			\put(144,59){$ \mbox{\footnotesize $T$} $}
			\put(178,72){$ \mbox{\footnotesize $T_{prod}$} $}
			\put(154,28){$ \mbox{\footnotesize $\gamma_TT$} $}
			
			\put(45,59){$ \mbox{\footnotesize $P$} $}
			\put(58,28){$ \mbox{\footnotesize $\gamma_PP$} $}
			\put(02,70){$ \mbox{\footnotesize $\alpha_{P}(P)$} $}
			
			\put(96,189){$ \mbox{\footnotesize $\eta_m(T)$} $}
			\put(96,148){$ \mbox{\footnotesize $\eta_e(T)$} $}

		\end{overpic}}
		\caption{A schematic view of the model of human thrombopoiesis. Solid lines denote fluxes, dashed lines terminating in solid circles denote positive feedback, and dashed lines ending with perpendicular lines denote negative feedback. Hematopoietic stem cells enter the megakaryocyte lineage as well as the other blood lines, undergo cell division, or are removed from the HSC pool through death. HSCs differentiated into the megakaryocyte lineage undergo cell divisions for $\tau_m$ days, after which they stop dividing and start endomitosis (nuclear division). These megakaryocytes undergo endomitosis for $\tau_e$ days until they finally start to shed platelets. Platelets remain in circulation until they are removed at random by degradation  or cleared by macrophages due to senescence, a platelet-dependent mechanism. Thrombopoietin is produced constitutively at a rate $T_{prod}$, and is removed from circulation either at random by degradation or by binding to receptors present on platelets and megakaryocytes}
		\label{schematic}
\end{figure}

In this section, we develop our mathematical model for the regulation of megakaryocyte, platelet and thrombopoietin dynamics in humans. We describe the
dynamics of the megakaryocytes as an age-structured model (Sect.~\ref{ssec:meg}), which is
divided in two stages: mitosis (Sect.~\ref{sssec:mitosis}) and endomitosis (Sect.~\ref{sssec:endo}). Platelet (Sect.~\ref{ssec:plate}) and thrombopoietin (Sect.~\ref{ssec:tpo}) dynamics are dealt with last. In the discussion of the development of the model, the reader may find Fig.~\ref{schematic} helpful.

\subsection{Megakaryocyte compartment}\label{ssec:meg}

\subsubsection{Mitosis}\label{sssec:mitosis}
	
We first model the megakaryocyte mitosis phase, starting from the moment the hematopoietic stem cells differentiate into the megakaryocytic lineage. These early cells, known as megakaryoblasts (or CFU-Meg in tissue culture), undergo mitosis (cell division) for some time until they stop and begin endomitosis.
	
Let $m_{m}(t,a)$ be the cell density of megakaryoblasts
as a function of time $t$ and age $a$, and $Q^*$ the equilibrium concentration
of hematopoietic stem cells at time (we assume here that the quiescent stem cells are at their normal steady state level throughout this paper).  We further assume that stem cells enter the megakaryoblasts compartment at a rate $\kappa_{P}$, and that a megakaryoblast proliferates for $\tau_{m}$ days at a thrombopoietin-dependent ($T(t)$) rate of $\eta_{m}(T(t))$. As discussed in Sect.~\ref{sec:physiol}, while TPO stimulates mitosis of megakaryoblasts, it is not necessary. Therefore, we assume a basal proliferation rate of megakaryoblasts even in absence of thrombopoietin. Based on these assumptions, we model the proliferation rate $\eta_{m}(T(t))$ as a Hill function
\begin{equation}
\eta_{m}(T)
=
\eta_{m}^{min}+(\eta_{m}^{max}-\eta_{m}^{min})\frac{T}{b_{m}+T},
\label{eq:mitosis_rate}
\end{equation}
where the parameter $\eta_{m}^{min}$ is the minimum effective rate of proliferation in absence of thrombopoietin, $\eta_{m}^{max}$ is the maximum effective rate of proliferation, and $b_{m}$ is the concentration of thrombopoietin at which the proliferation is half maximal.

Three comments are in order.  First, since we are not trying to model the details of megakaryoblast proliferation and apoptosis dynamics, equation \eqref{eq:mitosis_rate} simply gives an {\it effective} proliferation rate that includes both cellular birth and death.  Second, the choice of the Hill function in \eqref{eq:mitosis_rate} is taken to reflect the fact that TPO has a stimulatory, yet saturating, effect on the process. Third, in the absence of further experimental data, the choice of Hill coefficient is unclear and we have therefore opted for an estimate consistent with the qualitative observations by taking a Hill coefficient of $1$.

The dynamics of megakaryoblasts, then, is modeled by means a time-age evolution equation given by
\begin{equation}
\frac{\partial m_{m}}{\partial t}+\frac{\partial m_{m}}{\partial a}=\eta_{m}(T)m_{m},\qquad t\geq0,\qquad a\in\left[0,\tau_{m}\right].\label{eq:age_structure_mb}
\end{equation}
For the boundary condition, we take $m_{m}(t,0)=\kappa_{P}Q^{*}$, which is the rate hematopoietic stem cells enter the megakaryocyte lineage.
	
	We solve Eq.~\eqref{eq:age_structure_mb} using the method of characteristics
	to obtain
	\begin{equation} \label{eq:mb_density}
	m_{m}(t,a)=\kappa_{P}Q^{*}\exp\left[\int_{t-a}^{t}\eta_{m}(T(s)) \, \mathrm{d}s\right],\qquad { t\geq a},\qquad a\in[0,\tau_{m}],
	\end{equation}
	and
	\begin{equation} \label{eq:mb0_density}
	m_{m}(t,a)=m_{m}(0,a-t)\exp\left[\int_{0}^{t}\eta_{m}(T(s)) \, \mathrm{d}s\right],\qquad { t\in[0,a)},\qquad a\in[0,\tau_{m}].
	\end{equation}
	It is convenient to define an initial function $T(t)$ for $t\in[-\tau_m-\tau_e,0]$ and 
	following \eqref{eq:mb_density} let
	$$m_{m}(0,a)=\kappa_{P}Q^{*}\exp\left[\int_{-a}^{0}\eta_{m}(T(s))\, \mathrm{d}s\right], \qquad a\in(0,\tau_{m})$$
	so that equation \eqref{eq:mb0_density} reduces to \eqref{eq:mb_density}, and thus equation \eqref{eq:mb_density} can be applied for all $t		\geq0$.

	\subsubsection{Endomitosis}\label{sssec:endo}
	
	Next we consider the endomitosis (endoreplication) phase, starting from the moment megakaryocytes begin endomitosis until they start to  shed platelets.
	During this period, megakaryocytes no longer
	multiply, but rather grow in ploidy and size. Accordingly, we
	model the volume growth of megakaryocytes during endomitosis and we assume that megakaryocyte volume is an increasing function of megakaryocyte age so  the two may be simply related.
	
	Let $m_{e}(t,a)$ be the volume density of megakaryocytes
	in the endomitosis phase as a function of time $t$ and age $a$, $V_m$  the volume of a single megakaryocyte of ploidy 2N at age $a=0$.
	Suppose that a megakaryocyte undergoes endomitosis for $\tau_{e}$
	days and at a thrombopoietin-dependent rate of $\eta_{e}(T(t))$. As in the process of mitosis, TPO stimulates endomitosis in megakaryocytes but is not strictly essential. Thus we assume a basal endoreplication rate of megakaryocytes even in absence of thrombopoietin. Based on this fact, we model the proliferation rate $\eta_{e}(T(t))$ as a Hill function
	\begin{equation}
	\eta_{e}(T)=\eta_{e}^{min}+(\eta_{e}^{max}-\eta_{e}^{min})
	\frac{T}{b_{e}+T},
	\label{eq:endomitosis_rate}
	\end{equation}
	where the parameter $\eta_{e}^{min}$ is the minimum effective rate of endomitosis in absence of thrombopoietin, $\eta_{e}^{max}$ is the maximum effective rate of endomitosis, and $b_{e}$ is the concentration of thrombopoietin at which the endomitosis rate is half maximal. The comments relating to the choice of Hill function and coefficient following \eqref{eq:mitosis_rate} also apply here.

We model the volume growth of megakaryocytes, then, by means of a time-age structured equation given by	
	\begin{equation}
	\frac{\partial m_{e}}{\partial t}+\frac{\partial m_{e}}{\partial a}=\eta_{e}(T)m_{e},\qquad t\geq0,\qquad a\in\left[0,\tau_{e}\right].\label{eq:age_structure_mk}
	\end{equation}
	For the boundary condition, we take
	$$
	m_{e}(t,0)=V_{m}m_{m}(t,\tau_{m})=
	V_{m}\kappa_{P}Q^{*}\exp\left[\int_{t-\tau_{m}}^{t}\eta_{m}(T(s)) \, \mathrm{d}s\right],
	$$
	which is the product of the average volume of a megakaryocyte commencing endomitosis and the number of megakaryocytes at the end of the mitotic phase.
	
	As before we solve \eqref{eq:age_structure_mk} using the method of characteristics 
	and the initial function $T(t)$ for $t\in[-\tau_m-\tau_e,0]$
	to obtain
	\begin{equation}
	m_{e}(t,a)=V_{m}\kappa_{P}Q^{*}\exp\left[\int_{t-a-\tau_{m}}^{t-a}\eta_{m}(T(s)) \, \mathrm{d}s\right]\exp\left[\int_{t-a}^{t}\eta_{e}(T(s)) \, \mathrm{d}s\right],\quad t\geq0,\quad a\in[0,\tau_{e}].\label{eq:mk_density}
	\end{equation}
	The total megakaryocyte volume at time $t$ is
	\begin{equation}
	M_{e}\left(t\right)=\int_{0}^{\tau_{e}}m_{e}\left(t,a\right) \, \mathrm{d}a.\label{eq:mk_volume}
	\end{equation}
	
	\subsection{Platelet compartment}\label{ssec:plate}
	
	Platelet population dynamics are governed by the balance between
	platelet production and destruction. The platelet population is comprised of
	both platelets in circulation as well as those sequestered primarily in the spleen after their creation  from megakaryocytes at the end of the endomitosis stage. Platelets
	die at a random rate $\gamma_{P}$ proportional to platelet numbers. Platelets are also removed by senescence and cleared by macrophages \cite{Grozovsky2010} via a saturable mechanism, which we model via a saturable Hill function
	\begin{equation*}
   \alpha_P \frac{(P)^{n_{P}}}{(b_{P})^{n_{P}}+(P)^{n_{P}}},
	\end{equation*}
	where $\alpha_{P}$ is the maximal platelet-dependent removal rate, $b_{P}$ is the platelet concentration at which the removal rate is half its maximum and $n_{P}$ is the Hill coefficient modeling how steeply the platelet removal rate changes with platelet levels. We assume senescence will be reduced when platelet concentrations are low (the average age of platelets can be expected to be lower, since newly created platelets have age $0$, and there are few old platelets if the concentration is low), which implies $n_P>1$, and we choose $n_P=2$. Based on these considerations, we
	model the dynamics of platelets via the differential equation
	\begin{equation}
	\frac{\mathrm{d}P}{\mathrm{d}t}=\frac{D_{0}}{\beta_{P}}m_{e}(t,\tau_{e})-\gamma_{P}P
	-\alpha_{P}\frac{(P)^{n_{P}}}{(b_{P})^{n_{P}}+(P)^{n_{P}}},\label{eq:p}
	\end{equation}
	where $D_{0}$ is the fraction of megakaryocyte volume shed into platelets, and $\beta_{P}$ is the average volume of a platelet.

	\subsection{Thrombopoietin compartment}\label{ssec:tpo}
	
	Finally, as with platelets, we model TPO dynamics as the balance between
	production and destruction. We assume that TPO is produced
	at a constant rate $T_{prod}$ \citep{Kuter13}. As thrombopoietin is cleared mainly by receptors on megakaryocytes and circulating platelets, its endogenous removal rate is proportional to the total volume of megakaryocytes and circulating platelets. Since only a finite number of TPO receptors can clear thrombopoietin, we assume the endogenous removal rate is proportional to the saturable Hill function
	\begin{equation*}
	\frac{(T)^{n_{T}}}{(k_{T})^{n_{T}}+(T)^{n_{T}}},
	\end{equation*}
	where $k_{T}$ is the thrombopoietin concentration at which the removal rate is half the maximum removal rate and $n_{T}$ is the Hill coefficient modeling how steeply the TPO removal rate changes with TPO levels. Here, the Hill coefficient $n_T$ will be determined by the stoichiometry of TPO receptor interactions. We also assume a small renal clearance rate of $\gamma_{T}$ proportional to TPO levels. Thus, we model the dynamics of thrombopoietin with
	\begin{equation}
	\frac{\mathrm{d}T}{\mathrm{d}t}=T_{prod}-\gamma_{T}T
	-\alpha_{T}\left(M_{e}(t)+k_{S}\beta_{P}P\right)
	\frac{(T)^{n_{T}}}{(k_{T})^{n_{T}}+(T)^{n_{T}}}, \label{eq:tpo}
	\end{equation}
	where $\alpha_{T}$ is the maximum removal rate of thrombopoietin by internalization and $k_{S}$ is the average fraction of platelets circulating in the blood.
	
	\subsection{Model summary \label{subsec:summary}}
	
	As detailed above, our model of thrombopoiesis consists of two integro-differential equations with constant delays and an integral equation. The two differential equations model the dynamics of platelets and TPO, while the integral equation models the volume of megakaryocytes in the bone marrow. Thus, to summarize, our full model is given by
	\begin{equation}
	\label{eq:de_platelet}
	\frac{\mathrm{d}P}{\mathrm{d}t}=\frac{D_{0}}{\beta_{P}}m_{e}(t,\tau_{e})-\gamma_{P}P-\alpha_{P}\frac{(P)^{n_{P}}}{(b_{P})^{n_{P}}+(P)^{n_{P}}},
	\end{equation}
	\begin{equation}
	\label{eq:de_tpo}
	\frac{\mathrm{d}T}{\mathrm{d}t}=T_{prod}-\gamma_{T}T-\alpha_{T}\left(M_{e}(t)+k_{S}\beta_{P}P\right)\frac{(T)^{n_{T}}}{(k_{T})^{n_{T}}+(T)^{n_{T}}},
	\end{equation}
	where
	\begin{equation}
	\label{eq:platelet_prod_rate}
	m_{e}(t,a)=V_{m}\kappa_{P}Q^{*}\exp\left[\int_{t-a-\tau_{m}}^{t-a} \! \eta_{m}(T(s)) \, \mathrm{d}s\right]\exp\left[\int_{t-a}^{t}\eta_{e}(T(s)) \, \mathrm{d}s\right]
	\end{equation}
	and
	\begin{equation} \label{eq:Me}
	M_{e}(t)=\int_{0}^{\tau_{e}}\! m_{e}(t,a)da.
	\end{equation}
	The functions $\eta_{m}(T )$ and $\eta_{e}(T )$
	are given by
	\begin{equation}
	\label{eq:eta_m}
	\eta_{m}(T )=\eta_{m}^{min}+(\eta_{m}^{max}-\eta_{m}^{min})\frac{T}{b_{m}+T}
	\end{equation}
	and
	\begin{equation}
	\label{eq:eta_e}
	\eta_{e}(T )=\eta_{e}^{min}+(\eta_{e}^{max}-\eta_{e}^{min})\frac{T}{b_{e}+T}.
	\end{equation}
	
	All parameters are estimated in Appendix~\ref{app:parameter}, and the results of that estimation for a healthy human are given in Table~\ref{table:para-summary}. We show the existence and uniqueness of a positive stationary solution to our model in Appendix~\ref{app:EU}. Of particular note, owing to the lack of data specific to the HSC dynamics, to avoid issues of parameter identifiability $Q(t)=Q^*$ throughout.
	
	\begin{sidewaystable}[h!]
		\begin{center}
			\noindent\resizebox{\textwidth}{!}{%
				\begin{tabular}{|c|c|c|c|c|}
					\hline
					Name & Interpretation & Value & Units & References\tabularnewline
					\hline
					\hline
					$Q^*$ & HSCs density & $1.1$ & $10^6$ cells/kg & \citep{bernard2003oscillations}\tabularnewline
					\hline
					$\kappa_{p}$ & HSC differentiation rate into megakaryocyte line & $0.0072419$ & day\textsuperscript{-1} & \cite{ bernard2003oscillations,mackey2001cell}\tabularnewline
					\hline
					$V_{m}$ &  Volume of megakaryocyte of ploidy 2N & $\frac{4\pi(21)^{3}}{24}$ & fL & \citep{mcclatchey2002clinical}\tabularnewline
					\hline
					$\tau_{m}$ & MB proliferation duration & $ 8.09$ & days & Fit\tabularnewline
					\hline
					$\eta_{m}^{min}$ & Min MB proliferation rate & $0.38874$ & day\textsuperscript{-1} & Eq.~\eqref{eq:eta^min_m}\tabularnewline
					\hline
					$\eta_{m}^{max}$ & Max MB proliferation rate & $2.6828$ & day\textsuperscript{-1} & Eq.~\eqref{eq:eta^max_m}\tabularnewline
					\hline
					$b_{m}$ & TPO concentration for half max MB proliferation & $ 706$ & pg / ml & Fit\tabularnewline
					\hline
					$\tau_{e}$ & Endomitosis duration & 5.0 & days & \citep{finch1977kinetics,Kuter13}\tabularnewline
					\hline
					$\eta_{e}^{min}$ & Min endomitosis rate & $0.41022$ & day\textsuperscript{-1} & Eq.~\eqref{eq:eta^min_e}\tabularnewline
					\hline
					$\eta_{e}^{max}$ & Max endomitosis rate & $0.69335$ & day\textsuperscript{-1} & Eq.~\eqref{eq:eta^max_e}\tabularnewline
					\hline
					$b_{e}$ & TPO concentration for half max endomitosis  & $ 92.1$ & pg / ml & Fit\tabularnewline
					\hline
					$P^{*}$ & Normal platelet level & $31.071$ & $10^9$ platelets / kg & \citep{giles1981platelet}\tabularnewline
					\hline
					$\beta_{P}$ & Average volume of a platelet & 8.6 & fL & \citep{giles1981platelet}\tabularnewline
					\hline
					$D_{0}$ & Fraction megakaryocytes shedding  platelets & $0.21829$ & --- & Eq.~\eqref{eq:d_0}\tabularnewline
					\hline
					$\tau_{P}$ & Mean platelet survival time & 8.4 & days & \citep{tsan1984kinetics}\tabularnewline
					\hline
					$\alpha_{P}$ & Max platelet removal rate & $ 212.95$ & $10^9$ platelets / kg / day & Eq.~\eqref{eq:alpha_P}, \citep{tsan1984kinetics}\tabularnewline
					\hline
					$\gamma_{P}$ & Random loss rate of platelets & $0.05$ & day\textsuperscript{-1} & Fit\tabularnewline
					\hline
					$b_{P}$ & Platelet levels for half max removal & $308$ & $10^9$ platelets / kg & Fit\tabularnewline
					\hline
					$n_{P}$ & Hill coefficient for platelet removal & 2.0 & --- & \citep{hitchcock2014thrombopoietin}\tabularnewline
					\hline
					$T^{*}$ & TPO levels & 100 & pg / ml & \citep{Kuter13}\tabularnewline
					\hline
					$T_{prod}$ & TPO production rate & $61.6$ & pg / ml / day & Fit\tabularnewline
					\hline
					$\gamma_{T}$ & TPO renal clearance rate & $0.01$ & day\textsuperscript{-1} & Fit\tabularnewline
					\hline
					$k_{S}$ & Fraction of platelets circulating in the blood & 2/3 & --- & \citep{aster1966pooling}\tabularnewline
					\hline
					$\alpha_{T}$ & Maximum clearance rate of thrombopoietin & $144.87$ & $10^{-9}$pg kg / (fL ml day) & Eq.~\eqref{eq:alpha_T}\tabularnewline
					\hline
					$k_{T}$ & TPO half max clearance & $3180$ & pg / ml & Fit\tabularnewline
					\hline
					$n_{T}$ & TPO clearance Hill coefficient & 2.0 & --- & \citep{hitchcock2014thrombopoietin}\tabularnewline
					\hline
				\end{tabular}
			}
		\end{center}
		\caption{Summary of units and values of all model parameters.  HSC denotes hematopoietic stem cell, MB denotes megakaryoblast. All units have up to 5 significant digits}\label{table:para-summary}
	\end{sidewaystable}
	
	\clearpage
	
	\newpage
	\section{Model Analysis}\label{sec:analysis}
	
	The model presented in Eqs.~\eqref{eq:de_platelet}--\eqref{eq:eta_e} is a nonlinear system of two integro-differential equations that describes the process of thrombopoiesis. This section examines some of the mathematical results which can be derived from the model. We establish the existence of a unique positive equilibrium in Appendix~\ref{app:EU}. A local linear analysis about this equilibrium provides a complicated characteristic equation, which is studied numerically for stability and gives information on the parameter sensitivity for the model.  This local analysis provides the basis for examining Hopf bifurcations.
	
	The model from Sect.~\ref{subsec:summary} is condensed to two differential equations depending only on $P$ and $T$. The model equation for the platelets has the form
	\begin{equation}
	\frac{\mathrm{d}P}{\mathrm{d}t} = \frac{D_0V_m\kappa_PQ^*}{\beta_P} \exp\left[\int_{t - \tau_e - \tau_m}^{t - \tau_e} \! \eta_m(T(s)) \mathrm{d}s\right]
	\exp\left[\int_{t - \tau_e}^{t}\! \eta_e(T(s)) \, \mathrm{d}s\right] -\gamma_P P - F(P), \label{Pp}
	\end{equation}
	where
	\[
	F(P) = \alpha_P\frac{(P)^{n_P}}{(b_P)^{n_P} + (P)^{n_P}}.
	\]
	The model equation for the thrombopoietin is
	\begin{equation}
	\begin{split}
	\frac{\mathrm{d}T}{\mathrm{d}t} &= T_{prod} - \gamma_T T - \alpha_T\left(\int_0^{\tau_e} \! V_m\kappa_PQ^*\exp\left[\int_{t - a - \tau_m}^{t - a} \! \eta_m(T(s))\, \mathrm{d}s\right] \right.  \\
	&\quad \times\exp\left[\int_{t - a}^{t} \eta_e(T(s)) \, \mathrm{d}s\right] \, \mathrm{d}a + k_S\beta_P P\biggr)G(T),  \\
	\end{split} \label{Tp}
	\end{equation}
	where
	\[
	G(T) = \frac{(T)^{n_T}}{(k_T)^{n_T} + (T)^{n_T}}.
	\]
	
\subsection{Linearization about the single steady state} \label{subsec:local}
	
The study of a steady state solution begins by setting \eqref{Pp} and \eqref{Tp} equal to zero to determine the equilibrium solution $(P^*, T^*)$. The steady state solution of \eqref{Tp} readily gives $P^*$ depending on $T^*$ and is shown to be a function monotonically decreasing in $T^*$ from $+\infty$ to negative values for $T^* > 0$. This information is used in Eq.~\eqref{Pp},  where the decay terms are set equal to the production term. The monotonicity of the decay terms (decreasing in $T^*$) combined with the positively bounded monotonicity of the production terms (increasing in $T^*$) result in the existence of a unique positive equilibrium, $(P^*, T^*)$. Details of the proof are presented in Appendix~\ref{app:EU}.
	
The next step in the local analysis is linearizing Eqs.~\eqref{Pp} and \eqref{Tp} about the unique equilibrium $(P^*, T^*)$. See Appendix~\ref{app:linearization} for the details of this process. Let $x(t) = P(t)-P^*$ and $y(t) = T(t)-T^*$, and denote by $\partial_{P}$ and $\partial_{T}$ the partial derivatives with respect to the platelet and TPO variables, respectively. Linearizing Eq.~\eqref{Pp} about the equilibrium yields
\begin{equation}
\frac{\mathrm{d}x}{\mathrm{d}t} = A_2\left[\partial_{T}\eta_m(T^*)\int_{t - \tau_e - \tau_m}^{t - \tau_e} \! y(s) \, \mathrm{d}s
+ \partial_{T}\eta_e(T^*)\int_{t - \tau_e}^{t} \! y(s) \, \mathrm{d}s\right]
- \bigl(\gamma_P + \partial_{P}F(P^*)\bigr)x, \label{P2_lin}
\end{equation}
where
\begin{equation}
A_2 = \frac{D_0V_m\kappa_PQ^*}{\beta_P}\e^{\eta_m(T^*)\tau_m + \eta_e(T^*)\tau_e}.
\end{equation}
Linearizing Eq.~\eqref{Tp} about the equilibrium yields
\begin{equation}
\begin{split}
\frac{\mathrm{d}y}{\mathrm{d}t} &= -\alpha_T k_S\beta_PG(T^*)x - \bigl(\gamma_T + \alpha_T(A_1E_1 + k_S\beta_PP^*) \partial_{T}G(T^*)\bigr)y \\
&\quad- \alpha_TA_1G(T^*)\Biggl(\partial_{T}\eta_m(T^*)\int_0^{\tau_e} \! \e^{\eta_e(T^*)a}\left(\int_{t - a - \tau_m}^{t - a} \! y(s) \, \mathrm{d}s\right) \, \mathrm{d}a \\
&\quad+ \partial_{T}\eta_e(T^*)\int_0^{\tau_e} \! \e^{\eta_e(T^*)a}\left(\int_{t - a}^{t} y(s) \, \mathrm{d}s\right) \, \mathrm{d}a\Biggr), \\
\end{split}    \label{T2_lin}
\end{equation}
where
\begin{equation}
A_1 = V_m\kappa_PQ^*\e^{\eta_m(T^*)\tau_m} \qquad {\rm and} \qquad
E_1 = \frac{\e^{\eta_e(T^*)\tau_e} - 1}{\eta_e(T^*)}.
\end{equation}
	
\subsection{Characteristic equation}
\label{subsec:char_eqn}
	
The analysis above produced the linear functional equations in the variables $x(t)$ and $y(t)$, which are given by Eqs.~\eqref{P2_lin} and \eqref{T2_lin}. The linear functional equation is written as
\begin{equation}
\label{eq:lin_fde}
\frac{\mathbf{\mathrm{d}X}}{\mathrm{d}t} = \mathbf{L}(\mathbf{X}(t)), \qquad {\rm where} \qquad
\mathbf{X}(t) = \left(
\begin{array}{c}
x(t) \\
y(t) \\
\end{array}
\right).
\end{equation}
The characteristic equation is found by seeking solutions of the form
\[
\left(
\begin{array}{c}
x(t) \\
y(t) \\
\end{array}
\right) = \left(
\begin{array}{c}
c_1 \\
c_2 \\
\end{array}
\right)\e^{\lambda t}
\]
and inserting this into Eq.~\eqref{eq:lin_fde}. Using the results of Appendix~\ref{app:char_eqn} and dividing by $\e^{\lambda t}$, the linear system becomes
\[
\lambda \mathbf{I} \left(
\begin{array}{c}
c_1 \\
c_2 \\
\end{array}
\right) = \left(
\begin{array}{cc}
-L_1 & \phantom{-}L_2(\lambda) \\
\phantom{-}L_3 & -L_4(\lambda) \\
\end{array}
\right)\left(
\begin{array}{c}
c_1 \\
c_2 \\
\end{array}
\right).
\]
The coefficients $L_1$, $L_2(\lambda)$, $L_3$, and $L_4(\lambda)$ are given by
\begin{equation*}
\begin{split}
	L_1 &= \gamma_P + \partial_{P}F(P^*),\\
	L_2(\lambda) &= \frac{A_2}{\lambda}\left[\partial_{T}\eta_m(T^*)\e^{-\lambda \tau_e}
	\left(1 - \e^{-\lambda \tau_m}\right)
	+ \partial_{T}\eta_e(T^*)\left(1 - \e^{-\lambda \tau_e}\right)\right], \\
	L_3 &= -\alpha_T k_S\beta_PG(T^*), \rule[0.75cm]{0pt}{0pt} \\
	L_4(\lambda) &= C_1 + \frac{C_2}{\lambda}\Biggl[\partial_{T}\eta_m(T^*)
	\left(1 - \e^{-\lambda \tau_m}\right)
	\frac{\left(1 - \e^{-(\lambda-\eta_e(T^*)) \tau_e}\right)}{(\lambda-\eta_e(T^*))} \\
	&\quad+ \partial_{T}\eta_e(T^*)\left(\frac{\e^{\eta_e(T^*) \tau_e}-1}{\eta_e(T^*)} + \frac{\e^{-(\lambda-\eta_e(T^*))\tau_e}-1}{\lambda-\eta_e(T^*)}\right)\Biggr],
	\end{split}
\end{equation*}
where
\begin{equation*}
	C_1 = \gamma_T + \alpha_T(A_1E_1 + k_S\beta_PP^*)\partial_{T}G(T^*) \qquad {\rm and} \qquad C_2 = \alpha_T A_1 G(T^*).
\end{equation*}
	
Thus, the characteristic equation is 	
\begin{equation}
\label{ce}
\det\left|
\begin{array}{cc}
-L_1 - \lambda & L_2(\lambda) \\
L_3 & -L_4(\lambda) - \lambda \\
\end{array}
\right|
= (\lambda + L_1)(\lambda + L_4(\lambda)) - L_2(\lambda)L_3 = 0.
\end{equation}
Appendix~\ref{app:char_eqn} shows that this characteristic equation is a quartic in $\lambda$ with three distinct linear polynomials multiplying exponentials with $\lambda$ and the delays. This exponential polynomial is readily programmed with the model parameters, and numerical solutions to \eqref{ce} can be found. Specifically, we find the leading pair of complex eigenvalues, which allows for a stability analysis and to search for Hopf bifurcations.

\subsection{Parameter sensitivity of the model for healthy subjects}

Using the parameters from Table~\ref{table:para-summary} in the characteristic equation \eqref{ce}, the real and imaginary parts of the eigenvalues are found numerically. The leading pair of eigenvalues is given by $\lambda_1 = -0.058953 \pm 0.053015i$, which shows that the equilibrium state of the model is asymptotically stable.

Delay differential equations have characteristic polynomials with  infinitely many eigenvalues, and we proceeded to find the eigenvalues with the next largest real part, $\lambda_2 = -0.11375 \pm 0.3588i$. Later we show how this second pair of eigenvalues probably lead to the oscillations observed in the cyclic thrombocytopenia patients as parameters are varied.

To provide a measure of the parameter sensitivity of the eigenvalues of our model, we varied each model parameter by $\pm 10\%$ and computed how much the eigenvalues and equilibrium changed. See Tables~\ref{table:leading} and \ref{table:second} in Appendix~\ref{app:para-sense} for the eigenvalue and equilibrium computations for these parameter changes. The tables show that shifting any of the parameters by only 10\% cannot lead to a Hopf bifurcation. In fact, these small perturbations in the parameter values have very minimal effects on both the eigenvalues and the equilibrium. Thus, this model is extremely stable near the set of normal parameters.

Table~\ref{table:leading} of Appendix~\ref{app:para-sense} shows that the leading pair of eigenvalues $\lambda_1$ is most destabilized by (in descending order) increasing $b_P$, decreasing $\alpha_P$, decreasing $k_T$, increasing $\beta_P$, increasing $k_S$, increasing $b_m$, and decreasing $T_{prod}$. The greatest effect, however, only shifts the leading pair of eigenvalues by 11.3\%. Our study shows that changing these top seven parameters by 20\% only shifts the leading pair of eigenvalues to $\lambda_1 = -0.02555 \pm 0.06563i$, which still gives a stable equilibrium. It is surprising that varying the delays has little effect on the leading pair of eigenvalues $\lambda_1$.

The next largest eigenvalue, $\lambda_2$, are affected most by a different set of parameters as detailed in Table~\ref{table:second} of Appendix~\ref{app:para-sense}. A change of only 10\% in the parameters leads to at most a 6.3\% shift towards the loss of stability associated with the Hopf bifurcation. The most destabilizing changes for this pair of eigenvalues occur by (in descending order) increasing $\tau_e$, decreasing $b_m$, decreasing $k_S$, increasing $\beta_P$, increasing $\tau_m$, decreasing $b_P$, and increasing $\gamma_P$. Note here that the model delays are significant in changing the real part of the eigenvalues. A 20\% change in these top seven parameters shifts this pair of eigenvalues to $\lambda_2 = -0.09281 \pm 0.3091i$, which again yields a stable equilibrium. Interestingly, the frequency is moving closer to the frequencies observed in the oscillations in the cyclic thrombocytopenia patients.
	
\section{Application of the model to the study of cyclic thrombocytopenia}\label{sec:ctp}

Various modeling studies (see, e.g., \cite{apostu2008understanding,bernard2003analysis,colijn2005mathematicalI,colijn2005mathematicalII,colijn2007bifurcation,mahaffy1998hematopoietic,santillan2000regulation}) have associated oscillations in hematological diseases with a Hopf bifurcation induced by the change of one or more physiological parameters. In the context of CT, \citet{apostu2008understanding} found that changing the time for megakaryocyte maturity, reducing the relative growth rate of megakaryocytes, and increasing the random rate of destruction of platelets could generate platelet oscillations akin to those observed in CT. Their model, however, did not include an accurate description of the dynamics of thrombopoietin, megakaryoblasts, and megakaryocytes, and so it is unclear if their conclusions hold for the more physiologically realistic model presented here. In particular, the incorporation of a dynamic equation for thrombopoietin in our model could change these conclusions, as it is believed most platelet diseases, possibly including CT, arise due to disorders of TPO or its receptor \cite{hitchcock2014thrombopoietin}.

We revisit this issue here, and use our model to investigate the pathogenesis of CT and find for which parameters the model can generate oscillatory solutions similar to those observed in CT. We then use this knowledge to fit the model to various platelet and TPO data sets of patients with CT.

All but one of the patient data sets in our study were found to have statistically significant oscillations at the $\alpha=0.05$ confidence level or lower using the Lomb-Scargle periodogram technique in previous analyses \citep{apostu2008understanding,swinburne}. The one exception, the data from \citet{Connor11}, was published after \citep{apostu2008understanding} and \citep{swinburne}. Therefore, we performed our own Lomb-Scargle periodogram analysis and confirmed the presence of statistically significant oscillations at $\alpha=0.01$ (platelets) and $\alpha=0.05$ (TPO) confidence levels (data not shown).

\subsection{Parameter changes for generating periodic solutions} \label{subsec:generating_oscillations}

As discussed in Sect.~\ref{subsec:ctp_physio}, the clinical literature suggests that CT may be caused by immune-mediated platelet destruction (autoimmune CT), megakaryocyte deficiency and cyclic failure in platelet production (amegakaryocytic CT), or possible immune interference with or destruction of the TPO receptor. As a starting point for our analysis we identify the parameters of our model that, when modified, best reproduce these pathologies.

\begin{enumerate}
\item In the context of the model, we mimic an immune-mediated platelet destruction response by altering the parameters $\alpha_P$, which models the maximal platelet removal rate due to macrophages.

\item To replicate the effects of megakaryocyte deficiency and cyclic failure in platelet production, we change the value of $\tau_e$, the megakaryocyte proliferation duration, while keeping the total production of megakaryocytes, namely $\eta_e(T)\tau_e$, constant. Thus, whenever we scale $\tau_e$ by a factor of $a$, we scale $\eta_{e}^{min}$ and $\eta_{e}^{max}$ by a factor of $1/a$, thereby keeping $\eta_e(T)\tau_e$ constant. Increasing $\tau_e$ in this manner therefore amounts to reducing the rate of production of megakaryocytes, mimicking an ineffective rate of production of megakaryocytes.
\item Finally, changing $\alpha_T$ and $k_T$, the maximum clearance rate of thrombopoietin and TPO levels for half-maximal removal, respectively, replicate the possible interference with or destruction of the TPO receptor.
\end{enumerate}

In summary, based on clinical guidance we have identified the following four parameters as likely candidates for generating oscillations: $\alpha_P$,  $\tau_e$ (and indirectly $\eta_{e}^{min}$ and $\eta_{e}^{max}$), $\alpha_T$, and $k_T$.

\begin{figure}[h!]
	\centering
	\includegraphics[width=.45\textwidth]{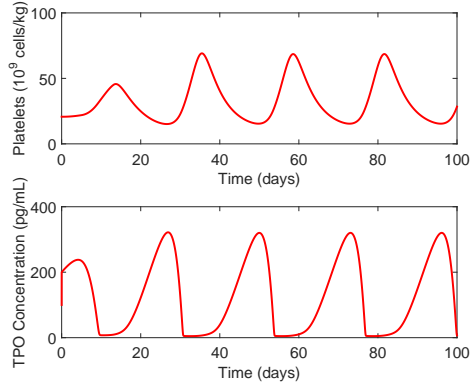}
	\caption{Oscillation in platelet counts (top) and thrombopoietin (bottom) generated by our model. All parameters are at normal, except for $\alpha_T$ and $k_T$ which are at $0.00075$ and $0.003$ times normal. The initial conditions for the model are $P(0)=P^*$ and $T(0)=T^* + 100$}
	\label{fig:alpha_t_and_k_t_only}
\end{figure}

Since most platelet diseases appear related to TPO or its receptor \cite{hitchcock2014thrombopoietin}, we first examined the effects of changing the values of $\alpha_T$ and $k_T$. We found that our model could generate oscillations when $\alpha_T$ and $k_T$ were significantly reduced. Oscillations were not generated when we kept $\alpha_T$ and $k_T$ at normal levels and changed $\alpha_P$ and $\tau_e$ alone. In Fig.~\ref{fig:alpha_t_and_k_t_only}, we show the oscillations generated by our model by setting $\alpha_T$ and $k_T$ to 0.075\% and 0.3\% their normal values, respectively. Alterations to the delay $\tau_e$ change the period of oscillations of both platelets and thrombopoietin, and modifying $\alpha_P$ changes the shape of oscillations of platelet and thrombopoietin levels (simulation data not shown).

\subsection{Fitting to platelet and thrombopoietin data} \label{subsec:fitting}

As discussed in the preceding section, our model can generate oscillations by significantly reducing the values of $\alpha_T$ and $k_T$. The shape and period of oscillations can be changed by modifying the values $\alpha_P$ and $\tau_e$. With this knowledge, we now show that our model can fit platelet and TPO patient data sets of patients with CT reported in the literature.

\begin{figure}[b!]
		\centering
		\label{fig:p_tpo_fits}
		\subfloat[][]{\label{fig:bruin_fit}\includegraphics[width=.42\textwidth]{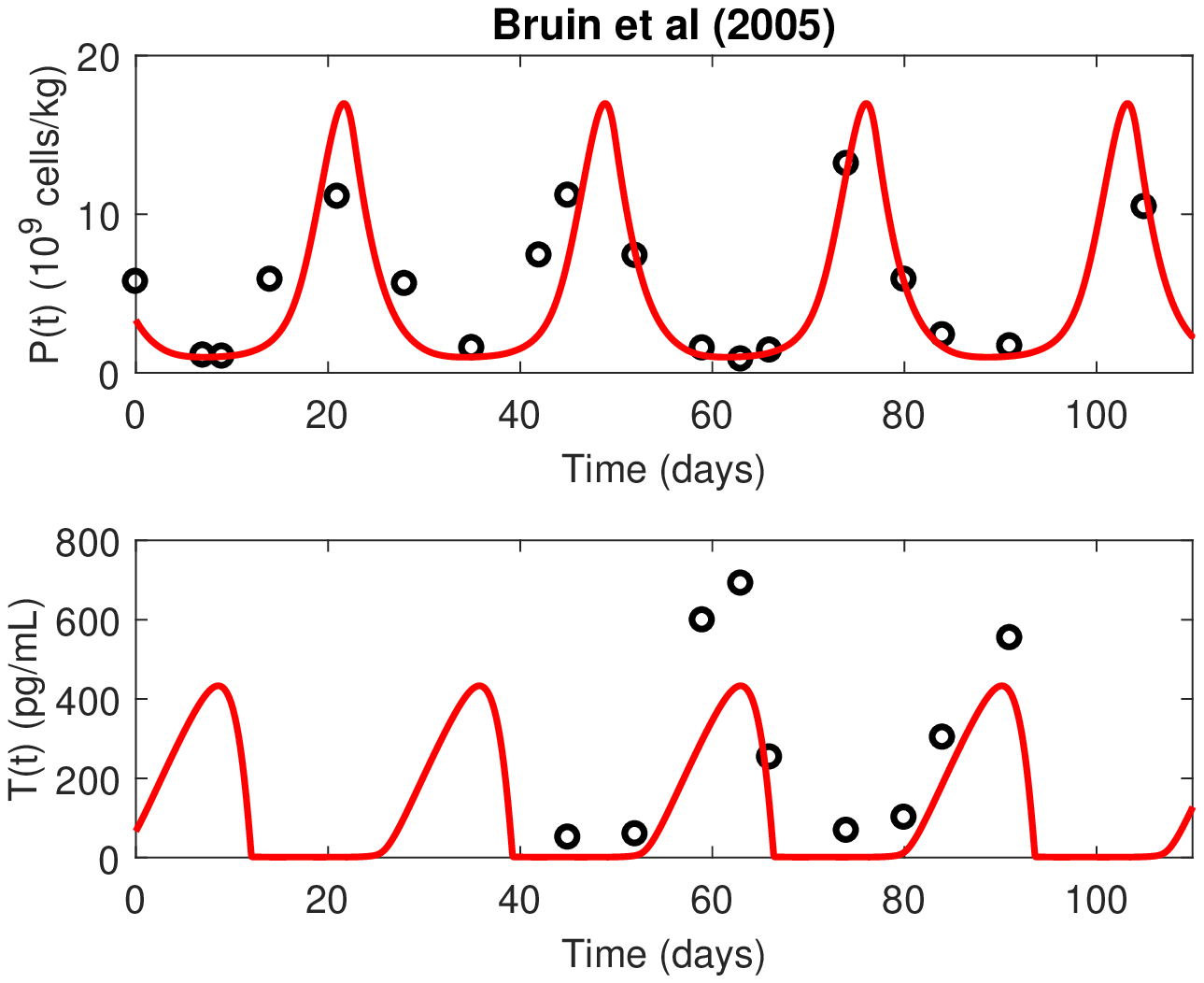}}
		\hspace{8pt}
		\subfloat[][]{\label{fig:connor_fit}\includegraphics[width=.42\textwidth]{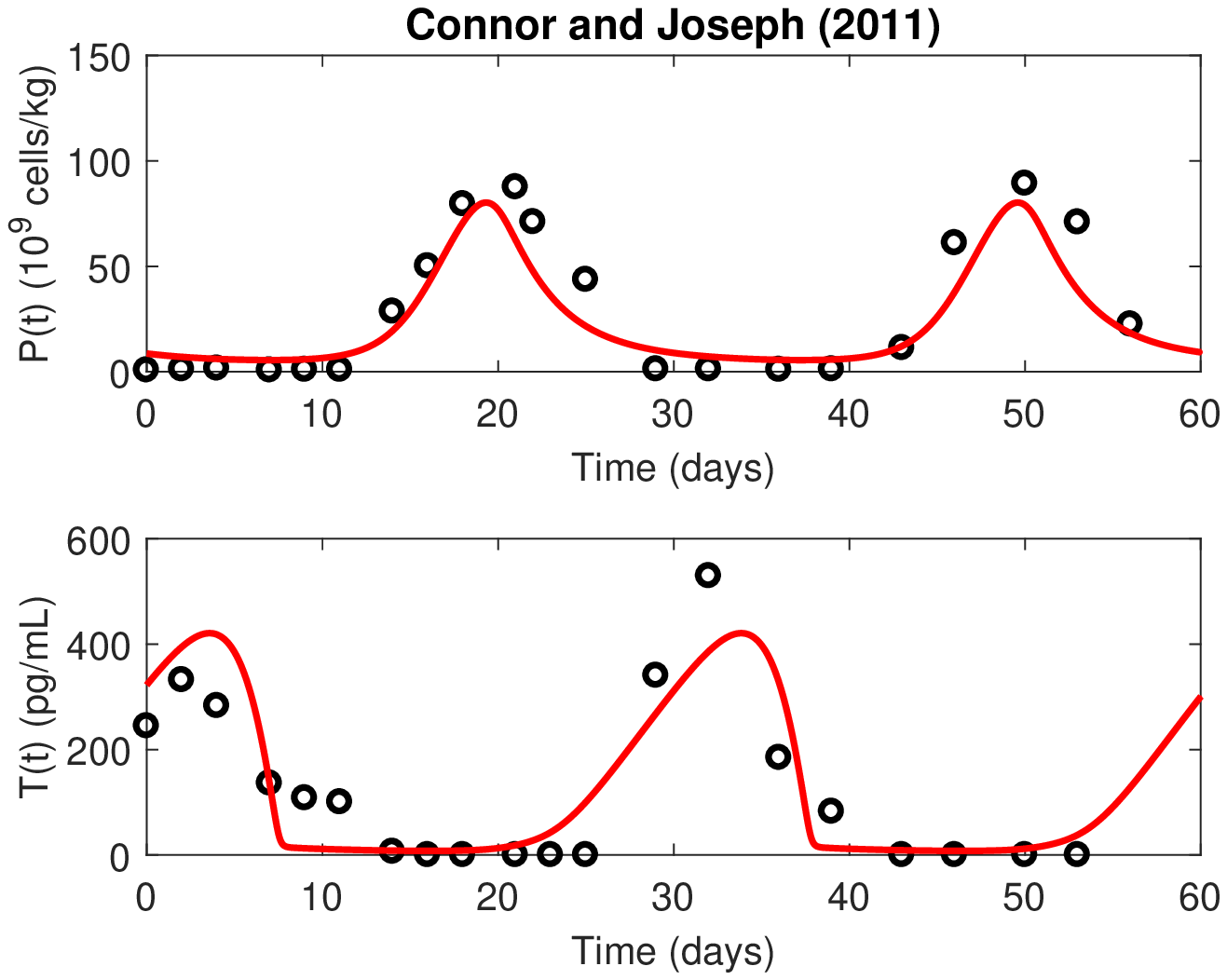}}
		
		\subfloat[][]{\label{fig:kimura_fit}\includegraphics[width=.42\textwidth]{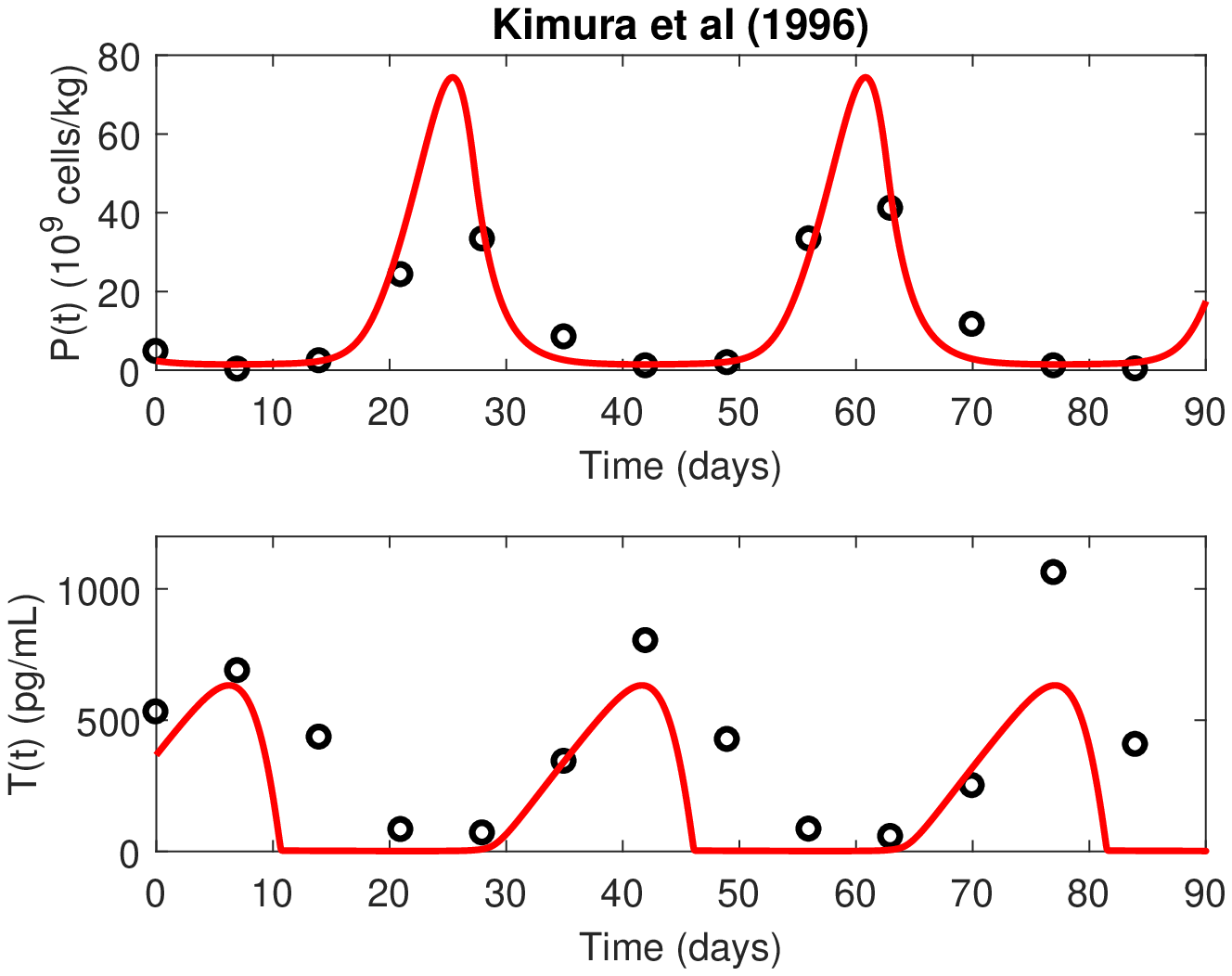}}
		\hspace{8pt}
		\subfloat[][]{\label{fig:zent_fit}\includegraphics[width=.42\textwidth]{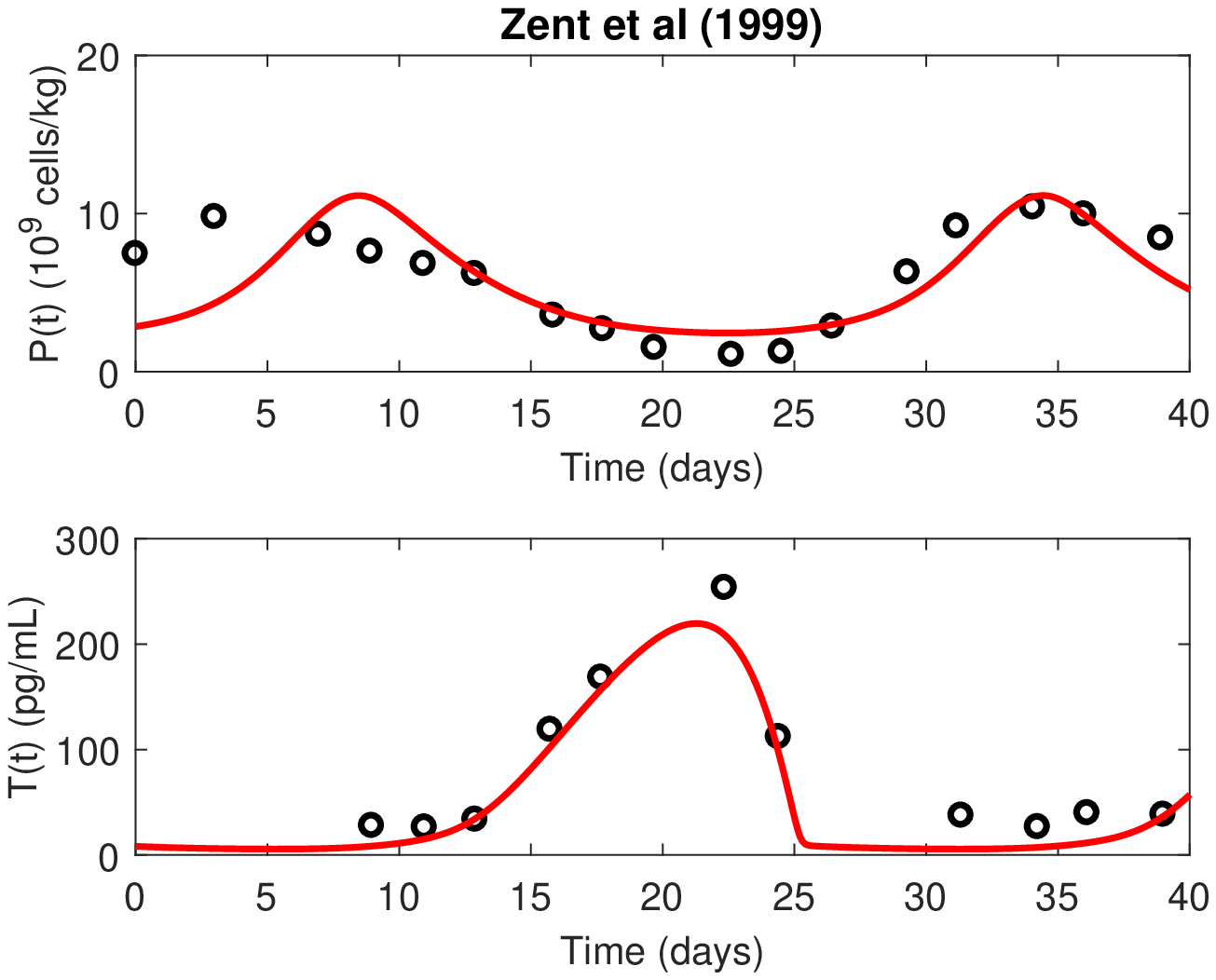}}
		\caption[Fits to Bruin, Connor, Kimura, and Zent data sets.]{Fits to the platelet and thrombopoietin data from:
			\subref{fig:bruin_fit} \citet{bruin};
			\subref{fig:connor_fit} \citet{Connor11};
			\subref{fig:kimura_fit} \citet{kimura96}; and
			\subref{fig:zent_fit} \citet{zent}   }
	\end{figure}

\begin{figure}
	\centering
	\includegraphics[scale=1]{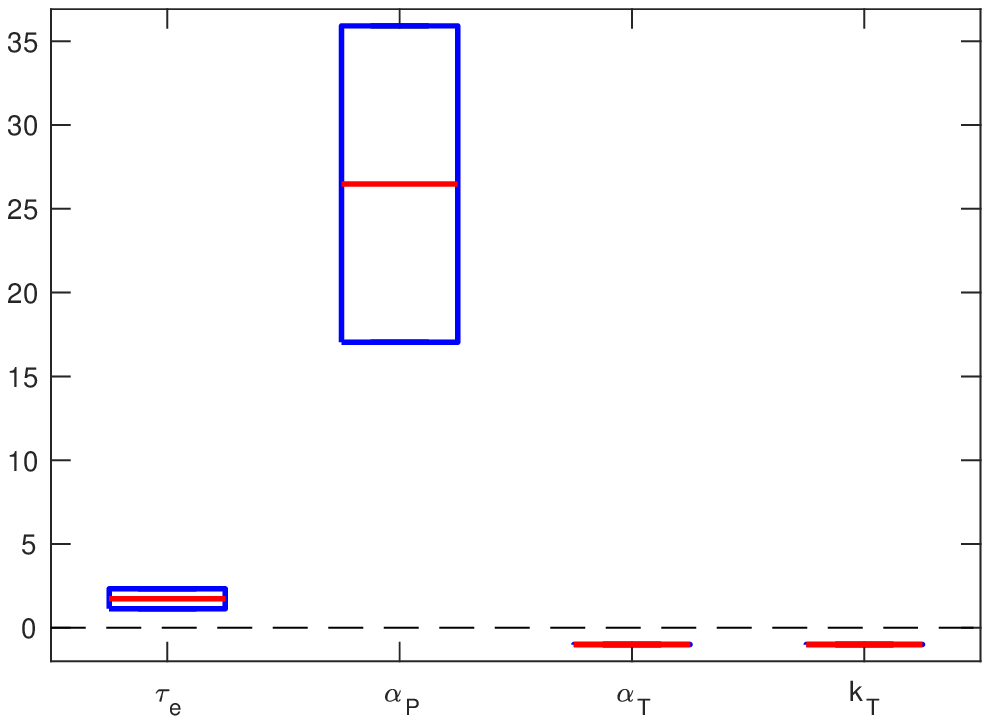}
	\caption{Box plots of the bootstrap confidence intervals (CIs) from fits of patients diagnosed with CT. If the boxplot of CI of the difference in the mean bootstrap estimate and 1 crosses the dashed line, we cannot reject the null hypothesis of no difference in mean relative error between the healthy and CT cases}
	\label{fig:Bootstraps}
\end{figure}

We fitted 15 patient data sets via a statistical procedure called the ABC method (see Appendix~\ref{app:ABC} for more details on the method). The fits are shown in Figs.~\ref{fig:bruin_fit}--\ref{fig:morley_fit}. The parameters changed to obtain these fits are shown in Tables~\ref{table:ct} and \ref{table:ct_relative}.

In every case, the parameters $\alpha_T$ and $k_T$ had to be decreased by a significant amount to obtain the fits (on average to 0.13512\% and 0.43521\% of the normal values of $\alpha_T$ and $k_T$, respectively). In all cases the maximal platelet removal rate had to be increased significantly (2140.9\% of normal, on average), with the delay $\tau_e$ also being increased but only by a moderate amount (236.42\% of normal, respectively).

\begin{sidewaystable}[h!]
	\begin{center}
		\noindent\resizebox{\textwidth}{!}{%
			\begin{tabular}{|c|c|c|c|c|c|c||c|c|c|c|}
				\hline
				Source & $\tau_e$ & $\alpha_P$ & $\alpha_T$ & $k_T$ & $P^*$ & $T^*$ & Diagnosis \tabularnewline
				\hline
				\hline
				Normal values & 5 & 213 & 144.9 & 3180 & 31.071 & 100 &--- \tabularnewline
				\hline
				\citet{bruin} & 10.552 & 13145 & 0.1365 & 3.8039 & 4.4547  & 90.92 & Amegakaryocytic CT \tabularnewline
				\citet{Connor11} & 12.595 & 726.41 & 0.0888 & 31.238 & 19.326 & 101.31 & CT \tabularnewline
				\citet{kimura96} & 16.491 & 5952.1 & 0.0165 & 8.2047 & 16.118 & 172.57 & Autoimmune CT  \tabularnewline
				\citet{zent} & 9.6100 & 2479 & 0.4082 & 13.366 & 5.1706 & 48.709 & Amegakaryocytic CT  \tabularnewline
				\hline
				\hline
				\citet{cohen} & 16.5105 & 5455.3 & 0.0888 & 15.228 & 6.805 & 91.332 & Amegakaryocytic CT \tabularnewline
				\citet{engstrom1966periodic} & 21.034 & 3303.7 & 0.041438 & 15.339 & 11.482 & 114.4 & Amegakaryocytic CT \tabularnewline
				\citet{helleberg1995cyclic} & 10.86 & 1253 & 0.33927 & 18.283 & 7.2045 & 51.727 & Autoimmune CT \tabularnewline
				\citet{Kosugi1994809} & 10.271 & 2955.4 & 0.55513 & 7.4199 & 3.7322 & 34.619 & Autoimmune CT \tabularnewline
				\citet{morley1969platelet} & 9.0350 & 212.95 & 0.2513 & 42.825 & 19.162 & 70.831 & Healthy \tabularnewline
				\citet{rocha1991danazol} & 7.8029 & 7058.8 & 0.15347 & 11.103 & 6.5286 & 97.635 & Autoimmune CT \tabularnewline
				\citet{vonschulthess1986} (Case 1) & 4.7713 & 1268.1 & 0.4565 & 8.2575 & 8.2781 & 60.142 & Healthy \tabularnewline
				\citet{vonschulthess1986} (Case 2) & 5.9465 & 81.666 & 0.2185 & 2.3984 & 24.211 & 69.391 & Healthy \tabularnewline
				\citet{skoog1957metabolic} & 10.32 & 9343.7 & 0.10981 & 6.3122 & 5.9655 & 101.16 & Autoimmune CT \tabularnewline
				\citet{wilkinson1966idiopathic} & 24.136 & 5517.8 & 0.039057 & 13.648 & 8.6759 & 111.29 & Amegakaryocytic CT \tabularnewline
				\citet{Yanabu1993155} & 7.381 & 9634.3 & 0.033121 & 10.174 & 13.358 & 177.63 & Autoimmune CT\tabularnewline
				\hline
				Average $\pm$ SD & $11.821 \pm 5.483$ & $4559.1 \pm 3929.2$ & $0.19576 \pm 0.17092$ & $13.840 \pm 10.625$ & $10.698 \pm 6.329$ & $92.911 \pm 41.103$ & --- \tabularnewline
				\hline
			\end{tabular}
		}
	\end{center}
	\caption{Parameter estimates for CT data.  The fits above the double line are for patients in which both platelet counts and thrombopoietin concentrations were available.  The two columns on the right for $P^*$ and $T^*$ are not fits but rather predicted values from the model. All numbers are displayed up to five significant digits}\label{table:ct}
	
	\bigskip
	\clearpage
	\begin{center}
		\noindent\resizebox{\textwidth}{!}{%
			\begin{tabular}{|c|c|c|c|c|c|c||c|c|c|c|}
				\hline
				Source & $\tau_e$ & $\alpha_P$ & $\alpha_T$ & $k_T$ & $P^*$ & $T^*$ & Diagnosis \tabularnewline
				\hline
				\hline
				\citet{bruin} & 2.1105 & 61.725 & 0.00094187 & 0.0011962 & 0.14337 & 0.9092 & Amegakaryocytic CT \tabularnewline
				\citet{Connor11} & 2.519 & 3.4111 & 0.00061293 & 0.0098234 & 0.62199 & 1.0131 & CT \tabularnewline
				\citet{kimura96} & 3.2981 & 27.95 & 0.00011414 & 0.0025801 & 0.51874 & 1.7257 & Autoimmune CT  \tabularnewline
				\citet{zent} & 1.922 & 11.641 & 0.0028172 & 0.0042033 & 0.16641 & 0.48709 &  Amegakaryocytic CT \tabularnewline
				\hline
				\hline
				\citet{cohen} & 3.3021 & 25.617 & 0.00061284 & 0.0047887 & 0.21901 & 0.91332 & Amegakaryocytic CT \tabularnewline
				\citet{engstrom1966periodic} & 4.2069 & 15.514 & 0.00028602 & 0.0048236 & 0.36954 & 1.144 & Amegakaryocytic CT \tabularnewline
				\citet{helleberg1995cyclic} & 2.172 & 5.8837 & 0.0023417 & 0.0057494 & 0.2319 & 0.51727 &  Autoimmune CT \tabularnewline
				\citet{Kosugi1994809} & 2.0543 & 13.878 & 0.0038317 & 0.0023308 & 0.12012 & 0.34619 & Autoimmune CT \tabularnewline
				\citet{morley1969platelet} & 1.807 & 1.0 & 0.0017347 & 0.013467 & 0.61671 & 0.70831 & Healthy \tabularnewline
				\citet{rocha1991danazol} & 1.5606 & 33.147 & 0.0010593 & 0.0034917 & 0.21011 & 0.97635 & Autoimmune CT \tabularnewline
				\citet{vonschulthess1986} (Case 1) & 0.95425 & 5.9549 & 0.0031508 & 0.0025967 & 0.26642 & 0.60142 & Healthy \tabularnewline
				\citet{vonschulthess1986} (Case 2) & 1.1893 & 0.38349 & 0.0015082 & 0.00075421 & 0.7792 & 0.69391 & Healthy \tabularnewline
				\citet{skoog1957metabolic} & 2.0639 & 43.876 & 0.00075797 & 0.001985 & 0.19199 & 1.0116 & Autoimmune CT \tabularnewline
				\citet{wilkinson1966idiopathic} & 4.8272 & 25.911 & 0.00026959 & 0.0042918 & 0.27922 & 1.1129 &  Amegakaryocytic CT \tabularnewline
				\citet{Yanabu1993155} & 1.4762 & 45.241 & 0.00022861 & 0.0031995 & 0.42992 & 1.7763 & Autoimmune CT\tabularnewline
				\hline
				Average $\pm$ SD & $2.3642 \pm 1.0965$ & $21.409 \pm 18.451$ & $0.0013512 \pm 0.0011797$ & $0.0043521 \pm 0.0033412$ & $0.34431 \pm 0.20371$ & $0.92911 \pm 0.41103$ & --- \tabularnewline
				\hline
			\end{tabular}
		}
	\end{center}
	\caption{Relative changes of parameters to normal values. All else as in Table \ref{table:ct}.  }\label{table:ct_relative}
\end{sidewaystable}

\clearpage
	\clearpage
	\begin{figure}[h]
		\centering
		\label{fig:p_fits1}
		\subfloat[][]{\label{fig:cohen_fit}\includegraphics[width=.46\textwidth]{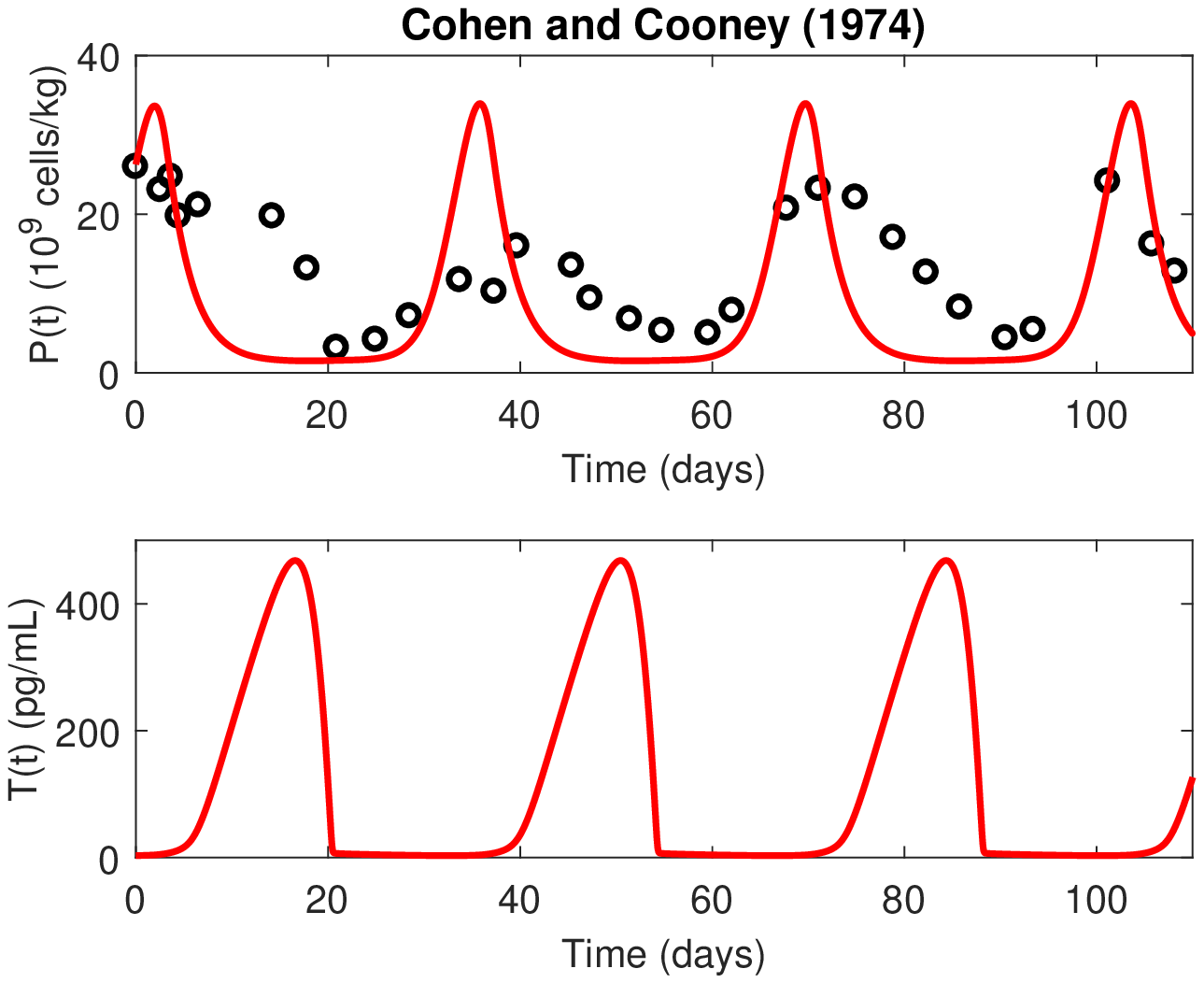}}
		\hspace{8pt}
		\subfloat[][]{\label{fig:engstrom_fit}\includegraphics[width=.46\textwidth]{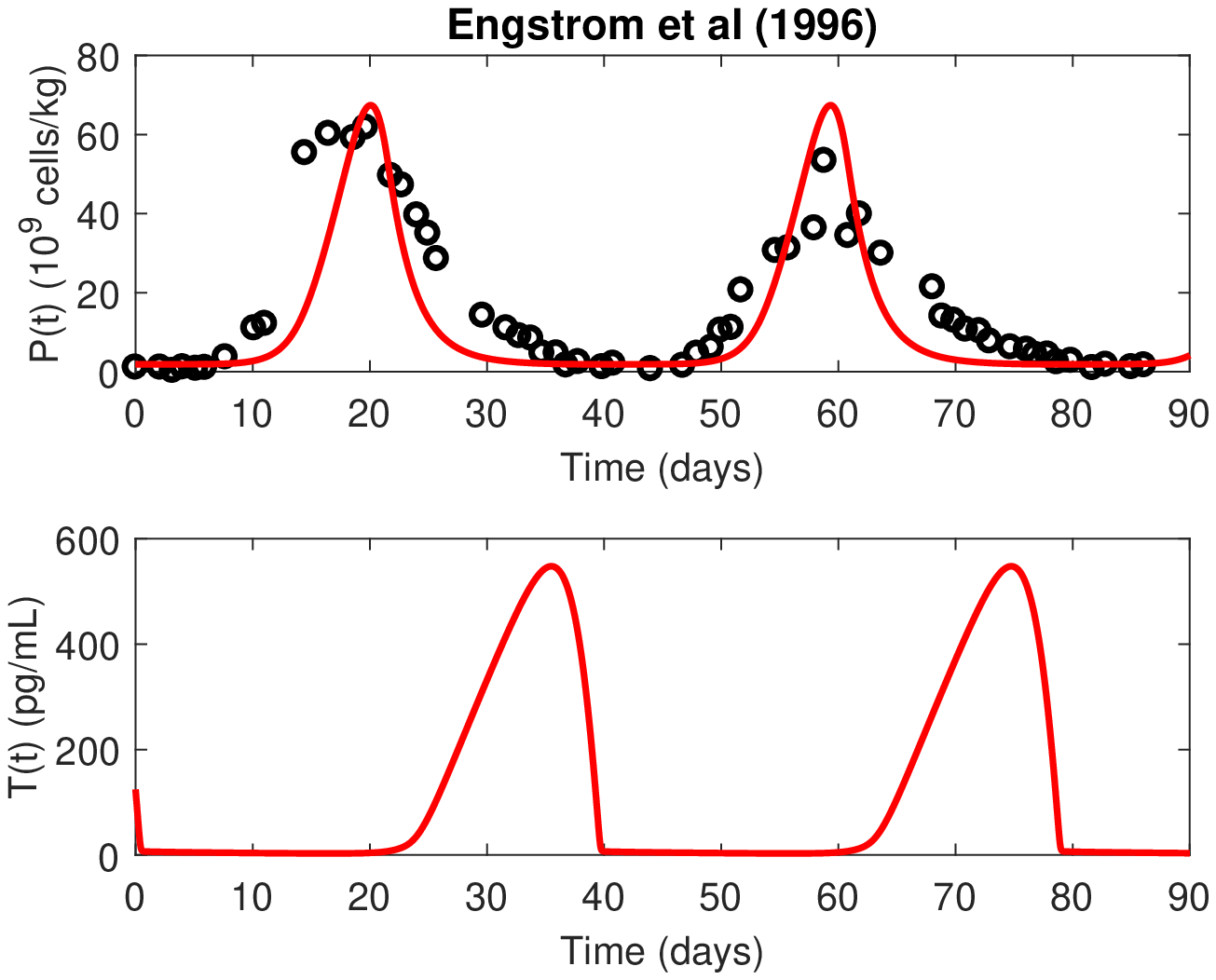}}
		
		\subfloat[][]{\label{fig:helleberg_fit}\includegraphics[width=.46\textwidth]{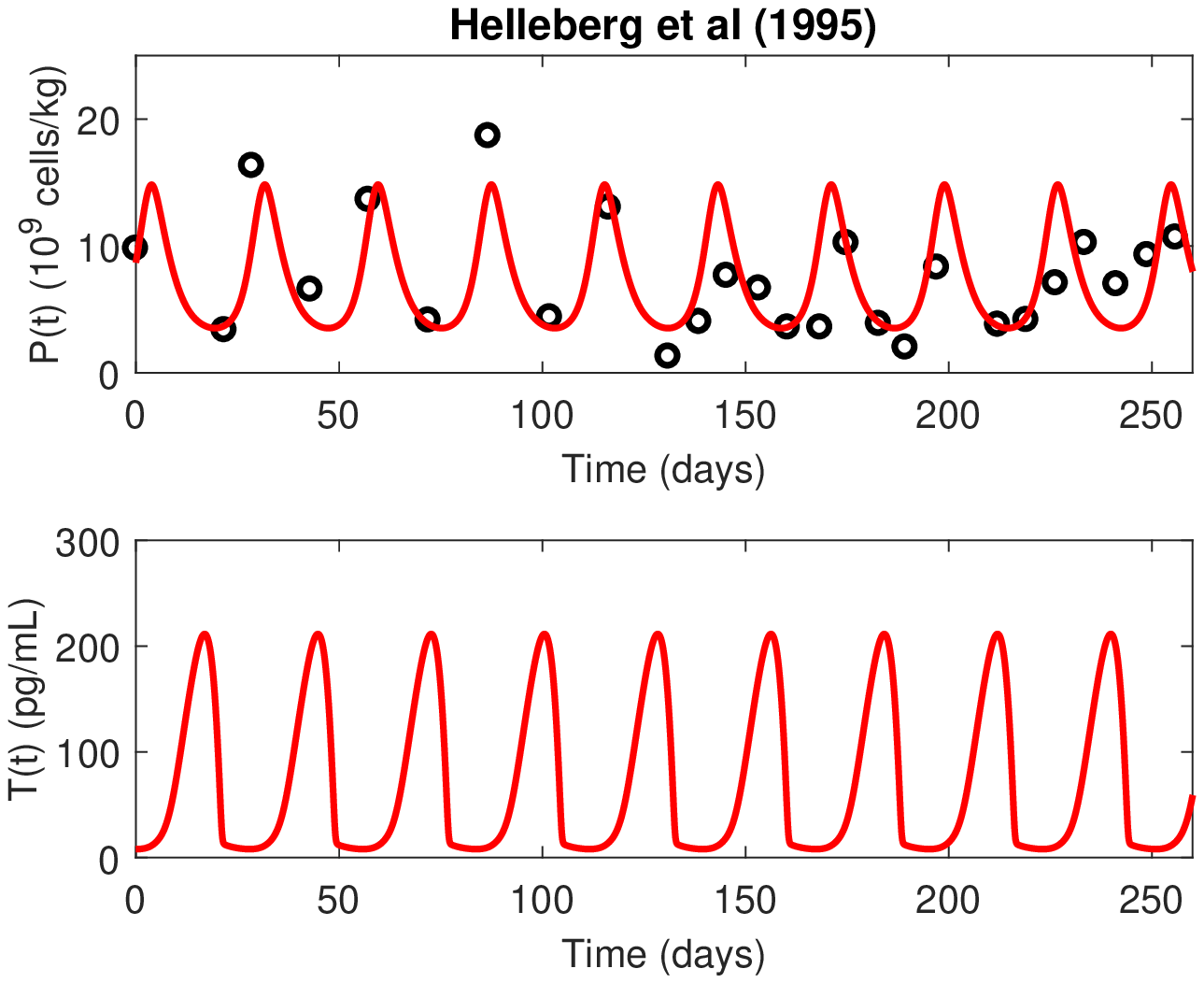}}
		\hspace{8pt}
		\subfloat[][]{\label{fig:kosugi_fit}\includegraphics[width=.46\textwidth]{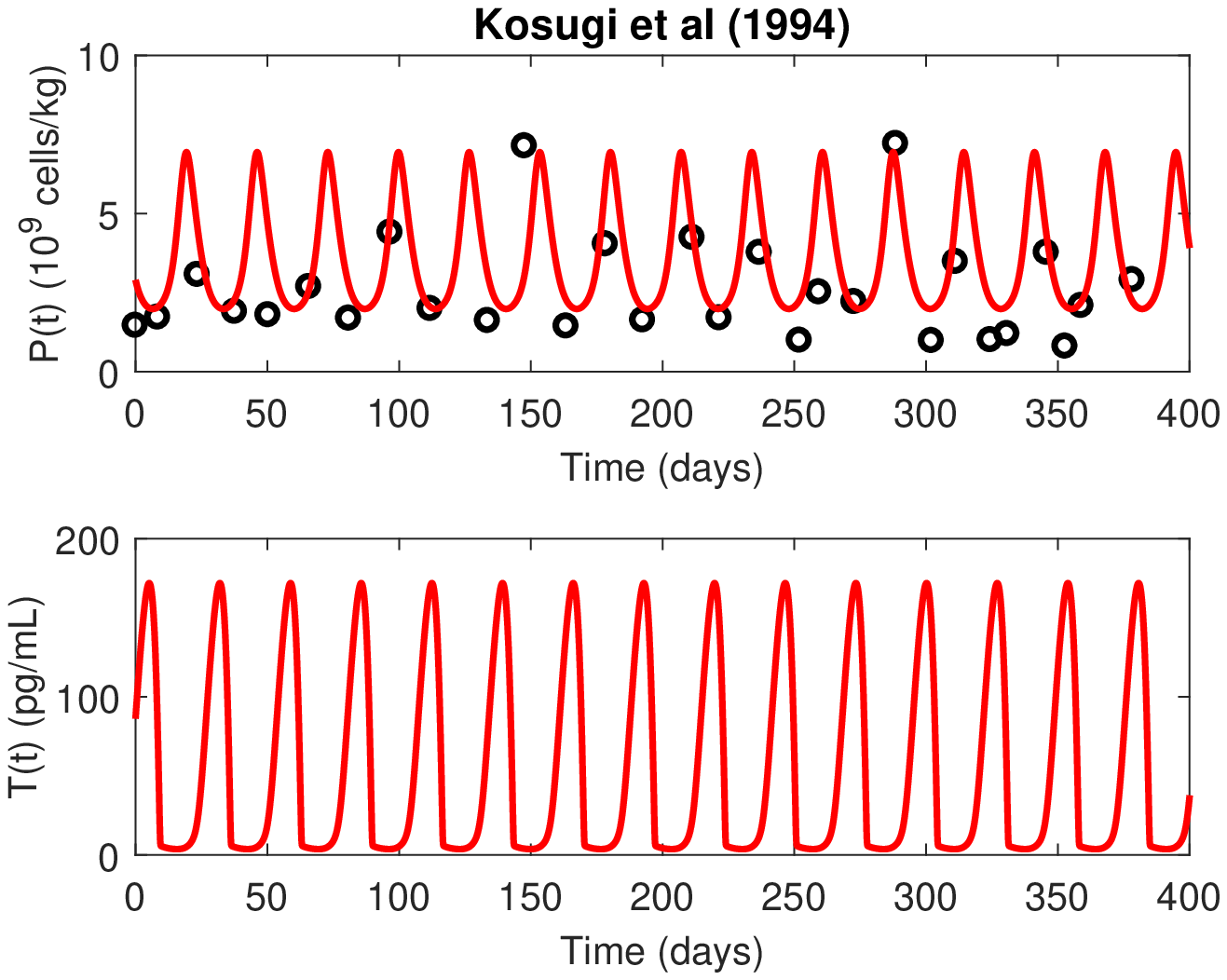}}
		
		\subfloat[][]{\label{fig:rocha_fit}\includegraphics[width=.46\textwidth]{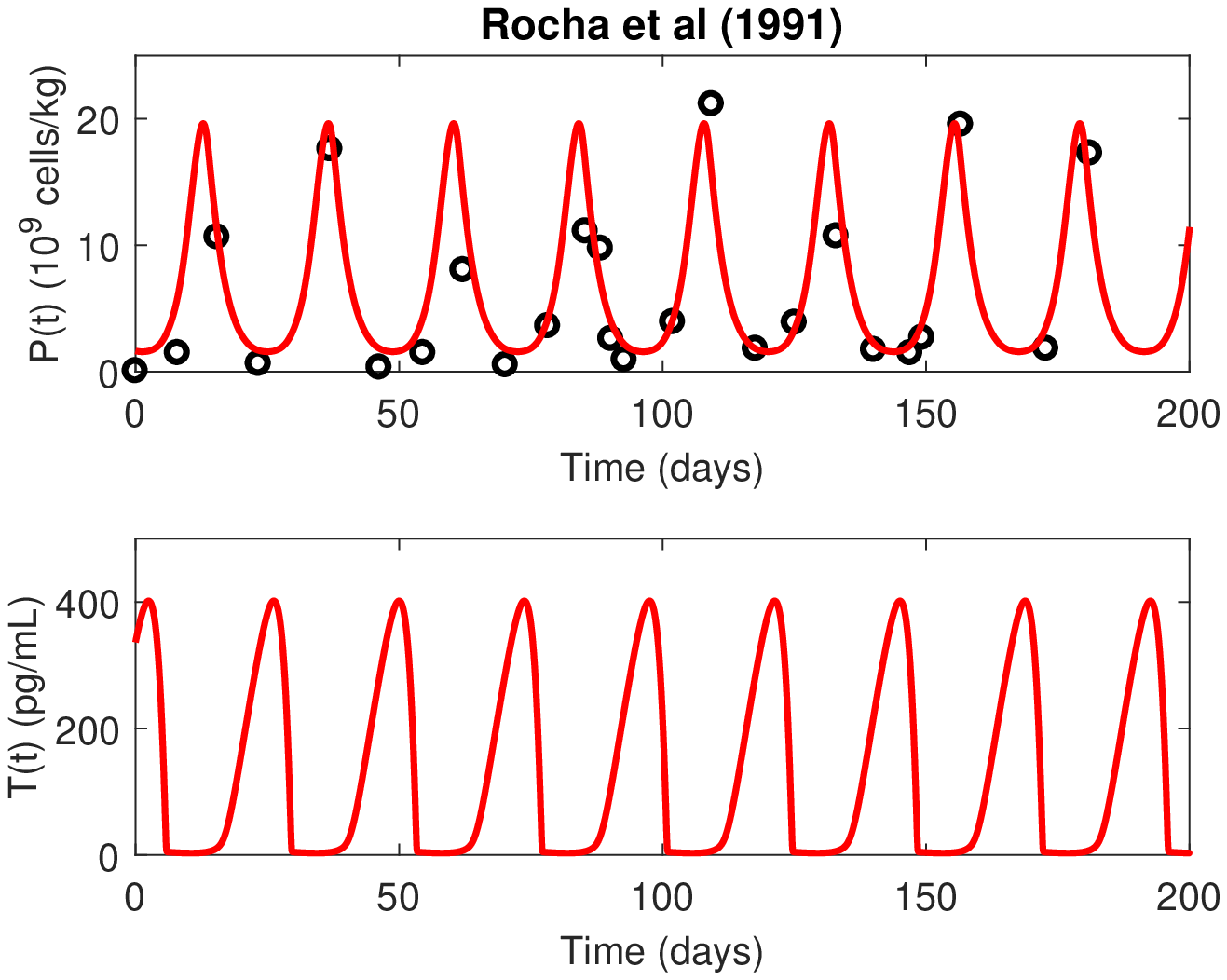}}%
		\hspace{8pt}%
		\subfloat[][]{\label{fig:skoog_fit}\includegraphics[width=.46\textwidth]{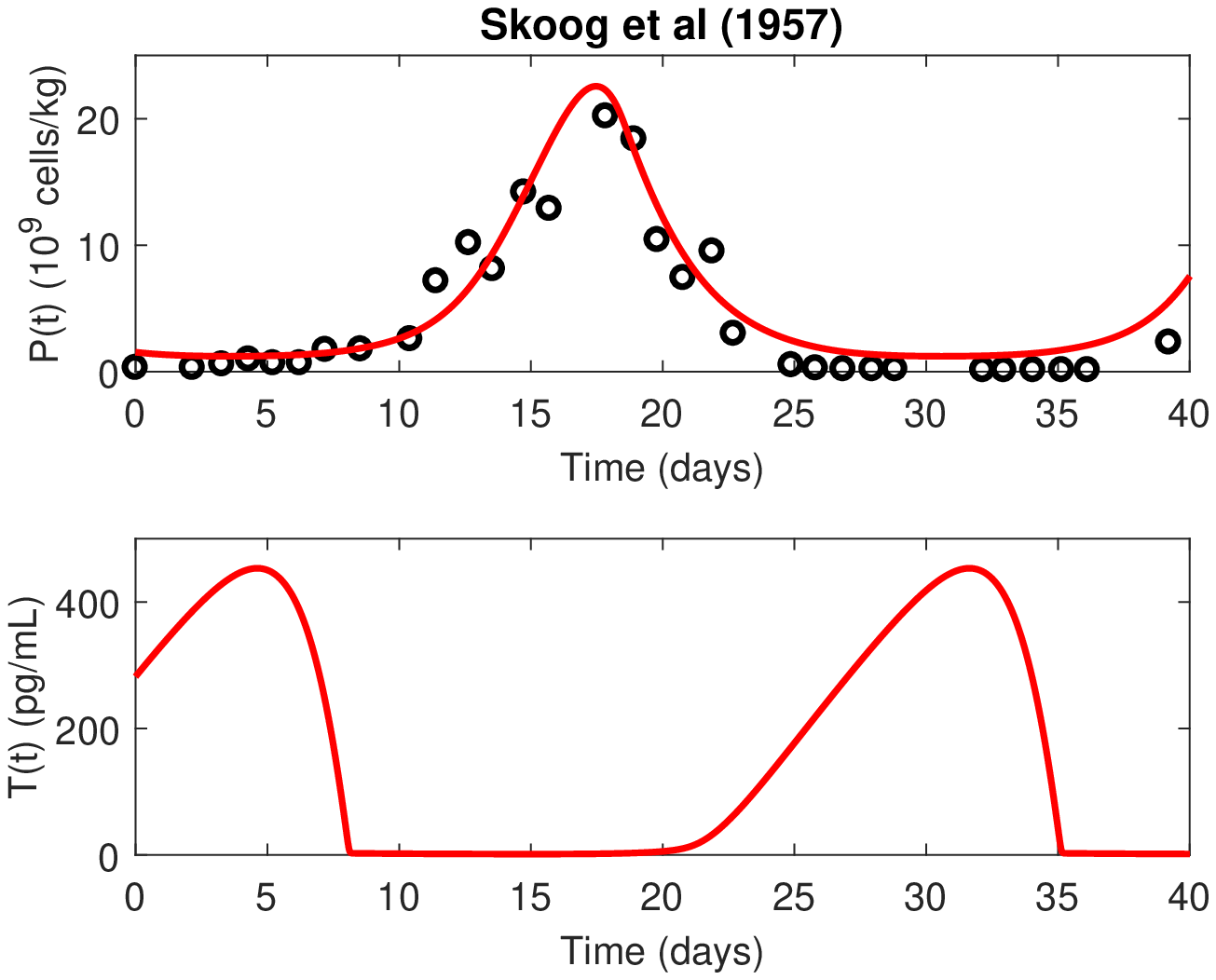}}%
		\caption[Fits to Cohen, Engstrom, Helleberg, Kosugi, Rocha, and Skoog data sets.]{Fits to the platelet data from:
			\subref{fig:cohen_fit} \citet{cohen};
			\subref{fig:engstrom_fit} \citet{engstrom1966periodic};
			\subref{fig:helleberg_fit} \citet{helleberg1995cyclic};
			\subref{fig:kosugi_fit} \citet{Kosugi1994809};
			\subref{fig:rocha_fit} \citet{rocha1991danazol}; and
			\subref{fig:skoog_fit} \citet{skoog1957metabolic}. Below each of the fitted platelet data we show the predicted behavior of the thrombopoietin levels (which were not available for these patients)}
	\end{figure}
\clearpage
	\begin{figure}[t!]
		\ContinuedFloat
		\centering
		\subfloat[][]{\label{fig:wilkinson_fit}\includegraphics[width=.48\textwidth]{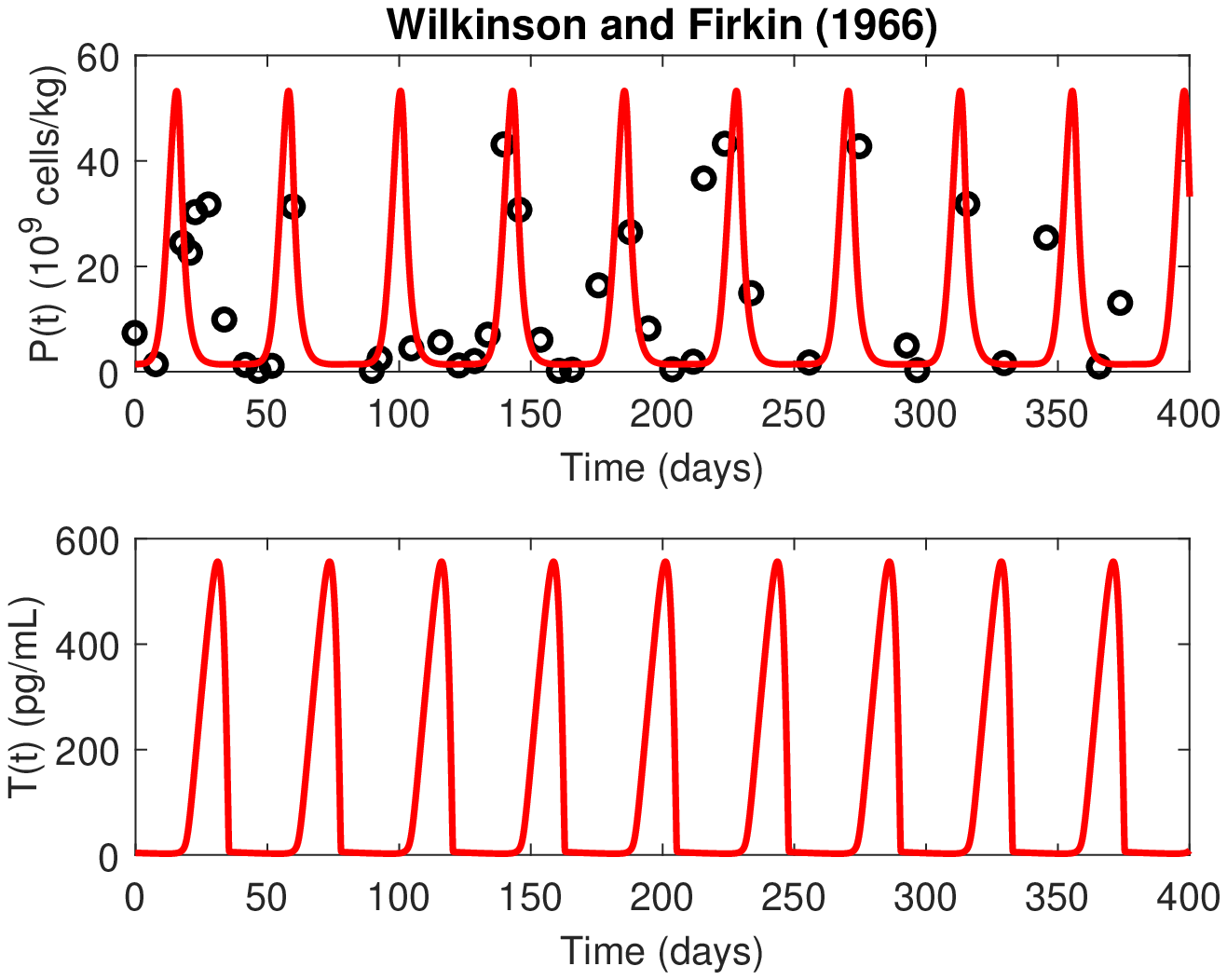}}%
		\subfloat[][]{\label{fig:yanabu_fit}\includegraphics[width=.48\textwidth]{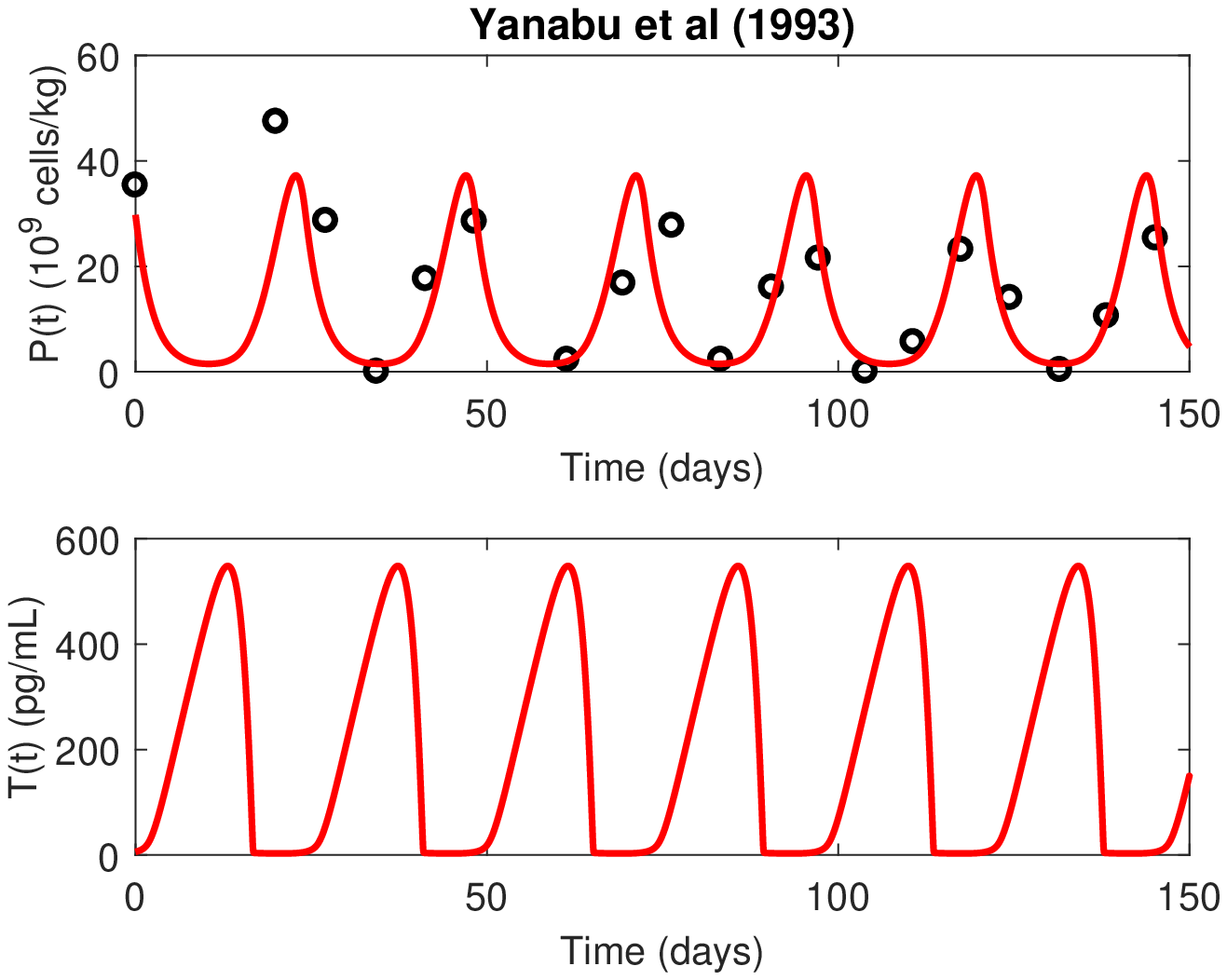}}%
		\caption[Fits to Wilkinson and Yanabu data sets.]{Fits to the platelet data from: %
			\subref{fig:wilkinson_fit} \citet{wilkinson1966idiopathic} and
			\subref{fig:yanabu_fit} \citet{Yanabu1993155}}%
	\end{figure}
	
To quantify the significance of the parameter changes required in the cases of patients diagnosed with CT, we used bootstrapping resampling techniques, which require no assumptions on the underlying distribution. To perform the bootstrapping, we used the \textit{bootci} function in MATLAB \cite{matlab}, which returns the sample estimates and computes (1-$\alpha$)\% bootstrap confidence intervals (CIs). CIs were computed on the difference in mean relative errors, as explained below. This construction implies that if a resulting CI contained 0, we fail to reject the null hypothesis that there is no difference in means. In this case, we conclude that there is no statistically significant difference in the parameter value for a healthy individual versus one with CT.

Using the average relative difference for each of the parameter values as given in Table~\ref{table:ct_relative}, we considered the difference between the reported value and 1 (since a relative change of 1 indicates no difference between the healthy individual and the CT case). We then generated 10000 bootstrap estimates and computed the bootstrap CI interval about the samples' mean relative differences minus 1 for each parameter of interest. The results of this analysis are given in Table~\ref{tab:Bootstraps}, alongside the difference in the average relative change of each parameter of both the fitting and bootstrap estimates and 1. In all cases, the value of the relative change for the estimates from the fitting procedure of Sect.~\ref{subsec:fitting} and the bootstrap samples are similar (Columns 2 and 3), indicating that a sufficient number of samples was generated. None of the CIs contain 0 and therefore we reject the null hypothesis and conclude that there are statistically significant differences at the $\alpha = 0.05$ level in all cases. The resulting bootstrap confidence intervals are also reflected in Fig.~\ref{fig:Bootstraps}, where the failure to reject the null corresponds to CIs which cross the x-axis. As evidenced by the results in Table~\ref{tab:Bootstraps} and Fig.~\ref{fig:Bootstraps}, both $\alpha_T$ and $k_T$ have particularly narrow bootstrap CIs, which suggest a higher degree of certainty in those cases. Since we reject the null hypothesis of no difference in means for these two parameters, the narrow CIs suggest that we are confident that there are significant differences between the CT and the healthy case. This leads us to believe that there may be an alteration in the TPO receptor or the interaction of TPO with the platelet lineage in patients with CT, but much more clinical investigation is required to substantiate this conclusion.

\begin{table}[h!]
	\centering
	\begin{tabular}{|c|c|c|l|}
		\hline
		& Difference of average & Difference of average& 95\% bootstrap CI\\
		&relative change and 1 (fit values) & relative change and 1 (bootstrap values)  &\\
		\hline
		$\tau_e$   & 1.6261 & 1.6269 & $[1.1418, 2.3241]$ \\
		$\alpha_P$ & 25.1496 &  25.1180 & $[17.0382, 25.9115]$ \\
		$\alpha_T$ & -0.9988& -0.9988 & $[-0.9993, -0.9980]$ \\
		$k_T$      & -0.9960 & -0.9960 & $[-0.9969, -0.9943]$ \\
		\hline
	\end{tabular}
	\caption{Differences of relative changes and 1 for parameter values from fits of patients diagnosed with CT and the bootstrap samples, and bootstrap 95\% confidence intervals. Column 2: For each parameter fit in the cyclic thrombocytopenic case, the difference in its relative change and 1 was calculated. Column 3: 10000 bootstrap samples were generated and the difference in the mean relative change and 1 were calculated. Column 4: the 95\% bootstrap confidence interval. CI: confidence interval}
	\label{tab:Bootstraps}
\end{table}

Based on our numerical experiments and the results, the platelet and thrombopoietin oscillations in the model occur due to a destabilization of the TPO control mechanism, in conjunction to an increased platelet-dependent removal rate and reduced megakaryocyte production. Though the relative change of the parameters $\alpha_T$ and $k_T$ with the normal parameters is very large, our results are nonetheless consistent with the clinical literature on CT.

\subsection{Platelet oscillations in healthy subjects}

\begin{figure}[h!]
	\centering
	\subfloat[][]{\label{fig:schulthess_case1_fit}\includegraphics[width=.48\textwidth]{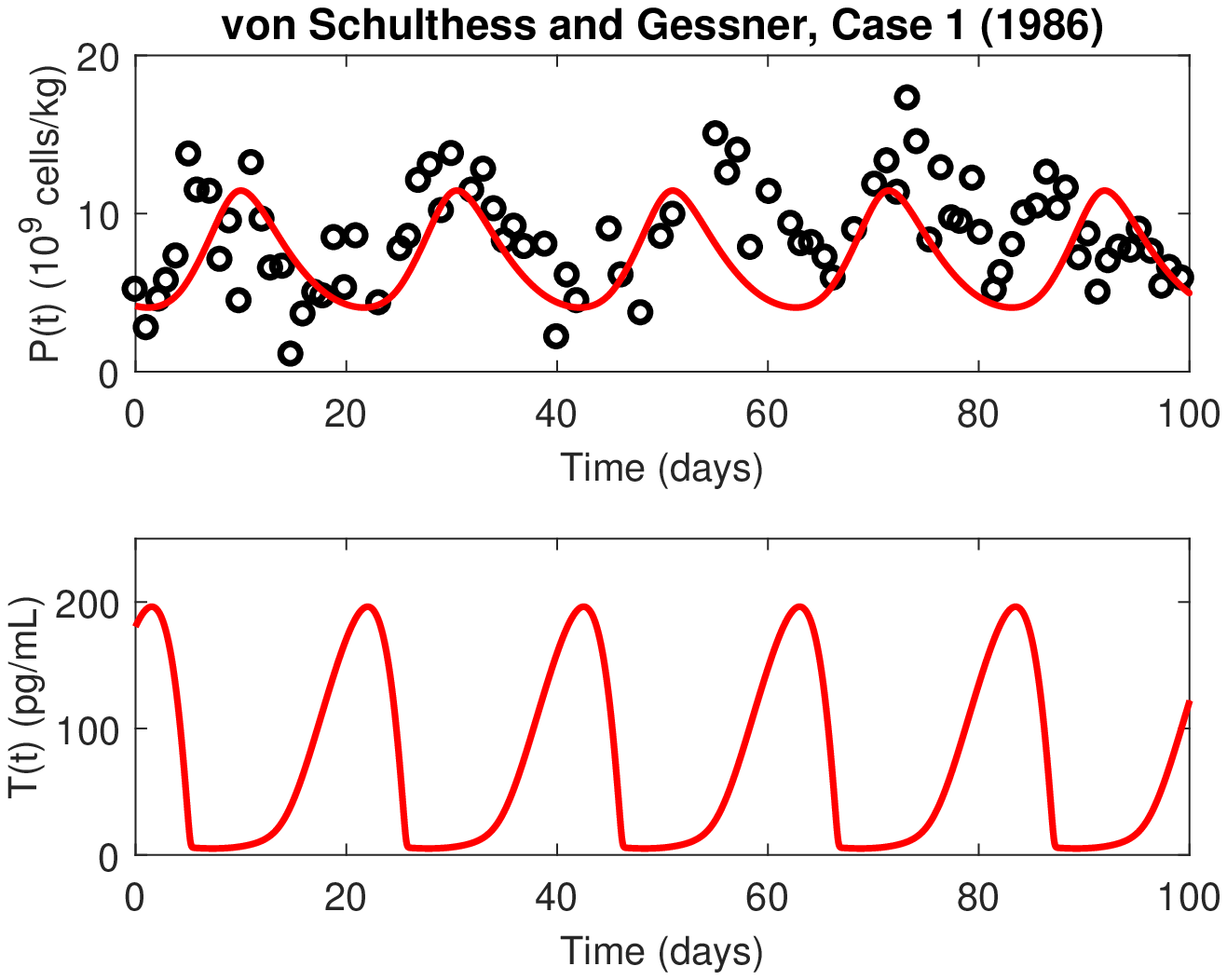}}%
	\hspace{8pt}%
	\subfloat[][]{\label{fig:schulthess_case2_fit}\includegraphics[width=.48\textwidth]{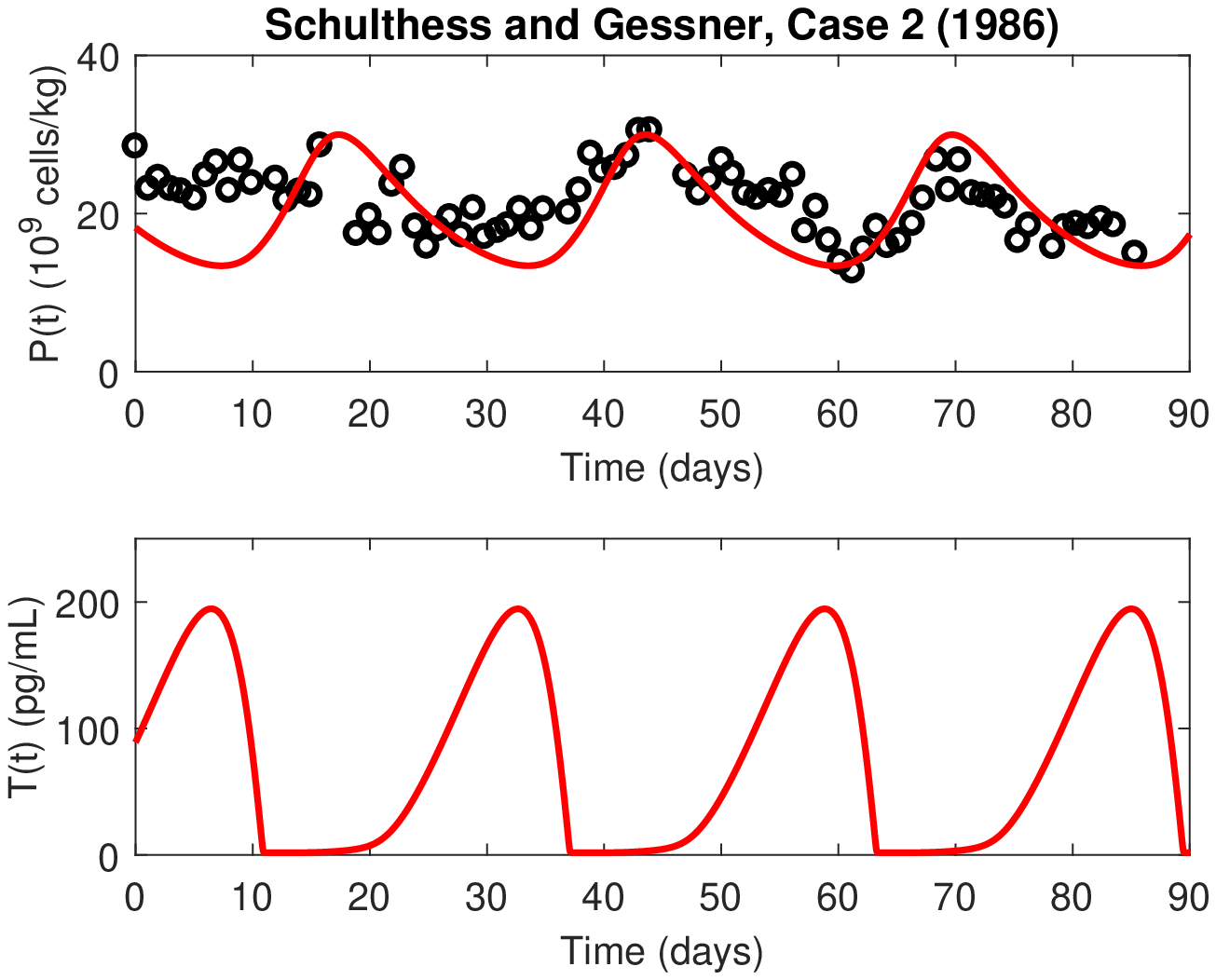}}%
	
	\subfloat[][]{\label{fig:morley_fit}\includegraphics[width=.48\textwidth]{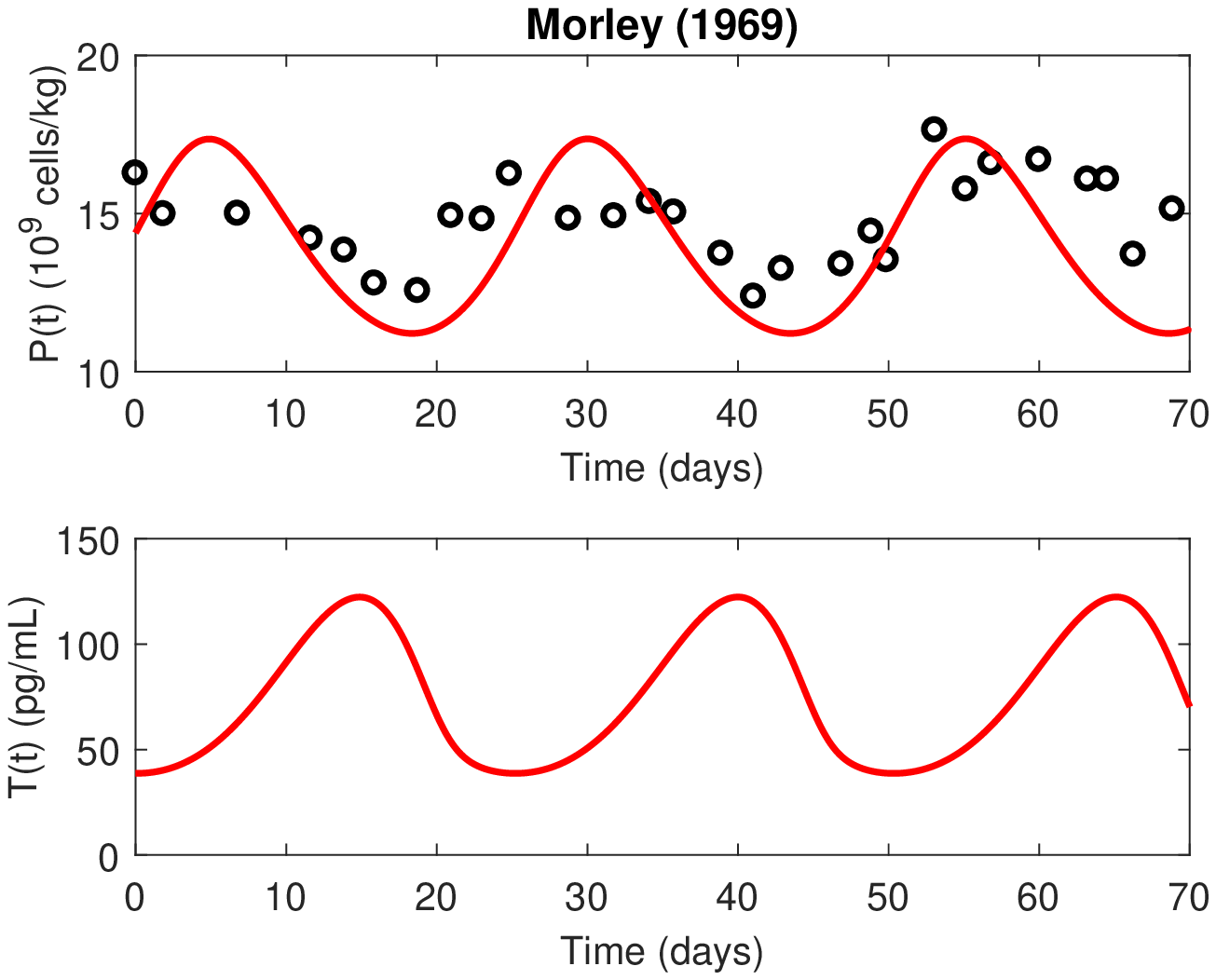}}%
	\caption[Fits to von Schulthess and Morley data]{Fits to the platelet data from:
		\subref{fig:schulthess_case1_fit} \citet{vonschulthess1986};
		\subref{fig:schulthess_case2_fit} \citet{vonschulthess1986};
		\subref{fig:morley_fit} \citet{morley1969platelet}.  Again we show the predicted TPO variation though the data were not available}\label{fig:normal_fits}
\end{figure}

We have also identified three published data sets indicating significant oscillations in platelets in apparently healthy male subjects without any obvious platelet pathology \citep{morley1969platelet}, \citep{vonschulthess1986}.  Interestingly in all three of these documented cases the oscillations are in the normal range of platelet levels.  We were able to fit the model to these data with changes in the parameters $\tau_e$, $\alpha_P$, $\alpha_T$, and $k_T$ (see Table~\ref{table:ct} and \ref{table:ct_relative}) and the results of our fits are shown in Fig.~\ref{fig:normal_fits}.

\begin{figure}
	\centering
	\includegraphics[scale=1]{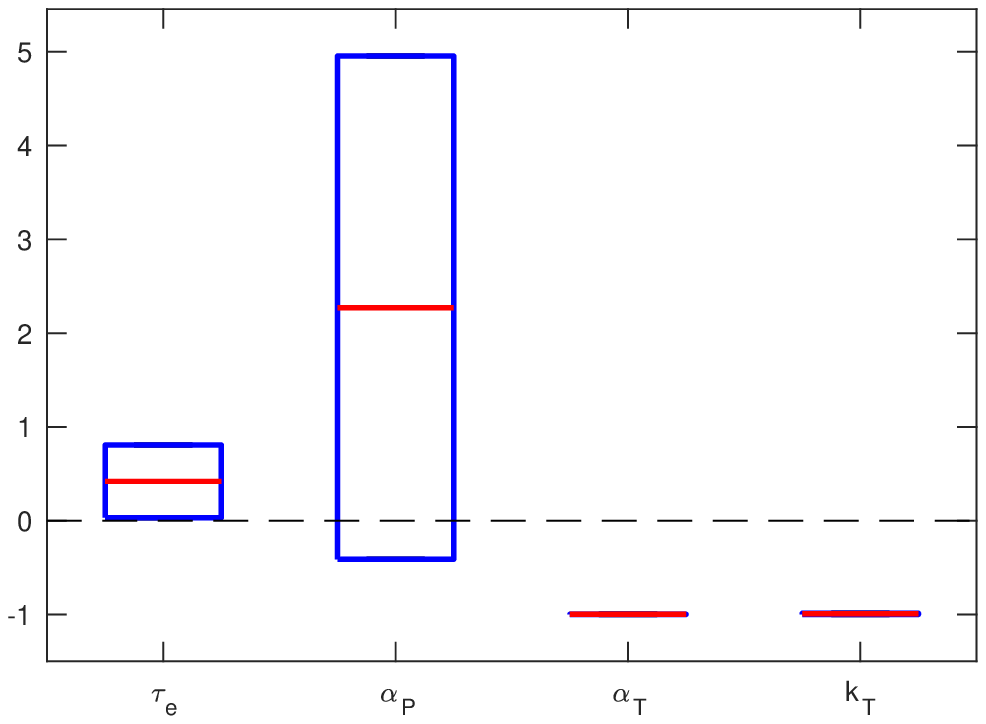}
	\caption{Box plots of the bootstrap confidence intervals (CIs) froms fits of healthy individuals displaying oscillations in circulating platelet levels. If the boxplot of CI of the difference in the mean bootstrap estimate and 1 crosses the x-axis, we cannot reject the null hypothesis of no difference in mean relative error between the healthy and oscillating cases}
	\label{fig:Bootstraps_normal}
\end{figure}

As we did in Sect.~\ref{subsec:fitting} with the patients diagnosed with CT, we used bootstrapping resampling techniques to assess the significance of parameter changes required in these three cases. The results of this analysis are given in Table~\ref{tab:Bootstraps_normal} and Fig. \ref{fig:Bootstraps_normal}. Only the CIs for $\alpha_P$ contains 0, and therefore we reject the null hypothesis and conclude that there are statistically significant differences at the $\alpha = 0.05$ level in all other cases. We believe that the lack of statistical significance of the changes to $\alpha_P$ in the healthy patient cases is likely related to small number of available datasets, as significant changes to $\alpha_P$ were required to fit the von Schulthess and Gessner cases. Nonetheless, we are unable to conclude that the change to $\alpha_P$ is statistically significant in the present study.  As in the bootstrap results from the patients diagnosed with CT, both $\alpha_T$ and $k_T$ have narrow bootstrap CIs, which suggests a higher degree of certainty in those cases. It is possible that these patients have an alteration in the TPO receptor or the interaction of TPO with the platelet lineage, just as in patients diagnosed with CT. We posit that it may be that cases of oscillating platelets which do not lead to pathological changes and that oscillations in the platelet lineage are far more common than the literature suggests. Further clinical investigation is required, however, to validate these hypotheses.

\begin{table}[h!]
	\centering
	\begin{tabular}{|c|c|c|l|}
		\hline
		& Difference of average & Difference of average& 95\% bootstrap CI\\
		&relative change and 1 (fit values) & relative change and 1 (bootstrap values)  &\\
		\hline
		$\tau_e$   & 0.3168 & 0.3173 & $[0.0326, 0.8070]$ \\
		$\alpha_P$ & 1.4461 &  1.4572 & $[-0.4110, 4.9549]$ \\
		$\alpha_T$ & -0.9979 & -0.9979 & $[-0.9985, -0.9968]$ \\
		$k_T$      & -0.9944 & -0.9944 & $[-0.9986, -0.9865]$ \\
		\hline
	\end{tabular}
	\caption{Differences of relative changes and 1 for parameter values from fits of patients displaying oscillations but diagnosed as healthy and the bootstrap samples, and bootstrap 95\% confidence intervals. Column 2: For each parameter fit, the difference in its relative change and 1 was calculated. Column 3: 10000 bootstrap samples were generated and the difference in the mean relative change and 1 were calculated. Column 4: the 95\% bootstrap confidence interval. CI: confidence interval}
	\label{tab:Bootstraps_normal}
\end{table}

\subsection{Hopf bifurcation for CT patients} \label{subsec:hopf_ct}

In Sect.~\ref{sec:analysis}, our linear analysis, including sensitivity to perturbation of the parameters, demonstrated a strong stability of our model for a healthy subject. The previous section provided fits to data for platelets and TPO in CT patients, but required shifts in four parameters with some changes being quite substantial. As the parameters are varied linearly between the two states, our numerical methods tracked the changes in the equilibrium and the pair of eigenvalues, resulting in a Hopf bifurcation leading to the cyclic behavior observed in the CT patients. As noted earlier, it is not the leading pair of eigenvalues for the normal parameter set, but rather the second leading pair that results in this bifurcation.

\begin{figure}[h!]
	\centering
	\subfloat[][]{\includegraphics[width=.48\textwidth]{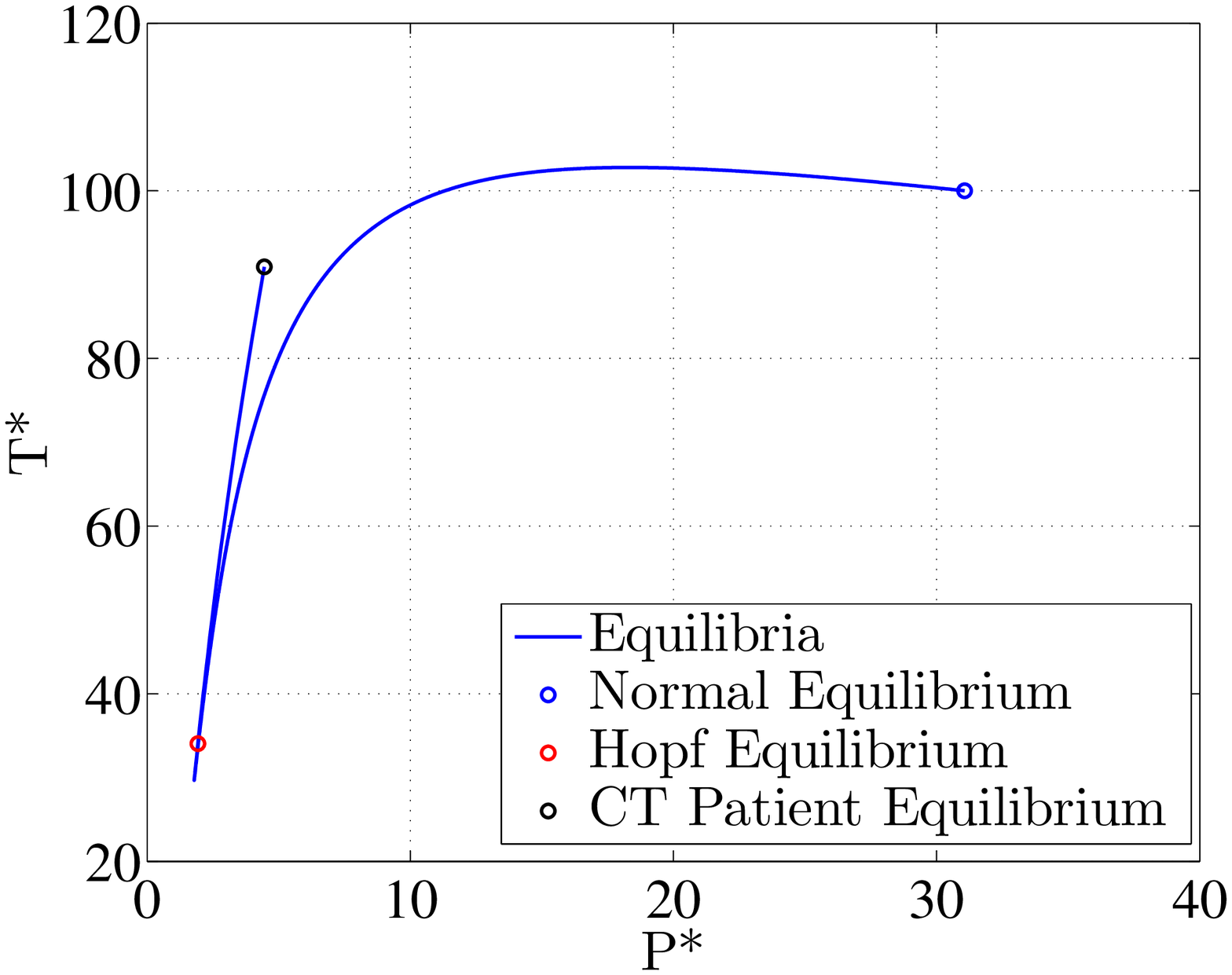}}%
	\hspace{8pt}%
	\subfloat[][]{\includegraphics[width=.48\textwidth]{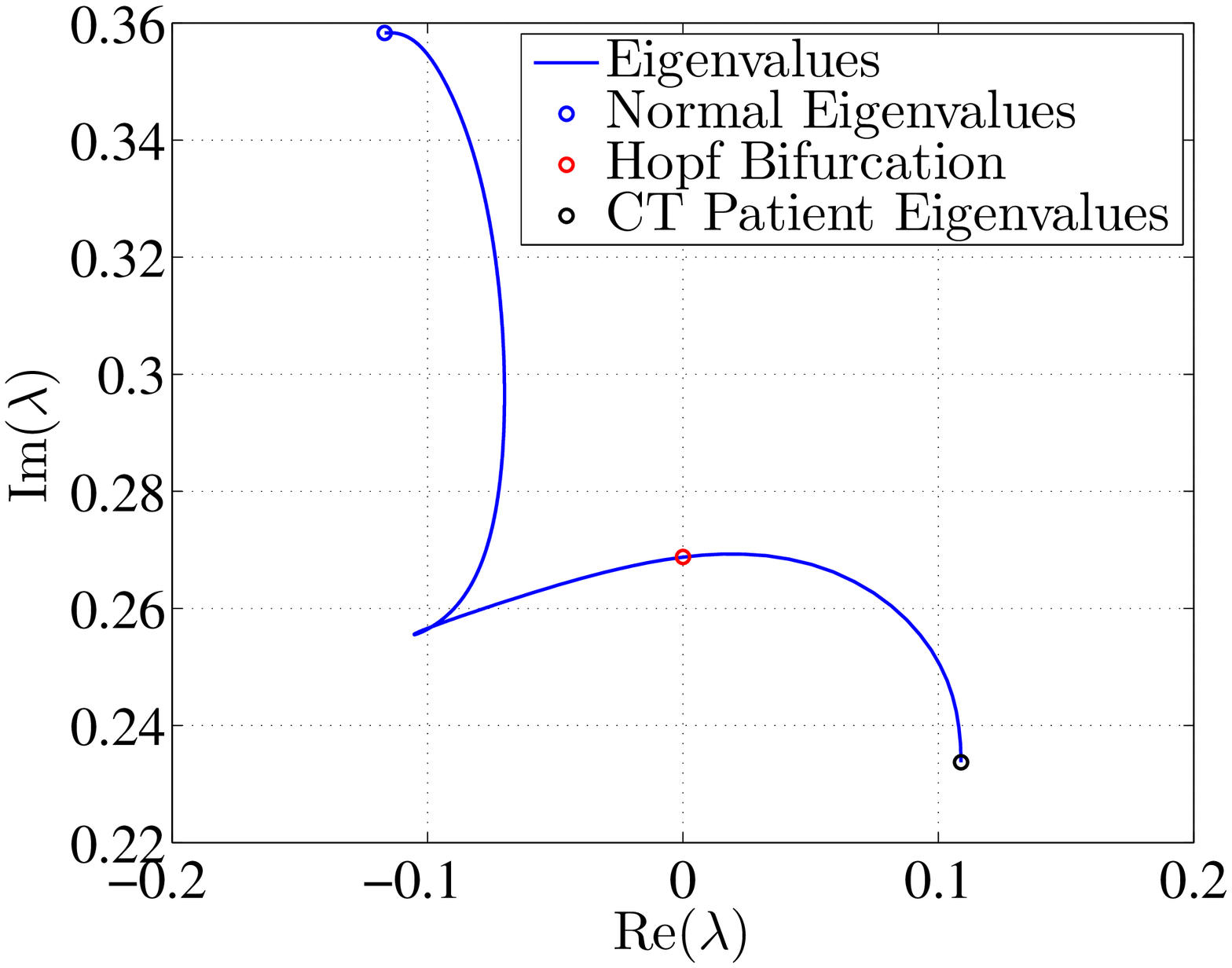}}%
	
	\subfloat[][]{\includegraphics[width=.48\textwidth]{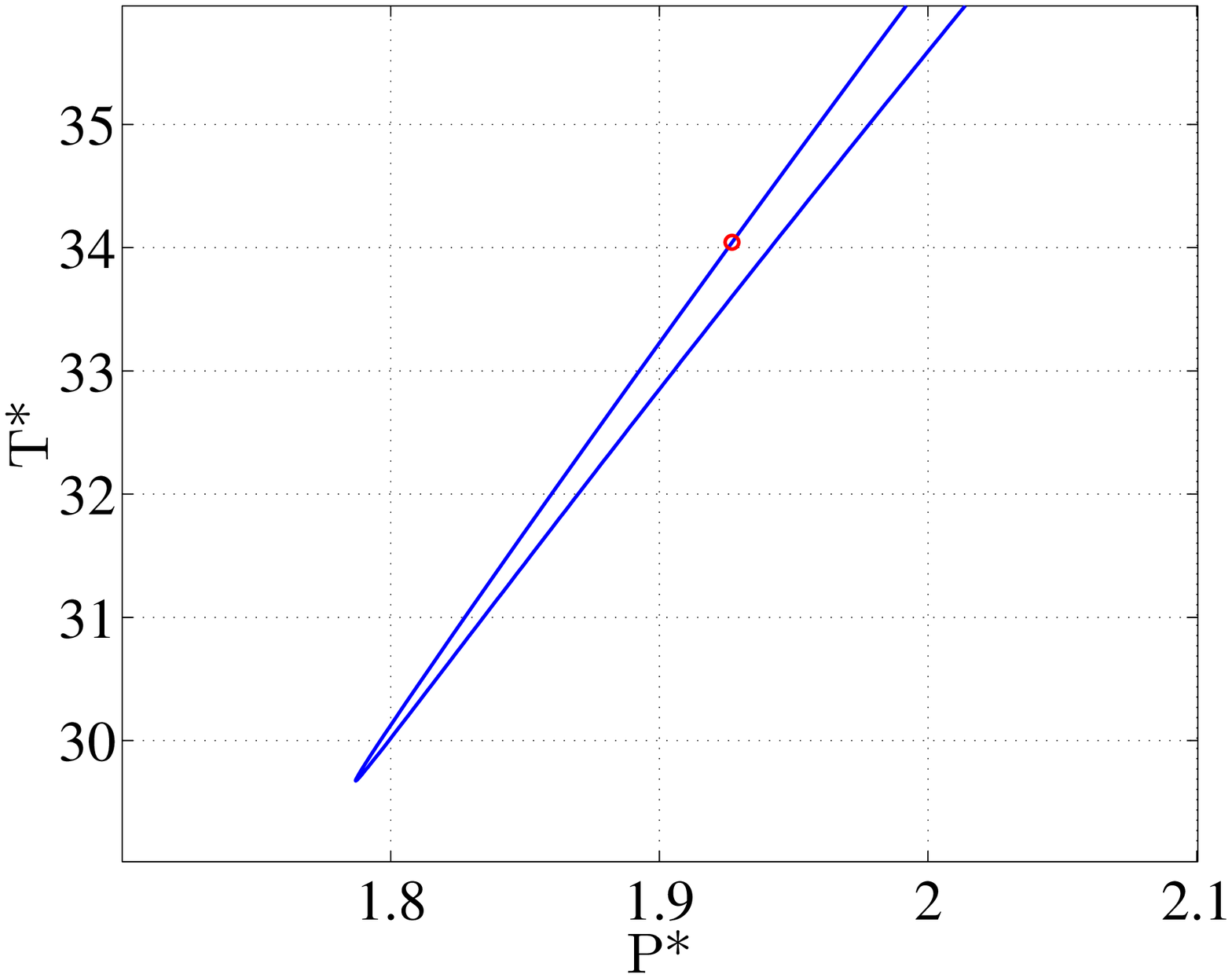}}%
	\hspace{8pt}%
	\subfloat[][]{\includegraphics[width=.48\textwidth]{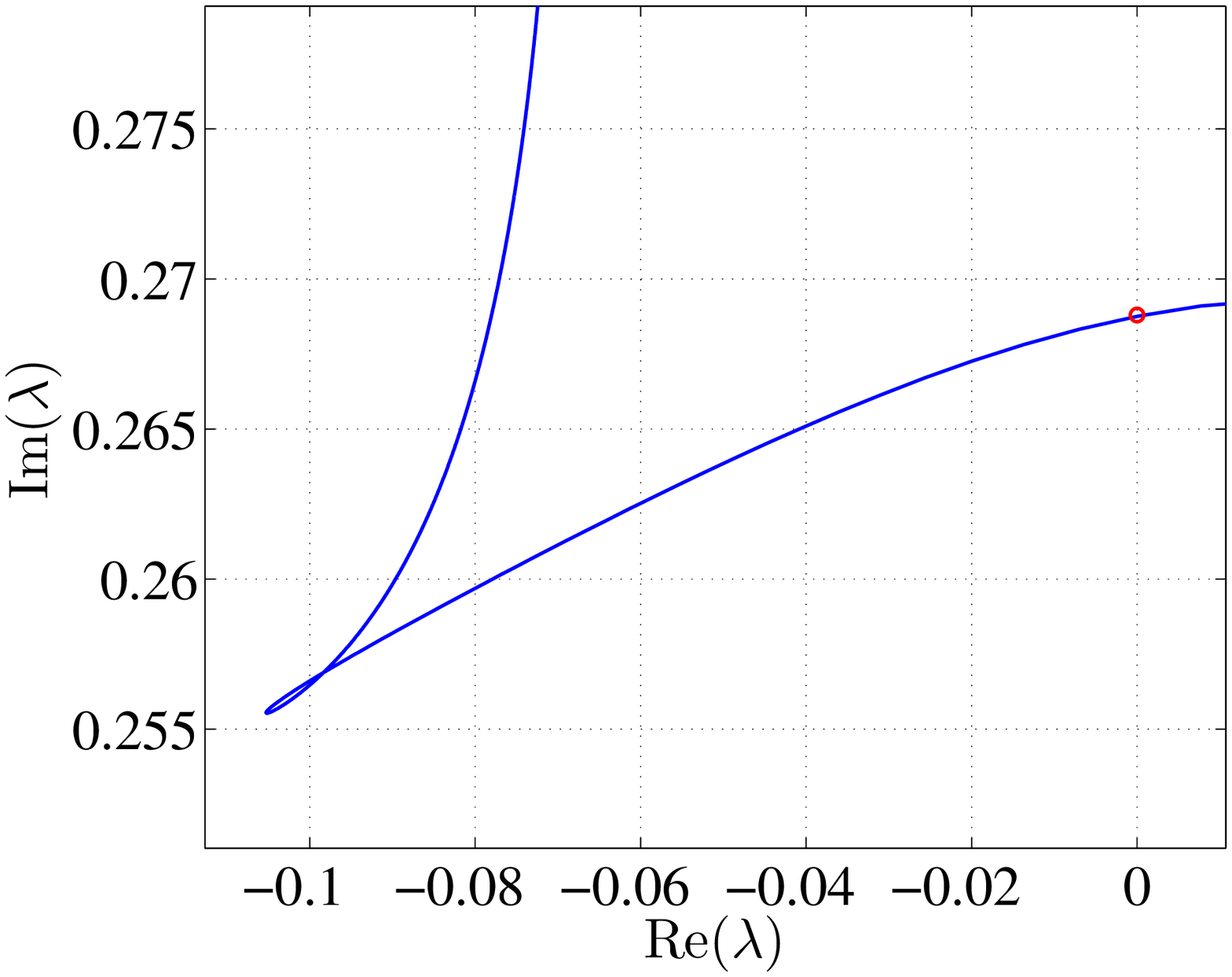}}%
	\caption{The curves on the left show the evolution of the equilibrium
		from healthy subject to CT patient as parameters vary. The curves on the
		right follow the eigenvalues. The second row shows the curves magnified}
	\label{fig:bruin_eq_ev}
\end{figure}

For this section we present details from the numerics for the CT patient of \citet{bruin}. Appendix~\ref{app:hopf_ct} includes details for the other three CT patients for which we have both platelet and thrombopoietin data. We used our analytic techniques to follow a hyperline in the 4D-parameter space from the normal parameter values to each of the parameter sets for the four CT patients with both platelet and TPO data, which are listed in Table~\ref{table:ct}. The program computes the equilibrium $(P^{*}$, $T^{*})$ at each set of parameters along with the corresponding eigenvalues. The eigenvalues are computed from the characteristic equation \eqref{ce} from Sect.~\ref{subsec:char_eqn}. Specifically, if $\boldsymbol \theta$ is the vector of parameters ($\tau_{e}$, $\alpha_{P}$, $\alpha_{T}$, $k_{T}$), $\boldsymbol \theta_{homeo}$ is the value of that vector of parameters at homeostasis, and $\boldsymbol  \theta_{patient}$ is the value of the vector of parameters for the CT patient, then
\begin{equation} \label{eq:hopf_change_parameters}
	\boldsymbol \theta=\boldsymbol \theta_{homeo}+(\boldsymbol \theta_{patient}-\boldsymbol \theta_{homeo})t, \qquad t \in [0,1].
\end{equation}
The results are displayed in Fig.~\ref{fig:bruin_eq_ev}.
	
The equilibrium for the normal parameters is $(P^{*}, T^{*}) = (31.071, 100)$, while the equilibrium for the CT patient is $(P^{*}, T^{*}) = (4.4547,  90.92)$. The graphs on the left of Fig.~\ref{fig:bruin_eq_ev} show the evolution of the equilibrium as the parameters vary linearly from normal to the values for the CT patient. The curve moves to the left, then starts heading toward the origin. The $T^*$ value reaches a minimum slightly below with $P^*$ dropping to approximately 1.8. This curve then smoothly doubles back and passes through $(P^{*}, T^{*}) = (1.927,34.043)$, where the Hopf bifurcation occurs and the model loses stability. Subsequently, the values of both $P^*$ and $T^*$ increase to the CT patient equilibrium with a low value of $P^*$ and $T^*$ around 90, which is similar to a healthy individual.
	
From numerically solving Eq.~\eqref{ce}, the eigenvalues for the normal case begin at $\lambda =  -0.11375 \pm 0.35888i$, producing an asymptotically stable equilibrium. (We note that the leading eigenvalue for this case is $\lambda =  -0.058953 \pm 0.053015i$, and it simply decreases in real and imaginary parts, becoming real along this change of parameters.) The eigenvalues create an arc with the imaginary part decreasing, while the real part first increases then decreases a little to a cusp-like region matching the similar region seen for the equilibrium. The eigenvalue curve actually crosses itself before the real part increases to the Hopf bifurcation at $\lambda = \pm 0.2688i$. The real part continues to increase slightly before arcing down to a lower frequency, and the real part increases to where the equilibrium of the CT patient is unstable with $\lambda = 0.1089 \pm 0.2337i$. This frequency is consistent with a period of approximately 26.9 days, which agrees well with the observed oscillations in the data.
	
Since four parameters are changing, it is hard to determine what kinetic effect is most influencing the loss of stability. However, it is clear from our simulations that the rapid shift in equilibrium results in a quick response of the eigenvalues. The cusp-like behavior observed is likely caused by one of the Hill functions governing the platelet model, which could rapidly transition to a different state in the equilibrium calculation. However, more detailed studies are needed of this phenomenon.

\section{Summary and discussion} \label{sec:discussion}

Motivated by recent laboratory and clinical findings on thrombopoiesis in humans, we have developed a model for the regulation of platelet production that, in contrast to previous models \cite{apostu2008understanding,santillan2000regulation}, incorporates the regulation mechanisms and dynamics of megakaryocytes and thrombopoietin. Our model of thrombopoiesis consists of two integro-differential equations with constant delays, describing the dynamics of platelets and thrombopoietin, and an integral equation of the dynamics of megakaryocytes in the bone marrow. As described in Sect.~\ref{sec:math_model} and Appendix~\ref{app:parameter}, we have estimated the parameters of the model as closely as possible from experimental and clinical data. The model has a unique positive steady state solution, which we demonstrated in Appendix~\ref{app:EU}. Furthermore, we have extended linear techniques to this complicated model and developed numerical methods for performing a stability analysis. This analysis has provided a tool to compare the sensitivity of the model to the many parameters and determine when stability changes occur.

To validate our approach to model development, we applied our model to the investigation of the pathogenesis of cyclic thrombocytopenia. The clinical literature speculates that CT may be caused by:
\begin{enumerate}
\item Immune-mediated platelet destruction (autoimmune CT).
\item Megakaryocyte deficiency and cyclic failure in platelet production (amegakaryocytic CT),
\item Possible immune interference with or destruction of the TPO receptor.
\end{enumerate}
The results presented in Sect.~\ref{sec:ctp} indicate that highly significant reductions (factor of 1000 and 100, respectively) in $\alpha_T$ and $k_T$, which are responsible for the platelet and megakaryocyte-dependent TPO removal rates, are necessary to induce oscillations roughly corresponding to those of CT. Those changes were also necessary to fit the data of \citet{morley1969platelet} and \citet{vonschulthess1986}, in which the apparently healthy subjects maintain platelet levels in the normal range in the face of statistically significant oscillations. In addition, changes in $\tau_e$ (which represents the duration of the  megakaryocyte maturation stage) as well as in $\alpha_P$ (which is responsible for the maximum removal rate of platelets due to macrophages) allow the accurate replication of clinical data on platelet and thrombopoietin dynamics. (The procedure we employed to fit the CT cases is described in detail in Appendix.~\ref{app:ABC}, and the numerics developed to simulate the model are presented in Appendix~\ref{app:numerics}.) These changes are consistent with the results from our bootstrapping results as well as the dependence of eigenvalue behavior that we have uncovered.  Whether the changes in $\alpha_T$ and $k_T$ are primary, with the changes $\tau_e$ as well as in $\alpha_P$ being secondary and due to an as yet unknown dynamic interconnection, we cannot say.

While it is believed that most platelet diseases, which may include CT, arise due to disorders in TPO or its receptor \cite{hitchcock2014thrombopoietin}, we are unsure why such a significant change in $\alpha_T$ and $k_T$ is needed to obtain oscillations. In the context of CT, our model suggests that a disorder in TPO or its receptor (destabilized TPO removal mechanism and decreased megakaryocyte production) along with an immune-mediated platelet destruction response are the main causes of CT. In contrast, \citet{apostu2008understanding} found that an increased random destruction of platelets (parameter $\gamma_P$ in this model) and decreased megakaryocyte production together could explain the onset of oscillations. Their model, however, did not accurately describe the dynamics of megakaryocytes and thrombopoietin. As such, our findings add further nuances to their results.

Given our current understanding of the regulation of thrombopoiesis, it is safe to say that there are unknown biological facets of the regulatory system that are not accounted for in our model and which await further elucidation by experimental biologists and clinicians. In addition, the mathematical analyses indicate that there remain details in the nonlinear model, which could be explored further and possibly give insight into the transitions from the stable normal state to the diseased state. At any rate, it is clear that a better understanding of the mechanisms implicated in the interaction of thrombopoietin and its receptor, specifically in patients with cyclic thrombocytopenia, will allow for further modeling refinements and a more precise picture of the origins of this dynamical disease, and thrombopoiesis in general.

\begin{acknowledgements}
	This research was supported by the NSERC (National Sciences and Engineering Research Council) of Canada through Discovery grants to JB, ARH, and MCM, and PGS-D program to MC. SRS thanks McGill University for a Science Undegraduate Research Award.
	GPL and MCM are especially grateful to Dr. Jayson Potts (UBC) for his initial contact that prompted the initiation of this research. We thank Prof.~Jiguo Cao (SFU) for introducing us to LW.	
\end{acknowledgements}

\begin{appendices}
	\addcontentsline{toc}{part}{\appendixname}
		
	\section{Parameter estimation and constraints}
	\label{app:parameter}
		
	This extensive appendix contains the details of the parameter estimation procedure for this model largely based on experimental data.
	First, in Sect.~\ref{subapp:homeo_relationships} we consider the model at homeostasis. We then use TPO-knockout experimental observations in Sect.~\ref{subapp:tpo_zero_relationships} to derive further parameter constraints. In Sect.~\ref{subapp:parameters_experimental} we provide estimates for other  parameters directly from experimental data. Finally, in Sect.~\ref{subapp:parameters_model} we calculate remaining parameters using experimental data and the relationships derived in Sects.~\ref{subapp:homeo_relationships} and \ref{subapp:tpo_zero_relationships}.
		
	\subsection{Homeostasis relationships}
	\label{subapp:homeo_relationships}
	Let $Q^{*}$ denote the stem cell concentration, $M_{e}^{*}$ the total
	megakaryocyte volume, $P^{*}$ the platelet concentration, $T^{*}$ the thrombopoietin concentration, $\eta_{m}^{*}$ and $\eta_{e}^{*}$, the rate of mitosis and endomitosis, respectively, and $\tau_{m}$ and $\tau_{e}$, the average time megakaryoblasts and megakaryocytes spend in the mitotic and endomitotic stages, respectively, at homeostasis. At this steady state, the equations for megakaryocyte production rate \eqref{eq:mb_density}, platelet production rate \eqref{eq:platelet_prod_rate}, megakaryocyte volume \eqref{eq:mk_volume}, platelet balance \eqref{eq:p}, and thrombopoietin balance \eqref{eq:tpo} become
	\begin{equation}
	\label{eq:mk_prod_rate_homeo}
	m_{m}^{*}(\tau_{m})=\kappa_{P}Q^{*}\e^{\eta_{m}^{*}\tau_{m}},
	\end{equation}
	\begin{equation}
	\label{eq:platelet_prod_rate_homeo}
	m_{e}^{*}(\tau_e)=V_m\kappa_PQ^*\e^{\eta^{*}_{m}\tau_m+\eta^{*}_{e}\tau_e},
	\end{equation}
	\begin{equation}
	\label{eq:mk_volume_homeo}
	M_{e}^{*}=V_{m}\kappa_{P}Q^{*}\e^{\eta_{m}^{*}\tau_{m}}\left(\frac{\e^{\eta_{e}^{*}\tau_{e}}-1}{\eta_{e}^{*}}\right),
	\end{equation}
	\begin{equation}
	\label{eq:platelet_balance_homeo}
	\frac{D_{0}}{\beta_{P}}m_{e}^{*}(\tau_e)=\gamma_{P}P^{*}+\alpha_{P}\frac{(P^{*})^{n_{P}}}{(b_{P})^{n_{P}}+(P^{*})^{n_{P}}},
	\end{equation}
	\begin{equation}
	\label{eq:tpo_balance_homeo}
	T_{prod}=\gamma_{T}T^{*}+\alpha_{T}(M_{e}^{*}+k_{S}\beta_{P}P^{*})\frac{(T^{*})^{n_{T}}}{(k_{T})^{n_{T}}+(T^{*})^{n_{T}}}.
	\end{equation}

	\subsection{TPO knock-out relationships}
	\label{subapp:tpo_zero_relationships}
	The elimination of TPO gene or its receptor in mice reduces megakaryocyte and platelet levels to approximately 10\% of normal \cite{de1996physiological}, a finding also observed in humans (Kaushansky, private communication). Therefore, the model must have a steady state solution at 10 \% normal platelet and megakaryocyte levels when the thrombopoietin production rate and level are both zero, giving
	\begin{equation*}
	\tau_{e}m_{m}^{*}(T=0,\tau_{m})=\frac{1}{10}\tau_{e}m_{m}^{*}(\tau_{m}),
	\end{equation*}
	and
	\begin{equation*}
	\frac{D_{0}}{\beta_{P}}m_{e}^{*}(T=0,\tau_{e})=\frac{1}{10}\gamma_{P}P^{*}+\alpha_{P}\frac{(P^{*})^{n_{P}}}{(10b_{P})^{n_{P}}+(P^{*})^{n_{P}}}.
	\end{equation*}
	Using Eqs.~\eqref{eq:mk_prod_rate_homeo} and \eqref{eq:platelet_prod_rate_homeo}, we rewrite these two relationships to obtain
	\begin{equation}
	\label{eq:mk_zero_tpo}
	\e^{\eta^{min}_{m}\tau_m}=\frac{1}{10}\e^{\eta^{*}_{m}\tau_m},
	\end{equation}
	\begin{equation}
	\label{eq:platelet_zero_tpo}
	\frac{D_{0}}{\beta_{P}}V_{m}\kappa_{P}Q^{*}\e^{\eta_{m}^{min}\tau_{m} + \eta_{e}^{min}\tau_{e}}=\frac{1}{10}\gamma_{P}P^{*}+\alpha_{P}\frac{\left(P^{*}\right)^{n_{P}}}{(10b_{P})^{n_{P}}+(P^{*})^{n_{P}}}.
	\end{equation}

	\subsection{Parameters estimated from experimental data}
	\label{subapp:parameters_experimental}
		
	\subsubsection{Megakaryocyte compartment}
		
	\citet{mackey2001cell} estimated the homeostatic concentration of HSCs using data from cats and mice, giving an estimate of $Q^{*}=1.1\times10^{6}\mbox{ cells/kg of body weight}$. We assume that humans have roughly this same number of stem cells per kg of body weight. We estimate the parameter $\kappa_{P}$, the
	rate at which stem cells commit to the megakaryocyte lineage, from
	the model of stem cells dynamics \citet{mackey2001cell} and \citet{bernard2003oscillations} developed,
	and the assumption that stem cells differentiate at an equal rate into
	all blood lineages. This gives, to four significant digits, an estimate
	of $\kappa_{P}\approx0.0072419\mbox{ cells/kg of body weight per day}$.
	See \citet{craig2015mathematical} for more details.
		
	\citet{tomer1996measurements} measured
	the diameters of megakaryocytes in the bone marrow of 10 healthy individuals. They
	found that megakaryocytes of ploidy 2N (those megakaryocytes which have not yet undergone endomitosis) had a mean diameter of $21 \pm 4\mbox{ \ensuremath{\mu}m}$,
	while the average megakaryocyte had a mean diameter of $37 \pm 4 \mbox{ \ensuremath{\mu}m}$.
	We set the average volume of a megakaryocyte of ploidy
	2N to be approximately that of a sphere with mean diameter $21\mbox{ \ensuremath{\mu}m}$,
	and hence, set $V_{m}=4\pi(21)^{3}/24\mbox{ fL}$. The parameter
	$\tau_{e}$, the time a megakaryocyte spends in endomitosis,
	is estimated by various sources to range from 5 to 7 days \citep{finch1977kinetics,Kuter13}.
	We take $\tau_{e}=5\mbox{ days}$.
		
	\subsubsection{Platelet compartment}
		
	\citet{giles1981platelet} measured the mean platelet count and volume
	in 1011 healthy human adult blood specimens. He found the mean platelet
	count to be $290\times10^{9}\mbox{ platelets / L of blood}$ and
	the mean platelet volume to be $8.6\mbox{ fL}$, so we set $\beta_{P}$ to be $8.6\mbox{ fL}$. Since, on average, one third of the total platelet mass in the body is sequestered
	by the spleen \citep{aster1966pooling}, we approximate the mean
	platelet count in the body is $1.5$ times this amount. The Hill coefficient for the platelet-dependent removal of platelets is assumed to take the value $n_P = 2$. Supposing that a healthy adult has roughly $5$ L of blood per $70$ kg of body weight gives $P^{*}=1.5\times(5/70)\times(290\times10^{9})\approx3.1071\times10^{10}\mbox{ platelets/kg of body weight}$.

\subsubsection{Thrombopoietin compartment}
		
Normal TPO concentrations in humans range from $50$ to $150\mbox{ pg/ml of blood}$ \citep{Kuter13}, and so we select the middle of the range as the homeostatic concentration,  $T^{*}=100\mbox{ pg/ml}$. Since only platelets in circulation, and not in the spleen, contribute to the binding of TPO, we set $k_{S}=2/3$, which is the fraction of the platelet mass in circulation \citep{aster1966pooling}. As there are two thrombopoietin binding sites on a TPO receptor \cite{hitchcock2014thrombopoietin}, we set $n_{T}=2$ for the binding coefficient of thrombopoietin.
		
		\subsection{Parameters calculated from experimental data and the model}
		\label{subapp:parameters_model}
		
		\subsubsection{Megakaryocyte and platelet compartments}
		
At equilibrium, the total megakaryocyte volume is
	\begin{equation}
	M_{e}^{*}=V_{m}\kappa_{P}Q^{*}\e^{\eta_{m}^{*}\tau_{m}}
	\left(\frac{\e^{\eta_{e}^{*}\tau_{e}}-1 }{\eta_{e}^{*}}\right), \label{eq:mk equil}
	\end{equation}
	wherein $\eta_{e}^{*} = \eta_e(T^*)$.  
		\citet{tomer1996measurements} found that the average megakaryocyte in humans has a mean diameter of $37 \pm 4 \ensuremath{\mu}m$, giving an approximate mean volume of $4\pi(18.5)^3/3 \mbox{ fL}$. Assuming the average megakaryocyte volume predicted by our model (total volume of megakaryocytes divided by the number of megakaryocytes) equals this value, we have from \eqref{eq:mk equil} that
		\begin{equation}
		\frac{M_{e}^{*}}{\tau_{e}m_{m}^{*}(\tau_{m})}=V_{m}\left(\frac{\e^{\eta_{e}^{*}\tau_{e}}-1}{\eta_{e}^{*}\tau_{e}}\right)\approx\frac{4\pi(37)^{3}}{24},
		\end{equation}
		which can be rearranged as
		\begin{equation}
		\frac{\e^{\eta_{e}^{*}\tau_{e}}-1}{\eta_{e}^{*}\tau_{e}}=\left(\frac{37}{21}\right)^{3}.\label{eq:eta^{*}_{e}}
		\end{equation}
		Using the MATLAB \cite{matlab} function \textit{fsolve}, which solves the equation $F(x) = 0$ for $x$ for some function $F$, we solved Eq.~\eqref{eq:eta^{*}_{e}} for $\tau_{e}\eta_{e}^{*}$, yielding $\tau_{e}\eta_{e}^{*} \approx 2.788$. Since $\tau_{e}$ is known, we have $\eta_{e}^{*} \approx 0.5576$.
		
		Using \textsuperscript{111}In-Oxine and \textsuperscript{111}In-tropolone
		(more reliable markers than the previously used \textsuperscript{51}Cr
		label), \citet{tsan1984kinetics} measured the mean platelet survival
		time $\tau_P$ to be $8.4 \pm 0.25$ days.
        We assume that the
		platelet production rate replenishes the full platelet population (those circulating and in the
		spleen) in about $\tau_{P}$ days. Therefore,
		\begin{equation}
		\frac{1}{\tau_{P}}P^{*}\approx\frac{D_{0}}{\beta_{P}}m_{e}^{*}(t,\tau_{e})=\frac{D_{0}}{\beta_{P}}V_{m}\kappa_{P}Q^{*}\e^{\eta_{m}^{*}\tau_{m}+\eta_{e}^{*}\tau_{e}},\label{eq:p_balance}
		\end{equation}
		and, in particular,
		\begin{equation}
		\frac{1}{\tau_{P}}P^{*}=\gamma_{P}P^{*}+\alpha_{P}\frac{(P^{*})^{n_{P}}}{(b_{P})^{n_{P}}+(P^{*})^{n_{P}}}.\label{eq:prod_p}
		\end{equation}
		Solving for $\alpha_{P}$, we have
		\begin{equation}
		\alpha_{P}=P^{*}\left(\frac{1}{\tau_{P}}-\gamma_{P}\right)\left(1+\left(\frac{b_{P}}{P^{*}}\right)^{n_{P}}\right).\label{eq:alpha_P}
		\end{equation}
		All parameters in \eqref{eq:alpha_P} except for $b_P$ and $\gamma_P$ are known. The rates of removal of the platelets from the blood should be positive, implying that $\gamma_P$ and $\alpha_P > 0$. The latter requires that
		\begin{equation}
		1/\tau_{P}-\gamma_{P}>0.
		\end{equation}
		
		One megakaryocyte sheds 1000-3000 platelets \citep{harker1969thrombokinetics}.
		Assuming, on average, that one megakaryocyte sheds 2000 platelets, then
		the rate of production of megakaryocytes times the number of platelets
		shed per megakaryocyte equals roughly the rate of production of platelets:
		\begin{equation}
		m_{m}^{*}(\tau_{m})\times\mbox{ 2000}=\kappa_{P}Q^{*}\e^{\eta_{m}^{*}\tau_{m}}\times2000=\frac{1}{\tau_{P}}P^{*},\label{eq:productions_equality}
		\end{equation}
		giving a rate of production of megakaryocytes of $1.85\times10^{6}\mbox{ megakaryocytes/kg/day}$, which is close to the value of $2 \times 10^{6} \mbox{ megakaryocytes/kg/day}$
		estimated to be the normal production rate of megakaryocytes \cite{finch1977kinetics}.
		
		The parameter $D_0$, the fraction of megakaryocytes shedding platelets, can be solved for in \eqref{eq:p_balance} by equating \eqref{eq:productions_equality} with \eqref{eq:p_balance} and using \eqref{eq:platelet_prod_rate_homeo}. This gives
		\begin{equation}
		D_{0}=2000\frac{\beta_{P}}{V_{m}}\e^{-\eta_{e}^{*}\tau_{e}}\approx0.21829.\label{eq:d_0}
		\end{equation}
		
		Rearranging \eqref{eq:productions_equality} and solving for $\eta_{m}^{*}\tau_{m}$, we get
		\[
		\eta_{m}^{*}\tau_{m}=\ln\left(\frac{1}{2000}\frac{1}{\kappa_{P}Q^{*}}\frac{1}{\tau_{P}}P^{*}\right)\approx5.4394.
		\]
		Substituting in Eq.~\eqref{eq:mk_zero_tpo} and solving for $\eta^{min}_{*}$, we have
		\begin{equation}	\eta_{m}^{min}=\frac{1}{\tau_{m}}\ln\left(\frac{1}{20000}\frac{1}{\kappa_{P}Q^{*}}\frac{1}{\tau_{P}}P^{*}\right).\label{eq:eta^min_m}
		\end{equation}
		
		Using Eq.~\eqref{eq:mitosis_rate} at homeostasis to solve for $\eta^{max}_m$ in Eq.~\eqref{eq:mk_zero_tpo} gives
		\begin{equation}
		\eta_{m}^{max}=\eta_{m}^{min}+\frac{\ln(10)}{\tau_{m}}\left(1+\frac{b_{m}}{T^*}\right).\label{eq:eta^max_m}
		\end{equation}
		
		We can now use the steady state equation for the platelet numbers in absence of thrombopoietin, Eq.~\eqref{eq:platelet_zero_tpo}, in combination with the expressions \eqref{eq:productions_equality} and \eqref{eq:d_0} to get
		\begin{equation*}
		\frac{1}{10\tau_{P}}P^{*}\e^{(\eta_{e}^{min}-\eta_{e}^{*})\tau_{e}}=\frac{1}{10}\gamma_{P}P^{*}+\alpha_{P}\frac{(P^{*})^{n_{P}}}{(10b_{P})^{n_{P}}+(P^{*})^{n_{P}}}.
		\end{equation*}
		Solving for $\eta_{e}^{min}-\eta_{e}^{*}$, we get
		\begin{equation}
		\eta_{e}^{min}=\eta_{e}^{*}+\frac{1}{\tau_{e}}\ln\left[\tau_{P}\gamma_{P}+10\tau_{P}\alpha_{P}\frac{(P^{*})^{n_{P}-1}}{(10b_{P})^{n_{P}}+(P^{*})^{n_{P}}}\right].\label{eq:eta^min_e}
		\end{equation}
		
		We can solve for $\eta^{max}_{e}$ via Eq.~\eqref{eq:endomitosis_rate} at steady state:
		\begin{equation}
		\eta_{e}^{max}=\eta_{e}^{min}+(\eta_{e}^{*}-\eta_{e}^{min})\left(1+\frac{b_{e}}{T^{*}}\right).\label{eq:eta^max_e}
		\end{equation}
		
		\subsubsection{Thrombopoietin compartment}
		
		Using Eq.~\eqref{eq:tpo_balance_homeo}, the homeostasis relationship for the thrombopoietin concentration, we can solve for $\alpha_T$, yielding
		\begin{equation}
		\alpha_{T}=\frac{T_{prod}-\gamma_{T}T^{*}}{M_{e}^{*}+\beta_{P}P^{*}}\left(1+\left(\frac{k_{T}}{T^{*}}\right)^{n_{T}}\right).\label{eq:alpha_T}
		\end{equation}
		
\subsection{Parameters fit from experimental data}
From the above calculations, it remains to estimate eight more parameters: $\tau_m$, $b_m$, $b_{e}$, $\gamma_P$, $b_P$, $T_{prod}$, $\gamma_T$, and $k_T$. The first five parameters pertain to megakaryocyte and platelet dynamics, while the last three pertain to thrombopoietin dynamics.

We digitized data from \citet{Wang10} for the circulating platelet and TPO levels in healthy patients following a $1\, \mu\mbox{g/kg \mbox{of bodyweight}}$ intravenous infusion of Romiplostim, a TPO mimetic with similar physiological activity to TPO. We then fitted the parameters by simulating the response of our model \eqref{eq:de_platelet}--\eqref{eq:Me} to an infusion of $1\, \mu\mbox{g/kg}$ of TPO and minimizing the squared error between data and simulation.

Specifically, the platelet and TPO data points (and error bars, when available) were interpolated and evaluated at 1000 points in each time interval (ranging from 0 to 42 days for the platelet data and 0 to 1 day for the TPO data), yielding the vectors of points $\mathbf P_{data}$ and $\mathbf T_{data}$. To simulate the response of our model to a $1\, \mu\mbox{g/kg}$ intravenous infusion of thrombopoietin, we ran the numerical algorithm described in Appendix~\ref{app:numerics} with initial history functions $(P_h, T_h) = (P^*, T^*)$ and initial conditions $P_0 = P^*$ and $T_0 = T_{data}(t=0)$. We then evaluated the solution of our model at the interpolated points $\mathbf P_{data}$ and $\mathbf T_{data}$ to obtain the model points $\mathbf P_{model}$ and $\mathbf T_{model}$, respectively.

\begin{figure}[h!]
	\centering\includegraphics[width=0.65\textwidth]{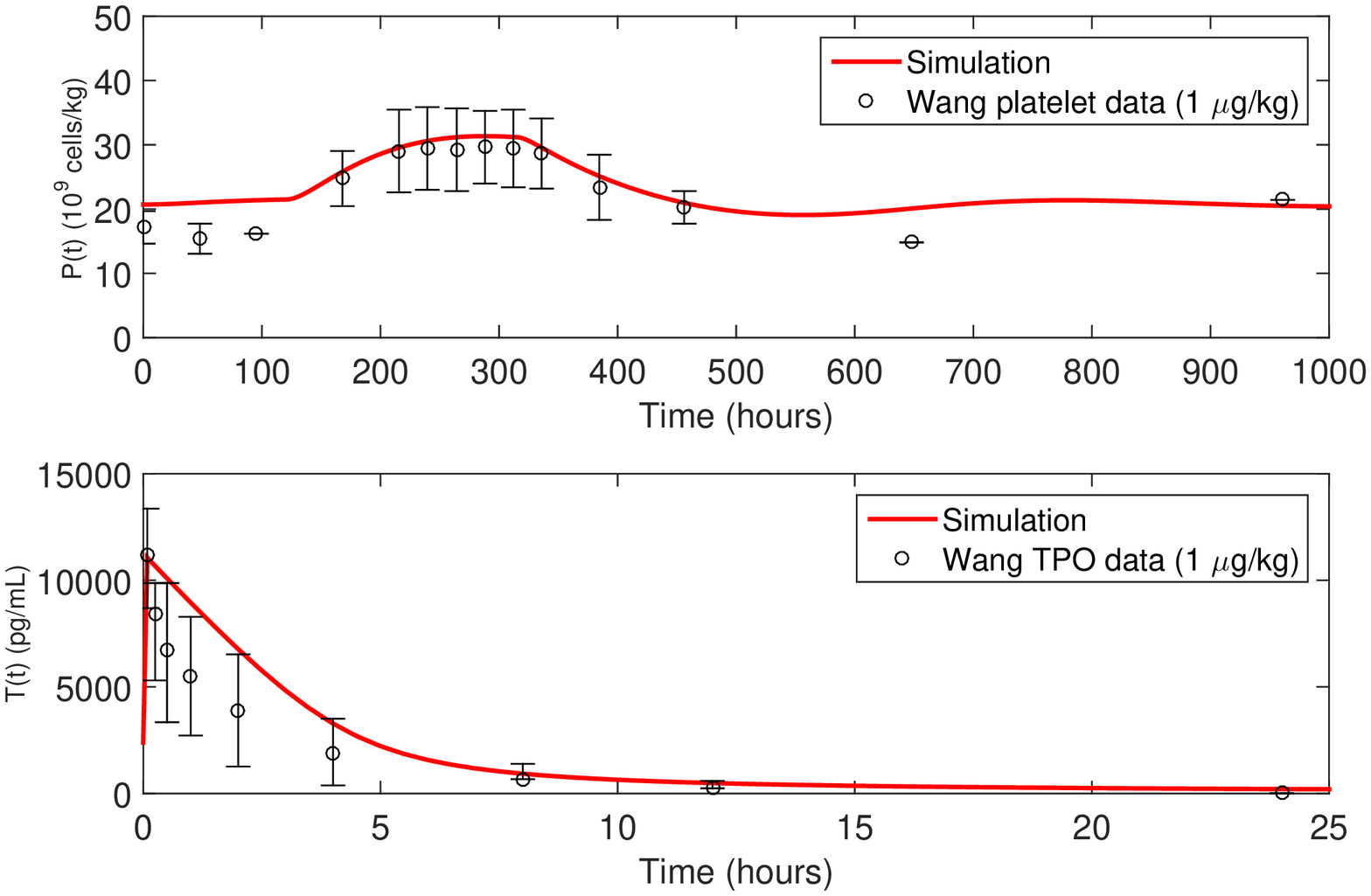}
		\caption{Fit of our model to the platelet and thrombopoietin data from \citet{wang2004pharmacodynamics} following a $1\, \mu\mbox{g/kg}$ intravenous infusion of TPO}
		\label{fig:platelet_fit}
\end{figure}

For parameter estimation, we minimized the fitting error
\begin{equation} \label{eq:Err}
Err = \frac{\|\frac{2}{3}\mathbf P_{model}(t) - \mathbf P_{data}\|_{2}}{\max(\mathbf P_{data})} + \frac{\|\mathbf T_{model}-\mathbf T_{data}\|_{2}}{\max(\mathbf T_{data})},
\end{equation}
by using the \textit{fmincon} function in MATLAB \cite{matlab} to find the set of parameters that minimizes \eqref{eq:Err}. (The factor of $2/3$ accounts for the fraction of platelets that circulate in blood in our model.) The fit obtained from this procedure is shown in Fig.~\ref{fig:platelet_fit}.

\section{Proof of existence and uniqueness of a positive steady state solution}
\label{app:EU} 

Here, we prove that the model has a unique, positive steady state solution by showing that Eqs.~\eqref{eq:p} and \eqref{eq:tpo} have a unique positive fixed point.

As discussed in Appendix~\ref{app:parameter}, Eqs.~\eqref{eq:p} and \eqref{eq:tpo} at steady state (homeostasis) are given by
\begin{equation} \label{eq:Pp}
\frac{D_{0}}{\beta_{P}}m_{e}^{*}(T^*,\tau_e)=\gamma_{P}P^{*}+\alpha_{P}\frac{(P^{*})^{n_{P}}}{(b_{P})^{n_{P}}+(P^{*})^{n_{P}}},
\end{equation}
and
\begin{equation} \label{eq:Tp}
T_{prod}=\gamma_{T}T^{*}+\alpha_{T}(M_{e}^{*}(T^*)+k_{S}\beta_{P}P^{*})\frac{(T^{*})^{n_{T}}}{(k_{T})^{n_{T}}+(T^{*})^{n_{T}}},
\end{equation}
where
\begin{equation*}
m_{e}^{*}(T^*,\tau_e)=V_m\kappa_PQ^*\e^{\eta^{*}_{m}(T^*)\tau_m+\eta^{*}_{e}(T^*)\tau_e},
\end{equation*}
\begin{equation*}
M_{e}^{*}(T^*)=V_{m}\kappa_{P}Q^{*}\e^{\tau_{m}\eta_{m}^{*}(T^*)}\left(\frac{\e^{\eta_{e}^{*}(T^*)\tau_{e}}-1}{\eta_{e}^{*}(T^*)}\right),
\end{equation*}
\begin{equation*}
\eta_{m}(T(t))=\eta_{m}^{min}+(\eta_{m}^{max}-\eta_{m}^{min})\frac{T}{b_{m}+T},
\end{equation*}
and
\begin{equation*}
\eta_{e}(T(t))=\eta_{e}^{min}+(\eta_{e}^{max}-\eta_{e}^{min})\frac{T}{b_{e}+T}.
\end{equation*}
Unlike most population and blood cell regulation models, notice that $(P^*= 0, T^*=0)$ is not an equilibrium of Eqs.~\eqref{eq:Pp} and \eqref{eq:Tp}. This is because of the nonzero constant production rate of thrombopoietin reflected in the first term of \eqref{eq:Tp}, and is analogous to infectious disease models for which there is a constant influx of susceptibles (see, for example, Sect.~2.1.2 of \citet{keeling2008modeling}).

Although finding $(P^*,T^*)$ involves solving two highly nonlinear equations in two unknowns, we can use Eq.~\eqref{eq:Tp} to solve for $P^*$ explicitly in terms of $T^*$, yielding
\begin{equation} \label{eq:P_eq}
\begin{split}
P^* &= \frac{1}{k_S\beta_P }\left[ \frac{T_{prod} - \gamma_T T^*}{\alpha_T}\left(1+\left(\frac{k_T}{T^*}\right)^{n_T}\right) \right. \\
& \left. \quad - V_{m}\kappa_{P}Q^{*}\e^{\eta_{m}^{*}(T^*)\tau_{m}}\left(\frac{\e^{\eta_{e}^{*}(T^*)\tau_{e}}-1}{\eta_{e}^{*}(T^*)}\right) \right].
\end{split}
\end{equation}
Denote the right-hand-side of Eq.~\eqref{eq:P_eq} by $\mathcal{L}(T^*)$. Writing
\begin{equation*}
\mathcal{L}_1(T^*) = \frac{T_{prod} - \gamma_T T^*}{\alpha_T}\left(1+\left(\frac{k_T}{T^*}\right)^{n_T}\right)
\end{equation*}
and
\begin{equation*}
\mathcal{L}_2(T^*)=V_{m}\kappa_{P}Q^{*}\e^{\eta_{m}^{*}(T^*)\tau_{m}}\left(\frac{\e^{\eta_{e}^{*}(T^*)\tau_{e}}-1}{\eta_{e}^{*}(T^*)}\right),
\end{equation*}
we can rewrite Eq.~\eqref{eq:P_eq} as
\begin{equation*}
\displaystyle P^* = \mathcal{L}(T^*)=\frac{1}{k_S\beta_P }\left[ \mathcal{L}_1(T^*) - \mathcal{L}_2(T^*)\right].
\end{equation*}
We now show that $\mathcal{L}$ is a monotone decreasing function of $T^*$, and thus defines an injective function of $P^*$. First, notice that $\mathcal{L}_1$ is a monotone decreasing function of $T^*$, approaching $+\infty$ as $T^* \to 0$, becoming negative-valued when $T^*~>~ T_{prod} / \gamma_T$, and approaching $-\infty$ as $T^* \to \infty$. Second, as both $\eta_m(T^*)$ and $\eta_e(T^*)$ are monotone increasing in $T^*$ (taking values in the intervals $(\eta_m^{min},\eta_m^{max})$ and $(\eta_e^{min},\eta_e^{max})$, respectively), the terms
\begin{equation*}
V_m\kappa_PQ^*\e^{\eta^{*}_{m}(T^*)\tau_m}\quad \mbox{and} \quad \frac{\e^{\eta^{*}_{e}(T^*)\tau_e}-1}{\eta^{*}_{e}(T^*)}
\end{equation*}
in $\mathcal{L}_2(T^*)$ are both monotone increasing whenever $\tau_e\eta^{*}_{e}(T^*) ~>~ 0$, which holds by definition in our model. As $\mathcal{L}_2$ is the product of two monotone increasing functions, it is also a monotone increasing function. Taken together, these results show that $\mathcal{L}$ defines an injective function of $P^*$.

Now consider Eq.~\eqref{eq:Pp} and denote its left-hand-side and right-hand-side by $g(T^*)$ and $h(P^*)$, respectively. The function $h$ is clearly monotone increasing in $P^*$, starting from $0$ and approaching $+\infty$ as $P^* \to \infty$. Using the previous argument on $\eta^{*}_m(T^*)$ and $\eta^{*}_e(T^*)$, we have that $g$ is a monotone increasing function of $T^*$ and uniformly bounded above and below by
\begin{equation*}
\frac{D_0V_m\kappa_PQ^*}{\beta_P} \e^{\eta_m^{min}\tau_m + \eta_e^{min}\tau_e}  \le g(T^*) \le  \frac{D_0V_m\kappa_PQ^*}{\beta_P} \e^{\eta_m^{max}\tau_m + \eta_e^{max}\tau_e}.
\end{equation*} Finding a positive stationary solution in $(P^*, T^*)$ therefore amounts to finding a value of $T^*$ satisfying
\begin{equation} \label{eq:g_h}
h(\mathcal{L}(T^*))=g(T^*)
\end{equation}
and then defining the corresponding value of $P^*$ being given by Eq.~\eqref{eq:P_eq}. As $h(\mathcal{L}(T^*))$ is monotone decreasing from $+\infty$ to negative values and $g(T^*)$ is monotone increasing between two positive values, there is a unique positive solution to Eq.~\eqref{eq:g_h}. That is, Eqs.~\eqref{Pp} and \eqref{Tp} have a unique positive steady state solution $(P^*,T^*)$.
		
		\section{Linearization of the thrombopoiesis equations and bifurcation analysis}
		\label{app:linearization}
		
		We take the model in Sect.~\ref{sec:analysis}, which has a differential equation for the platelets, \eqref{Pp}, and one for the thrombopoietin, \eqref{Tp}, depending only on $P$ and $T$. From Appendix.~\ref{app:EU}, the model has a unique positive equilibrium, $(P^*, T^*)$. Linearizing about the equilibrium, we let $x(t) = P(t)-P^*$ and $y(t) = T(t)-T^*$. We use Taylor expansions for both the exponential function and $\eta_m(T(s))$ to obtain the following linear approximation:
		\begin{equation*}
		\begin{split}
		\exp\left[\int_{t - \tau_e - \tau_m}^{t - \tau_e} \! \eta_m(T(s)) \, \mathrm{d}s\right] &\approx
		\exp\left[\int_{t - \tau_e - \tau_m}^{t - \tau_e} (\eta_m(T^*) + \partial_{T}\eta_m(T^*)y(s)) \, \mathrm{d}s\right] \\
		& = \e^{\eta_m(T^*)\tau_m}\exp\left[\partial_{T}\eta_m(T^*)\int_{t - \tau_e - \tau_m}^{t - \tau_e} \! y(s) \, \mathrm{d}s\right] \\
		& \approx \e^{\eta_m(T^*)\tau_m}\left(1 + \partial_{T}\eta_m(T^*)\int_{t - \tau_e - \tau_m}^{t - \tau_e} \! y(s) \, \mathrm{d}s\right). \\
		\end{split}
		\end{equation*}
		Similarly,
		\begin{equation*}
		\exp\left[\int_{t - \tau_e}^{t} \eta_e(T(s)) \, \mathrm{d}s\right] \approx
		\e^{\eta_e(T^*)\tau_e}\left(1 + \partial_{T}\eta_e(T^*)\int_{t - \tau_e}^{t} \! y(s) \, \mathrm{d}s\right),
		\end{equation*}
		\begin{equation*}
		\exp\left[\int_{t - a - \tau_m}^{t - a} \! \eta_m(T(s)) \, \mathrm{d}s\right] \approx
		\e^{\eta_m(T^*)\tau_m}\left(1 + \partial_{T}\eta_m(T^*)\int_{t - a - \tau_m}^{t - a} \! y(s) \, \mathrm{d}s\right),
		\end{equation*}
		and
		\begin{equation*}
		\exp\left[\int_{t - a}^{t} \eta_e(T(s)) \, \mathrm{d}s\right] \approx
		\e^{\eta_e(T^*)a}\left(1 + \partial_{T}\eta_e(T^*)\int_{t - a}^{t} y(s) \, \mathrm{d}s\right).
		\end{equation*}
		
		Linearizing the integral product in Eq.~\eqref{Tp}, we obtain the following approximation:
		\begin{equation*}
		\begin{split}
		&\int_0^{\tau_e}\exp\left[\int_{t - a - \tau_m}^{t - a} \! \eta_m(T(s)) \, \mathrm{d}s\right]\exp\left[\int_{t - a}^{t} \eta_e(T(s)) \, \mathrm{d}s\right] \, \mathrm{d}a \\
		&\qquad \approx \int_0^{\tau_e}\! \e^{\eta_m(T^*)\tau_m + \eta_e(T^*)a}
		\left(1 + \partial_{T}\eta_m(T^*)\int_{t - a - \tau_m}^{t - a} \! y(s) \, \mathrm{d}s\right)\left(1 + \partial_{T}\eta_e(T^*)\int_{t - a}^{t} \! y(s) \, \mathrm{d}s\right) \, \mathrm{d}a \\
		&\qquad \approx \e^{\eta_m(T^*)\tau_m}\left[\frac{\e^{\eta_e(T^*)\tau_e} - 1}{\eta_e(T^*)}
		+ \partial_{T}\eta_m(T^*)\int_0^{\tau_e}\e^{\eta_e(T^*)a}\left(\int_{t - a - \tau_m}^{t - a} y(s) \, \mathrm{d}s\right) \, \mathrm{d}a \right. \\
		&\qquad \quad  + \partial_{T}\eta_e(T^*)\int_0^{\tau_e} \! \e^{\eta_e(T^*)a}\left(\int_{t - a}^{t} y(s) \, \mathrm{d}s\right) \, \mathrm{d}a\Biggr].
		\end{split}
		\end{equation*}
		
		These results can be used to find the linearization of the platelet and thrombopoietin equations. The platelet equation with only the constant and linear terms (higher order terms dropped) is given by
		\begin{equation}
		\begin{split}
		\frac{\mathrm{d}x}{\mathrm{d}t} &=  A_2\left[1 + \partial_{T}\eta_m(T^*)\int_{t - \tau_e - \tau_m}^{t - \tau_e} \! y(s) \, \mathrm{d}s
		+ \partial_{T}\eta_e(T^*)\int_{t - \tau_e}^{t} \! y(s) \, \mathrm{d}s\right] \\
		&\quad  - \gamma_P(x + P^*) - \bigl(F(P^*) + \partial_{P}F(P^*)x\bigr),
		\end{split} \label{P_lin}
		\end{equation}
		where
		\[
		A_2 = \frac{D_0V_m\kappa_PQ^*}{\beta_P}\e^{\eta_m(T^*)\tau_m+\eta_e(T^*)\tau_e}.
		\]
		The thrombopoietin equation can be written as
		\begin{equation*}
		\begin{split}
			\frac{\mathrm{d}y}{\mathrm{d}t} &= T_{prod} - \gamma_T(y + T^*) - \alpha_T\left(A_1\left[E_1
			+ \partial_{T}\eta_m(T^*)\int_0^{\tau_e} \! \e^{\eta_e(T^*)a}\left(\int_{t - a - \tau_m}^{t - a} \! y(s) \, \mathrm{d}s\right) \, \mathrm{d}a \right.\right. \\
			&\quad + \partial_{T}\eta_e(T^*)\int_0^{\tau_e} \! \e^{\eta_e(T^*)a}\left(\int_{t - a}^{t} y(s) \, \mathrm{d}s\right) \, \mathrm{d}a\biggr] + k_S\beta_P(x + P^*)\Biggr)(G(T^*) + \partial_{T}G(T^*)y),
		\end{split}
		\end{equation*}
		where
		\[
		A_1 = V_m\kappa_PQ^*\e^{\eta_m(T^*)\tau_m} \qquad {\rm and} \qquad
		E_1 = \frac{\e^{\eta_e(T^*)\tau_e} - 1}{\eta_e(T^*)}.
		\]
		The thrombopoietin equation with only the constant and linear terms (higher order terms dropped) is given by:
		\begin{equation}
		\begin{split}
		\frac{\mathrm{d}y}{\mathrm{d}t} &= T_{prod} - \gamma_T(y + T^*) - \alpha_T\Biggl[(A_1E_1 + k_S\beta_PP^*)G(T^*)
		+ (A_1E_1 + k_S\beta_PP^*)\partial_{T}G(T^*)y \\
		& \quad + k_S\beta_PG(T^*)x + A_1G(T^*)\Biggl(\partial_{T}\eta_m(T^*)\int_0^{\tau_e}
		\! \e^{\eta_e(T^*)a}\left(\int_{t - a - \tau_m}^{t - a} \! y(s) \, \mathrm{d}s\right) \, \mathrm{d}a \\
		& \quad + \partial_{T}\eta_e(T^*)\int_0^{\tau_e} \! \e^{\eta_e(T^*)a}\left(\int_{t - a}^{t} y(s) \, \mathrm{d}s\right) \, \mathrm{d}a\Biggr)\Biggr],  \label{T_lin}
		\end{split}
		\end{equation}
		
		By the definition of an equilibrium, the constant terms in Eqs.~\eqref{P_lin} and \eqref{T_lin} sum to zero, yielding the linear equations for platelets, \eqref{P2_lin}, and thrombopoietin, \eqref{T2_lin}, given in Sect.~\ref{subsec:local}.
		
		\subsection{Details for the characteristic equation}
		\label{app:char_eqn}
		
		We examine the integral terms in Eqs.~\eqref{P2_lin} and \eqref{T2_lin}, using the exponential form for $y(t) = \e^{\lambda t}$. There are four integrals, which we evaluate below:
		\[
		\int_{t - \tau_e - \tau_m}^{t - \tau_e} \! \e^{\lambda s} \, \mathrm{d}s
		= \frac{\e^{\lambda (t-\tau_e)}}{\lambda}\left(1 - \e^{-\lambda \tau_m}\right),
		\]
		\[
		\int_{t - \tau_e}^{t} \! \e^{\lambda s} \, \mathrm{d}s
		= \frac{\e^{\lambda t}}{\lambda}\left(1 - \e^{-\lambda \tau_e}\right),
		\]
		\[
		\int_0^{\tau_e} \! \e^{\eta_e(T^*)a}\left(\int_{t - a - \tau_m}^{t - a}
		\e^{\lambda s} \, \mathrm{d}s\right) \, \mathrm{d}a
		= \frac{\e^{\lambda t}\left(1 - \e^{-\lambda \tau_m}\right)}{\lambda(\lambda - \eta_e(T^*))}
		\left(1 - \e^{\eta_e(T^*)\tau_e-\lambda \tau_e}\right),
		\]
		and
		\[
		\int_0^{\tau_e} \! \e^{\eta_e(T^*)a}\left(\int_{t - a}^{t}\e^{\lambda s} \, \mathrm{d}s\right) \, \mathrm{d}a
		= \frac{\e^{\lambda t}}{\lambda}\left(\frac{\e^{\eta_e(T^*)\tau_e}-1}{\eta_e(T^*)} + \frac{\e^{-(\lambda-\eta_e(T^*))\tau_e}-1}{\lambda - \eta_e(T^*)}\right).
		\]
		These expressions are used in the terms $L_2(\lambda)$ and $L_4(\lambda)$ in the characteristic equation for the linear functional Eq.~\eqref{eq:lin_fde}.
		
		If Eq.~\eqref{ce} is multiplied by $\lambda(\lambda - \eta_e)$, then the terms in the denominator can be eliminated (at the expense of introducing the roots $\lambda = 0$ and $\eta_e$). The first polynomial piece becomes
		\[
		(\lambda + L_1)(\lambda + C_1)\lambda(\lambda - \eta_e) =
		\lambda^4 + (L_1 + C_1 - \eta_e)\lambda^3 + \left(L_1C_1
		- (L_1+C_1)\eta_e\right)\lambda^2 - L_1C_1\eta_e\lambda.
		\]
		Next we consider the portion $L_4(\lambda) - C_1$ in \eqref{ce}
		\begin{equation*}
		\begin{split}
			(L_4(\lambda) - C_1)\lambda(\lambda - \eta_e) &=
			C_2\partial_{T}\partial_{T}\eta_m(T^*)\left(1 - \e^{-\lambda \tau_m}\right)
			\left(1 - \e^{-(\lambda-\eta_e) \tau_e}\right) \\
			&\quad + C_2\partial_{T}\eta_e(T^*)\left(\frac{\e^{\eta_e \tau_e}-1}{\eta_e}(\lambda-\eta_e) + (\e^{-(\lambda-\eta_e)\tau_e}-1)\right) \\
			&= C_2\partial_{T}\eta_m(T^*)\left(1 - \e^{-\lambda \tau_m} - \e^{\eta_e \tau_e}\e^{-\lambda \tau_e}
			+ \e^{\eta_e \tau_e}\e^{-\lambda(\tau_e+\tau_m)}\right) \\
			&\quad + C_2\partial_{T}\eta_e(T^*)\left(\frac{\e^{\eta_e \tau_e}-1}{\eta_e}\right)(\lambda-\eta_e)
			+ C_2\partial_{T}\eta_e(T^*)(\e^{\eta_e \tau_e}\e^{-\lambda \tau_e} - 1).
		\end{split}
		\end{equation*}
		We take the previous expression and multiply by $\lambda + L_1$ and define
		\[
		D_1 = \frac{C_2\partial_{T}\eta_e(T^*)\left(\e^{\eta_e \tau_e} - 1\right)}{\eta_e}, \qquad
		D_2 = C_2\e^{\eta_e \tau_e}(\partial_{T}\eta_e(T^*)-\partial_{T}\eta_m(T^*)), \qquad {\rm and} \qquad
		D_3 = C_2\e^{\eta_e \tau_e}\partial_{T}\eta_m(T^*).
		\]
		The results are
		\begin{equation*}
		\begin{split}
			(\lambda + L_1)(L_4(\lambda) - C_1)\lambda(\lambda - \eta_e) &= (\lambda + L_1)
			\left(C_2\partial_{T}\eta_m(T^*) + D_1(\lambda - \eta_e) - C_2\partial_{T}\eta_e(T^*)\right) \\
			&\quad- (\lambda + L_1)C_2\partial_{T}\eta_m(T^*)\e^{-\lambda \tau_m} + (\lambda + L_1)D_2\e^{-\lambda \tau_e}
			+ (\lambda + L_1)D_3\e^{-\lambda (\tau_e + \tau_m)}
		\end{split}
		\end{equation*}
		and
		\begin{equation*}
		L_2(\lambda)L_3\lambda(\lambda-\eta_e) = L_3(\lambda-\eta_e)A_2
		\left(\partial_{T}\eta_m(T^*)\e^{-\lambda \tau_e} - \partial_{T}\eta_m(T^*)\e^{-\lambda (\tau_e + \tau_m)}
		+ \partial_{T}\eta_e(T^*) - \partial_{T}\eta_e(T^*)\e^{-\lambda \tau_e}\right).
		\end{equation*}
		Multiplying by $\lambda(\lambda-\eta_e)$ produces a quartic exponential polynomial in the eigenvalues, which can be analyzed using techniques we have developed earlier \cite{mahaffy1982}. The characteristic equation can be written as
		\begin{eqnarray*}
			\lambda^4 + K_3\lambda^3 + K_2\lambda^2 + K_1\lambda + K_0
			+ (\alpha_1\lambda  + \alpha_0)\e^{-\lambda \tau_m} & & \\
			+ (\beta_1\lambda  + \beta_0)\e^{-\lambda \tau_e}
			+ (\gamma_1\lambda  + \gamma_0)\e^{-\lambda (\tau_e + \tau_m)} & = & 0.
		\end{eqnarray*}
		We examine the terms above and obtain the following coefficients:
		\begin{equation*}
		\begin{split}
			K_0 & =  L_1C_2(\partial_{T}\eta_m(T^*) - \partial_{T}\eta_e(T^*)) - D_1L_1\eta_e + L_3A_2\partial_{T}\eta_e(T^*)\eta_e, \\
			K_1 &= -L_1C_1\eta_e + C_2(\partial_{T}\eta_m(T^*) + \partial_{T}\eta_e(T^*) + D_1(L_1 - \eta_e)
			- L_3A_2\partial_{T}\eta_e(T^*), \\
			K_2 &= L_1C_1 - (L_1+C_1)\eta_e + D_1, \\
			K_3 &= L_1 + C_1 - \eta_e, \\
			\alpha_0 & = -C_2\partial_{T}\eta_m(T^*)L_1, \\
			\alpha_1 &= -C_2\partial_{T}\eta_m(T^*), \\
			\beta_0 & = D_2L_1 - L_3A_2\eta_e(\partial_{T}\eta_e(T^*)-\partial_{T}\eta_m(T^*)), \\
			\beta_1 & = D_2 + L_3A_2(\partial_{T}\eta_e(T^*)-\partial_{T}\eta_m(T^*)), \\
			\gamma_0 & = D_3L_1 + L_3A_2\eta_e\partial_{T}\eta_m(T^*) \\
			\gamma_1 & = D_3 - L_3A_2\partial_{T}\eta_m(T^*).
		\end{split}
		\end{equation*}
\newpage
\section{Parameter sensitivity of the model for healthy subjects}
\label{app:para-sense}
The characteristic equation \eqref{ce} from the linear analysis is used to study the sensitivity of each parameter near its normal value.  We developed MATLAB programs from the linear analysis to compute the leading eigenvalues of the model.  With the normal parameters of Table~\ref{table:para-summary} we obtained the equilibrium $(P^*, T^*) = (31.071, 100)$, which has the leading pair of eigenvalues $\lambda_1 = -0.058953 \pm 0.053015i$. It follows that the model with these eigenvalues is asymptotically stable. It should be noted that the frequency is equivalent to a period of 118.5 days, which is far from the observed oscillation periods in either the healthy subjects or patients with cyclic thrombocytopenia.

To analyze the sensitivity of the model to the various parameters, we used our MATLAB code to find the new equilibrium and leading eigenvalues as we varied each parameter by $\pm 10$\%. Table~\ref{table:leading} gives a complete listing of how the equilibrium changes and leading eigenvalues shifts with all of the individual parameter changes. We are most interested in stability changes, so the lowest ratio of the real part of the leading eigenvalues gives the largest change in the direction of a Hopf bifurcation.
\begin{table}[h!]
	\begin{center}
		\begin{tabular}{|c|c|c|c|c|c|c|c|}  \hline
		
		&  & $P^*$ & $T^*$ & Real & Imag & ratio Re & ratio Im  \\ \hline
		Normal &  & 31.071 & 100.0 & -0.058953 & 0.053015 &  &   \\ \hline
		$b_P$ & -10\% & 29.0196 & 100.9495 & -0.067296 & 0.043747 & 1.142 & 0.825  \\ \hline
		$b_P$ & +10\% & 32.8346 & 99.3905 & -0.052283 & 0.058441 & 0.887 & 1.102  \\ \hline
		$\alpha_P$ & -10\% & 32.0273 & 99.7183 & -0.055079 & 0.056355 & 0.934 & 1.063  \\ \hline
		$\alpha_P$ & +10\% & 30.0968 & 100.5067 & -0.062686 & 0.049258 & 1.063 & 0.929  \\ \hline
		$\gamma_P$ & -10\% & 31.7203 & 99.8432 & -0.058229 & 0.053277 & 0.988 & 1.005  \\ \hline
		$\gamma_P$ & +10\% & 30.3240 & 100.4136 & -0.059747 & 0.052683 & 1.013 & 0.994  \\ \hline
		$\kappa_P$ & -10\% & 30.0012 & 102.4043 & -0.057024 & 0.055546 & 0.967 & 1.048  \\ \hline
		$\kappa_P$ & +10\% & 31.9549 & 98.1063 & -0.06075 & 0.050488 & 1.030 & 0.952  \\ \hline
		$\beta_P$ & -10\% & 33.3576 & 100.5375 & -0.062871 & 0.049083 & 1.066 & 0.926  \\ \hline
		$\beta_P$ & +10\% & 29.0111 & 99.7651 & -0.055601 & 0.055916 & 0.943 & 1.055  \\ \hline
		$\alpha_T$ & -10\% & 32.2845 & 102.8172 & -0.060871 & 0.052074 & 1.033 & 0.982  \\ \hline
		$\alpha_T$ & +10\% & 29.9044 & 97.7452 & -0.057327 & 0.053726 & 0.972 & 1.013  \\ \hline
		$k_T$ & -10\% & 28.6234 & 94.9109 & -0.055442 & 0.054430 & 0.940 & 1.027  \\ \hline
		$k_T$ & +10\% & 33.3536 & 105.0196 & -0.062445 & 0.051155 & 1.059 & 0.965  \\ \hline
		$\gamma_T$ & -10\% & 31.0302 & 100.1744 & -0.058994 & 0.053038 & 1.001 & 1.000  \\ \hline
		$\gamma_T$ & +10\% & 30.9910 & 100.0910 & -0.058940 & 0.052996 & 1.000 & 1.000  \\ \hline
		$T_{prod}$ & -10\% & 29.7700 & 97.4513 & -0.056887 & 0.053955 & 0.965 & 1.018  \\ \hline
		$T_{prod}$ & +10\% & 32.1795 & 102.5986 & -0.060925 & 0.052003 & 1.033 & 0.981  \\ \hline
		$k_S$ & -10\% & 31.5216 & 101.2177 & -0.062364 & 0.045469 & 1.058 & 0.858  \\ \hline
		$k_S$ & +10\% & 30.5245 & 99.0902 & -0.055813 & 0.058817 & 0.947 & 1.109  \\ \hline
		$b_e$ & -10\% & 31.4498 & 99.4983 & -0.059528 & 0.052684 & 1.010 & 0.994  \\ \hline
		$b_e$ & +10\% & 30.6166 & 100.7105 & -0.058451 & 0.053328 & 0.991 & 1.006  \\ \hline
		$b_m$ & -10\% & 33.1719 & 95.6242 & -0.062519 & 0.051101 & 1.060 & 0.964  \\ \hline
		$b_m$ & +10\% & 29.2024 & 104.2848 & -0.056023 & 0.054177 & 0.950 & 1.022  \\ \hline
		$\tau_m$ & -10\% & 31.0106 & 100.1327 & -0.057789 & 0.060694 & 0.980 & 1.145  \\ \hline
		$\tau_m$ & +10\% & 31.0106 & 100.1327 & -0.059777 & 0.045867 & 1.014 & 0.865  \\ \hline
		$\tau_e$ & -10\% & 31.7286 & 101.6540 & -0.059127 & 0.057713 & 1.003 & 1.089  \\ \hline
		$\tau_e$ & +10\% & 30.3550 & 98.7243 & -0.058709 & 0.048716 & 0.996 & 0.919  \\ \hline
		\end{tabular}
	\end{center}
\caption{Parameter sensitivity to changes in the normal parameters of the leading pair of eigenvalues $\lambda_1$}\label{table:leading}
\end{table}

For the sensitivity analysis, if we only consider the movement of the leading pair of eigenvalues toward the imaginary axis, then the smallest values in the $7^{th}$ column (ratio Re) give the greatest shift toward instability. The parameter changes that destabilize the model most are (in descending order) increasing $b_P$, decreasing $\alpha_P$, decreasing $k_T$, increasing $\beta_P$, increasing $k_S$, increasing $b_m$, and decreasing $T_{prod}$.

As an experiment to extend this analysis, we chose to increase or decrease all seven of these parameters by 20\% to see what happened to the equilibrium and the leading eigenvalues. The result of all seven changes resulted in the equilibrium $(P^*, T^*) = (20.772, 85.117)$ and leading eigenvalues $\lambda_1 = -0.02555 \pm 0.06563i$. We note that this more than halves the distance of the real part toward the imaginary axis, and also the frequency shifts the period to 95.74 days.

The analysis shows that near normal, there are several parameters to which the model is very insensitive. Surprisingly, this includes all of the delay parameters, $\tau_e$ and $\tau_m$. It is also quite insensitive to changes in $\gamma_P$, $\gamma_T$, and $b_e$. However, our numerical study shows that $\tau_e$ and $\tau_m$ can have effects on the imaginary part.

This analysis shows that the leading eigenvalues have the wrong frequency for the observed oscillations in cyclic thrombocytopenia patients. This suggests the need to examine the next, second leading pair of eigenvalues for this model. Its frequency is closer to the range of interest and provides a starting point for a Hopf bifurcation study of our cyclic thrombocytopenia patients. Again, with the normal parameters the second eigenvalues are $\lambda_2 = -0.11375 \pm 0.35888i$, which gives a quasiperiod near 17.5 days. Table~\ref{table:second} shows the effects on this pair of eigenvalues as the parameters are changed by $\pm 10\%$.
\begin{table}[h!]
\begin{center}
	\begin{tabular}{|c|c|c|c|c|c|c|c|}  \hline
		
		&  & $P^*$ & $T^*$ & Real & Imag & ratio Re & ratio Im  \\ \hline
		Normal &  & 31.071 & 100 & -0.11375 & 0.35888 &  &   \\ \hline
		$b_P$ & -10\% & 29.0196 & 100.9495 & -0.11375 & 0.3588 & 0.974 & 1.001  \\ \hline
		$b_P$ & +10\% & 32.8346 & 99.3905 & -0.11949 & 0.3577 & 1.024 & 0.998  \\ \hline
		$\alpha_P$ & -10\% & 32.0273 & 99.7183 & -0.11826 & 0.35796 & 1.013 & 0.999  \\ \hline
		$\alpha_P$ & +10\% & 30.0968 & 100.5067 & -0.11536 & 0.35853 & 0.988 & 1.001  \\ \hline
		$\gamma_P$ & -10\% & 31.7203 & 99.8432 & -0.11799 & 0.35806 & 1.011 & 0.999  \\ \hline
		$\gamma_P$ & +10\% & 30.324 & 100.4136 & -0.11551 & 0.35847 & 0.990 & 1.001  \\ \hline
		$\kappa_P$ & -10\% & 30.0012 & 102.4043 & -0.11561 & 0.35761 & 0.990 & 0.998  \\ \hline
		$\kappa_P$ & +10\% & 31.9549 & 98.1063 & -0.11776 & 0.35886 & 1.009 & 1.002  \\ \hline
		$\beta_P$ & -10\% & 33.3576 & 100.5375 & -0.11524 & 0.35855 & 0.987 & 1.001  \\ \hline
		$\beta_P$ & +10\% & 29.0111 & 99.7651 & -0.11809 & 0.358 & 1.012 & 0.999  \\ \hline
		$\alpha_T$ & -10\% & 32.2845 & 102.8172 & -0.11411 & 0.35789 & 0.978 & 0.999  \\ \hline
		$\alpha_T$ & +10\% & 29.9044 & 97.7452 & -0.11912 & 0.35858 & 1.020 & 1.001  \\ \hline
		$k_T$ & -10\% & 28.6234 & 94.9109 & -0.12202 & 0.35892 & 1.0453 & 1.002  \\ \hline
		$k_T$ & +10\% & 33.3536 & 105.0196 & -0.11204 & 0.35756 & 0.960 & 0.998  \\ \hline
		$\gamma_T$ & -10\% & 31.0302 & 100.1744 & -0.11662 & 0.35831 & 0.999 & 1.000  \\ \hline
		$\gamma_T$ & +10\% & 30.991 & 100.091 & -0.11684 & 0.35823 & 1.001 & 1.000  \\ \hline
		$T_{prod}$ & -10\% & 29.77 & 97.4513 & -0.11874 & 0.35543 & 1.017 & 0.992  \\ \hline
		$T_{prod}$ & +10\% & 32.1795 & 102.5986 & -0.11491 & 0.3607 & 0.984 & 1.007  \\ \hline
		$k_S$ & -10\% & 31.5216 & 101.2177 & -0.112 & 0.35762 & 0.959 & 0.998  \\ \hline
		$k_S$ & +10\% & 30.5245 & 99.0902 & -0.12118 & 0.35899 & 1.038 & 1.002  \\ \hline
		$b_e$ & -10\% & 31.4498 & 99.4983 & -0.11717 & 0.35822 & 1.004 & 1.000  \\ \hline
		$b_e$ & +10\% & 30.6166 & 100.7105 & -0.11632 & 0.35828 & 0.996 & 1.000  \\ \hline
		$b_m$ & -10\% & 33.1719 & 95.6242 & -0.11169 & 0.36068 & 0.957 & 1.007  \\ \hline
		$b_m$ & +10\% & 29.2024 & 104.2848 & -0.12133 & 0.35606 & 1.039 & 0.994  \\ \hline
		$\tau_m$ & -10\% & 31.0106 & 100.1327 & -0.12172 & 0.37874 & 1.043 & 1.057  \\ \hline
		$\tau_m$ & +10\% & 31.0106 & 100.1327 & -0.11224 & 0.33985 & 0.962 & 0.949  \\ \hline
		$\tau_e$ & -10\% & 31.7286 & 101.654 & -0.12517 & 0.37101 & 1.072 & 1.036  \\ \hline
		$\tau_e$ & +10\% & 30.355 & 98.7243 & -0.10939 & 0.34619 & 0.937 & 0.966  \\ \hline
	\end{tabular}
\end{center}
\caption{Parameter sensitivity to changes in the normal parameters of the second leading pair of eigenvalues $\lambda_2$}\label{table:second}
\end{table}

For the sensitivity analysis, we examine the movement of the second leading pair of eigenvalues, $\lambda_2$, toward the imaginary axis. Again, the smallest values in the $7^{th}$ column (ratio Re) give the greatest shift toward instability. The most destabilizing changes for this pair of eigenvalues occur by (in descending order) increasing $\tau_e$, decreasing $b_m$, increasing $k_S$, increasing $\beta_P$, increasing $\tau_m$, decreasing $b_P$, and decreasing $\gamma_P$. We note that the delays $\tau_e$ and $\tau_m$ affect the movement of the real part of these eigenvalues.

To extend this analysis, we chose to increase or decrease all seven of these parameters by 20\% to see what happened to the equilibrium and eigenvalues. The result of all seven changes resulted in the equilibrium $(P^*, T^*) = (26.505, 87.236)$ and  eigenvalues $\lambda_2 = -0.09281 \pm 0.3091i$. Note that the distance of the real part toward the imaginary axis is slightly more than for the leading pair with its most significant parameters, and also the frequency shifts the period to 20.33 days.

\section{Hopf bifurcation for CT patients} \label{app:hopf_ct}

This appendix continues the studies of the CT patients from Sect.~\ref{subsec:hopf_ct}. The four parameters are varied linearly between the normal state and the best fitting parameters for several CT patients. Again, our numerical methods tracked the changes in the equilibria and the pairs of eigenvalues, which result in Hopf bifurcations leading to the cyclic behavior observed in the CT patients.

\begin{figure}[htb!]
	\centering
	\begin{tabular}{ccc}
		\includegraphics[scale=0.3]{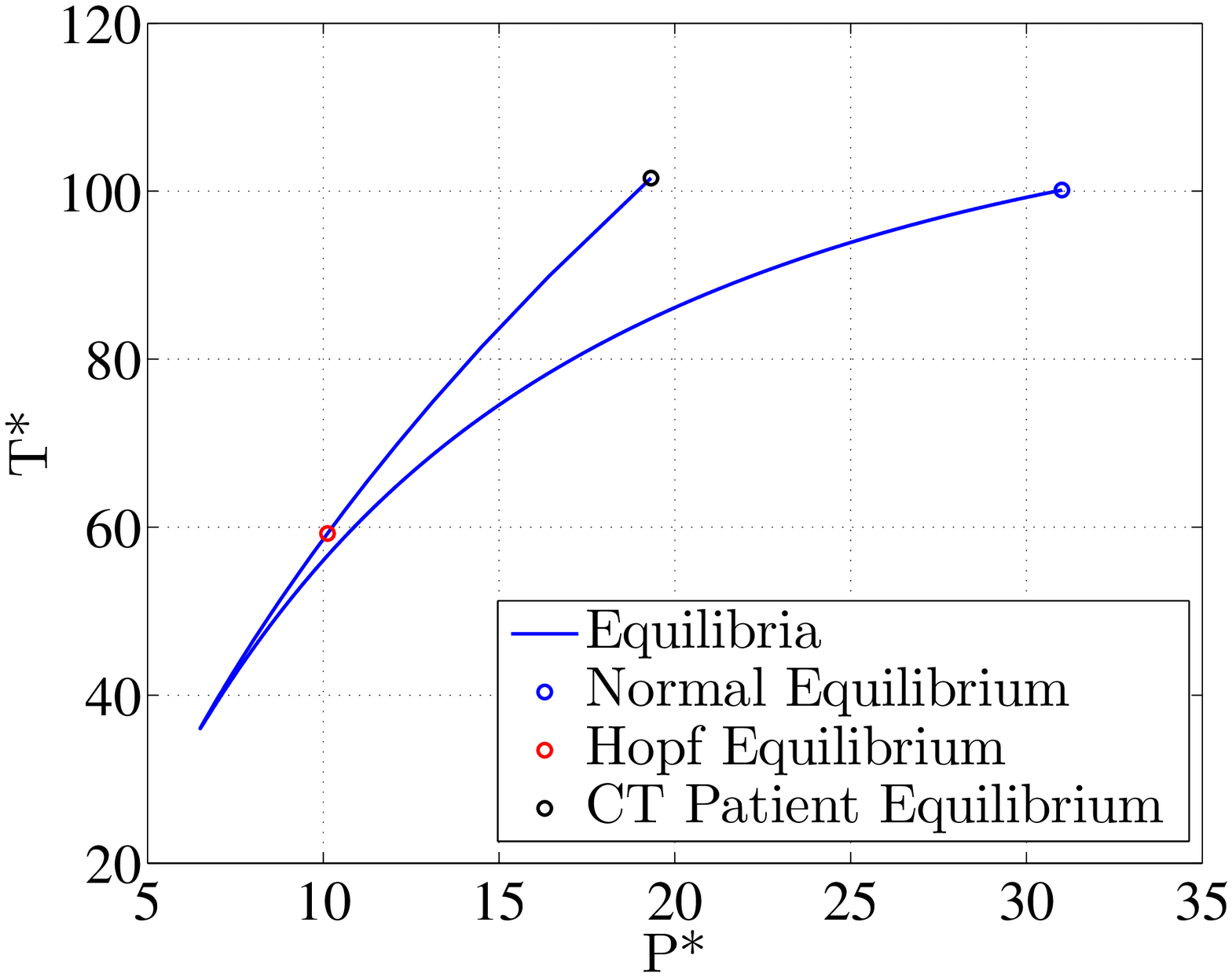}
		& \ &
		\includegraphics[scale=0.3]{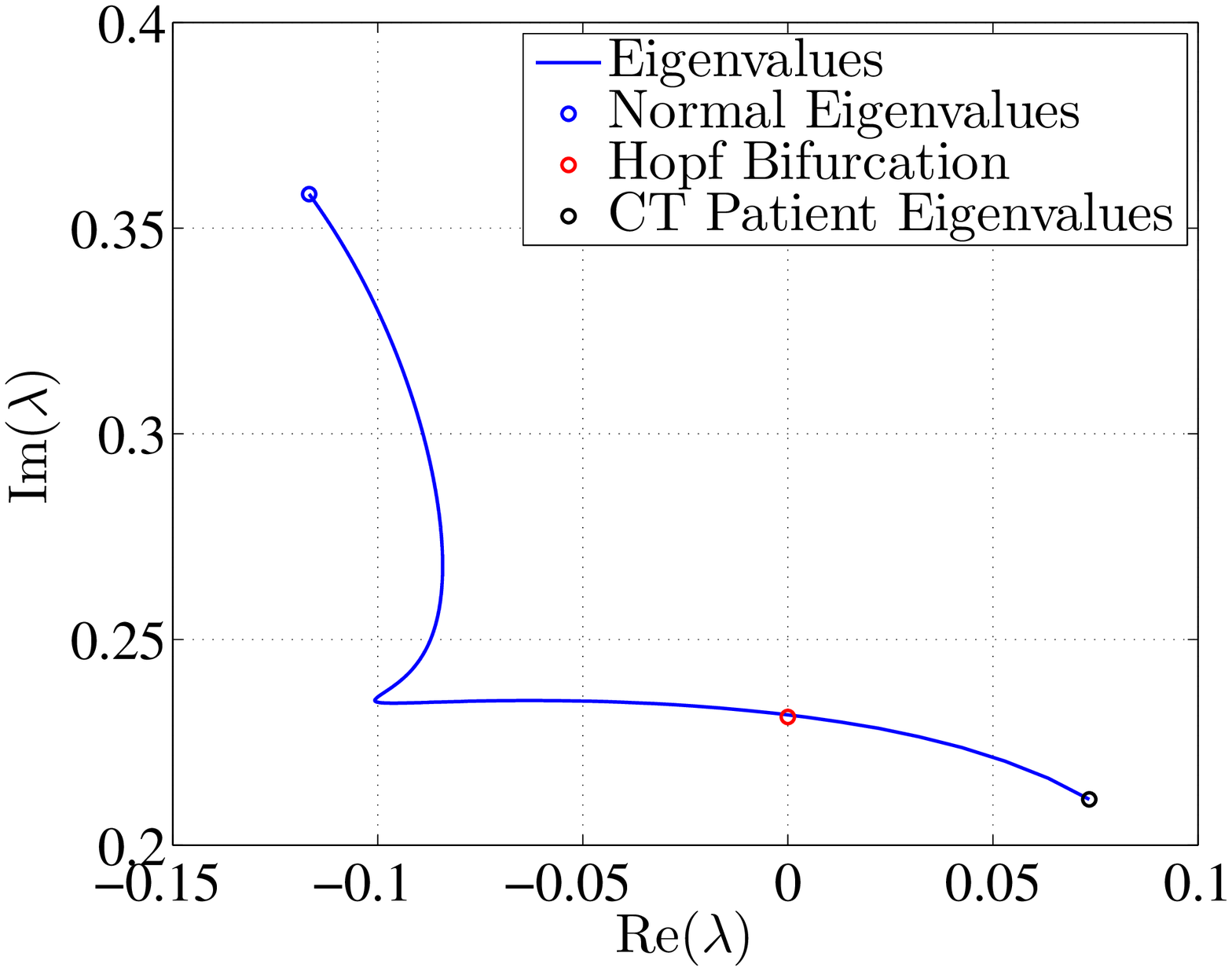} \\
		\includegraphics[scale=0.3]{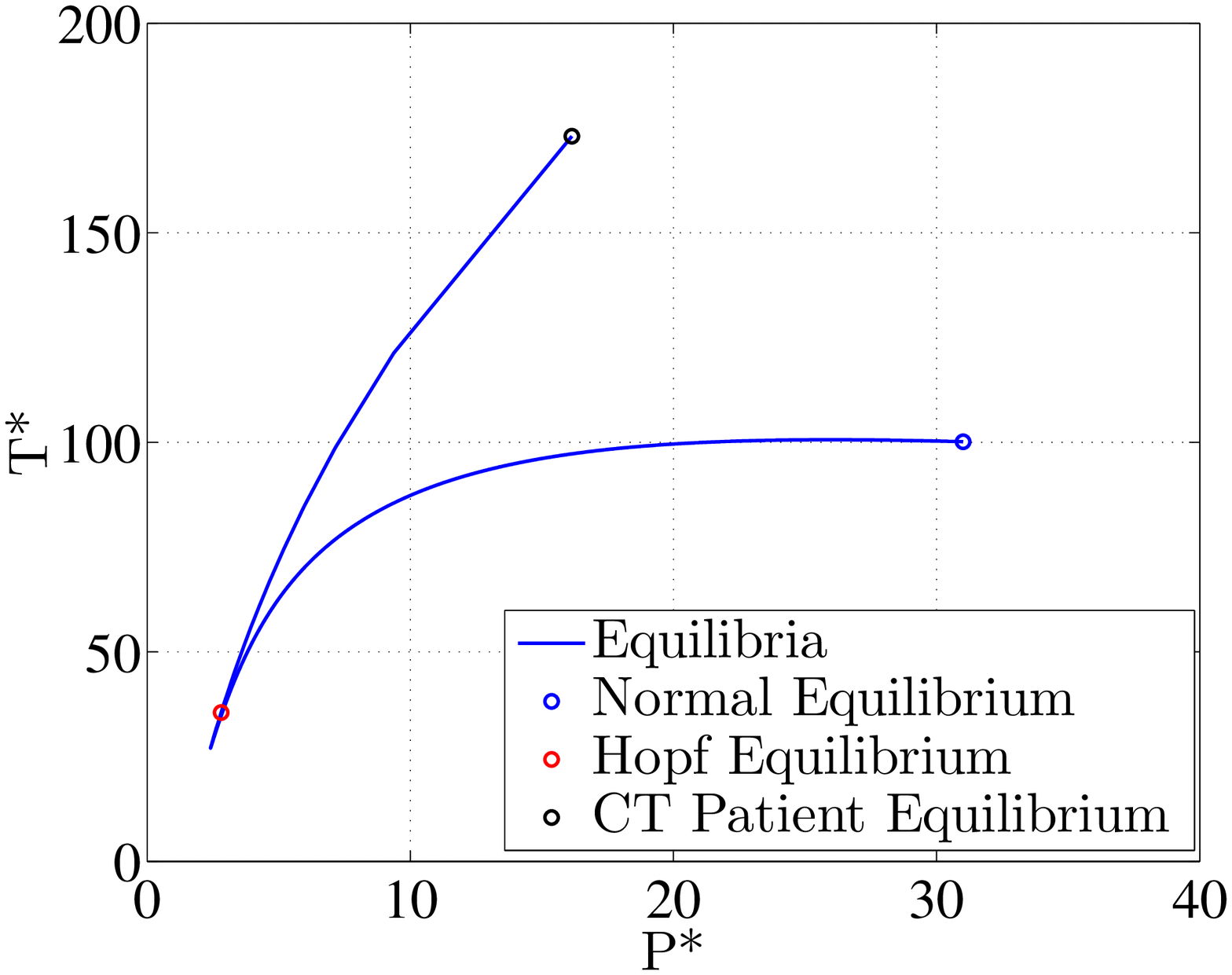}
		& \ &
		\includegraphics[scale=0.3]{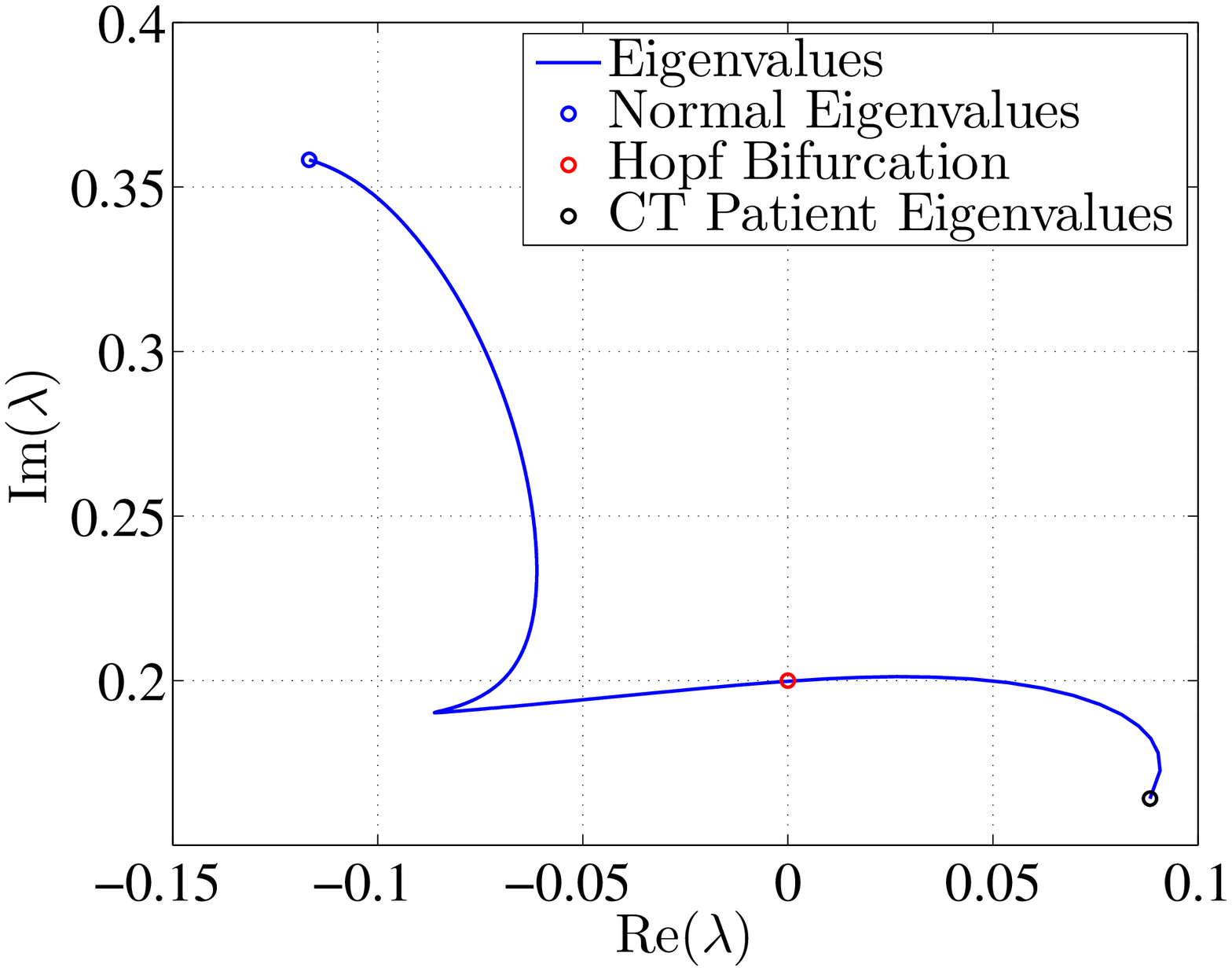} \\
		\includegraphics[scale=0.3]{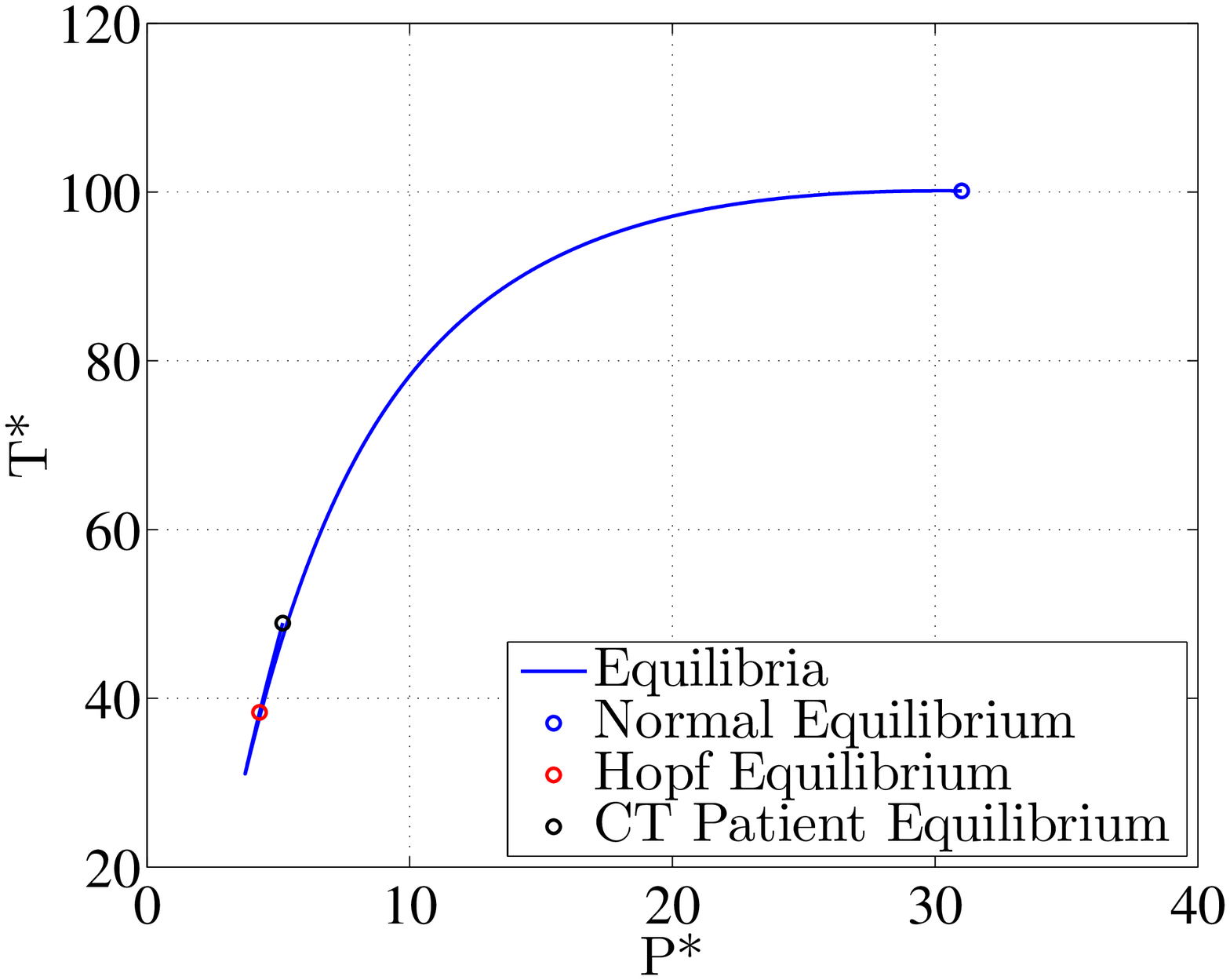}
		& \quad &
		\includegraphics[scale=0.3]{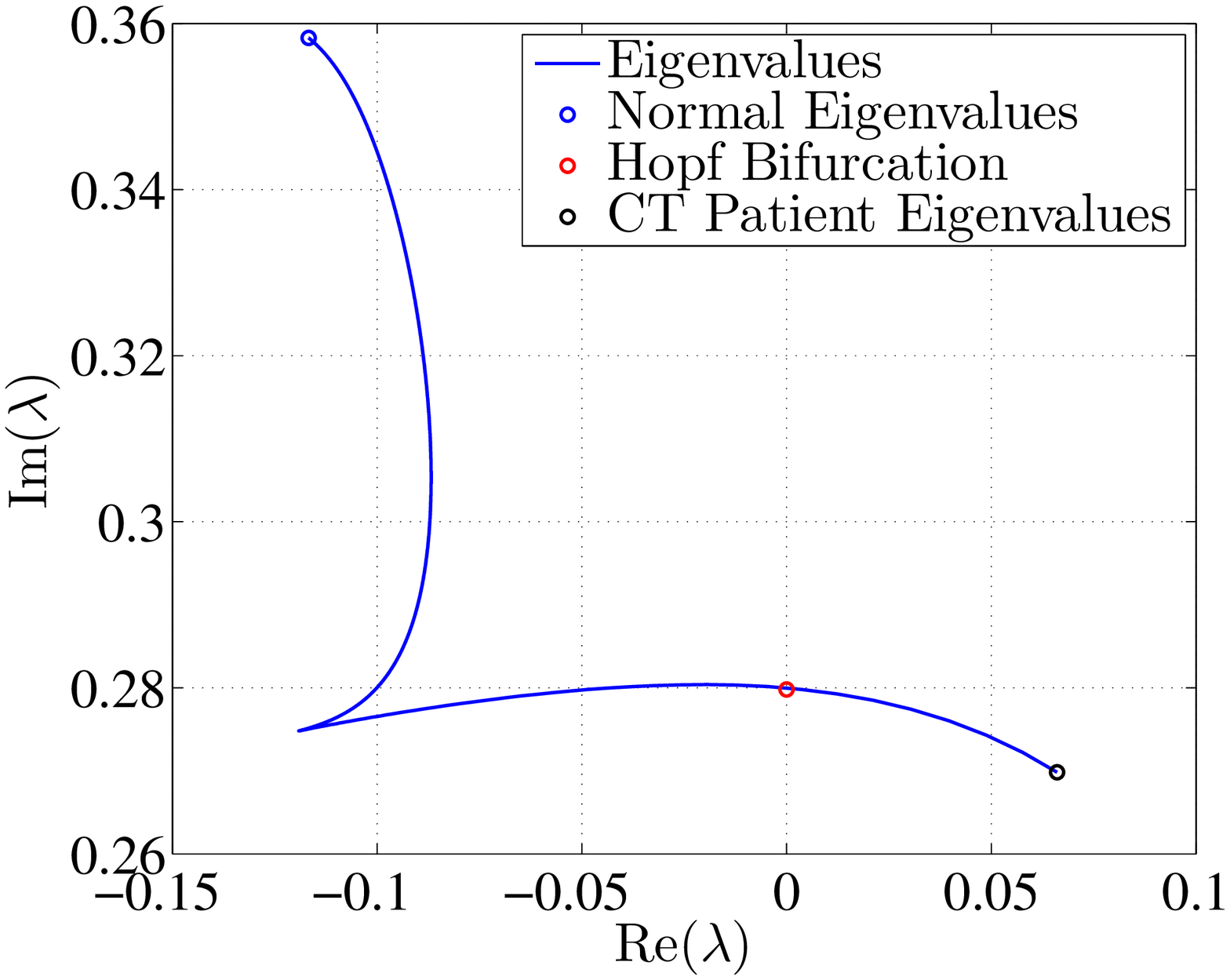}
	\end{tabular}
	\caption{The curves on the left show the evolution of the equilibrium
		from healthy subject to CT patient as parameters vary. The curves on the
		right follow the eigenvalues. The rows represent the evolution from healthy
		subject to the CT patients of \citet{Connor11}, \citet{kimura96}, and \citet{zent}, respectively}
		\label{fig:app_CT_eq_ev}
\end{figure}

Table~\ref{table:ct} in Sect.~\ref{subsec:fitting} shows the best parameter fit to $\tau_{e}$, $\alpha_{P}$, $\alpha_{T}$, and $k_{T}$ for the CT patients of \citet{Connor11}, \citet{kimura96}, and \citet{zent}. As was done with the CT patient of \citet{bruin} in Sect.~\ref{subsec:hopf_ct}, a hyperline in the 4D-parameter space from the normal parameter values to each of the parameter sets for these three CT patients was followed, and the numerical values of the equilibrium $(P^{*}$, $T^{*})$ and eigenvalues $\lambda$ were tracked at each set of parameter values. As before, the eigenvalues tracked were the ones from the second leading pair of the healthy subject, which is the pair that undergoes a Hopf bifurcation as the hyperline extends to any of the CT patients. The results are displayed in Fig.~\ref{fig:app_CT_eq_ev}.

The equilibrium for the normal parameter set is $(P^{*}, T^{*}) = (31.071, 100.00)$. As the parameter sets move along each of the hyperlines, the equilibria first shift slowly in an arc toward the origin. For all of our examples, this first arc of equilibria takes over 98\% of the hyperline, that is, in Eq.~\eqref{eq:hopf_change_parameters} if the arc is created by $t \in [0, t_1)$, then $t_1 > 0.98$. For an unknown reason (likely a transition in one of the Hill functions), the equilibrium rapidly shifts away from this slow path toward the origin along a different trajectory. This new direction roughly doubles back, but heads to the different states of equilibria for each of the different patients. It is along this rapidly evolving path that the Hopf bifurcation occurs. Fig.~\ref{fig:app_CT_eq_ev} (and Fig.~\ref{fig:bruin_eq_ev}) shows that the cusp-like behavior is similar in all cases, but the evolving paths are distinct for each patient. This complicates the interpretation of how the cyclic thrombocytopenia is explained through the parameters. Table~\ref{table:app_eq_ev} gives the values for the equilibria of the Hopf bifurcation along with the equilibria for the best fitting parameters of the different CT patients.  

\begin{table}[htb]
	\centering
	\noindent
	\begin{small}
		\begin{tabular}{|c||c|c||c|c|}
			\hline
			& $(P^*_h, T^*_h)$ & $\lambda_h$ & $(P^*_e, T^*_e)$ & $\lambda_e$  \tabularnewline
			\hline
			Connor & (10.118, 59.244) & $\pm  0.2311i$ & (19.326, 101.31) & $0.07348 \pm 0.2111i$ \tabularnewline
			\hline
			Kimura & (2.809, 35.484) & $\pm 0.2000i$ & (16.118, 172.57) & $0.08832 \pm 0.1641i$ \tabularnewline
			\hline
			Zent & (4.286, 38.329) & $\pm 0.2798i$ & (5.1706, 48.709) & $0.06605 \pm  0.2698i$ \tabularnewline
			\hline
		\end{tabular}
	\end{small}
	\caption{The second and third columns give the equilibria and eigenvalues at the Hopf bifurcation. The fourth and fifth columns give the equilibria and eigenvalues for the specific CT patient}
	\label{table:app_eq_ev}
\end{table}

We numerically solve Eq.~\eqref{ce}, starting at the eigenvalues for the normal case with $\lambda = -0.11375 \pm 0.3588i$. In all cases, the eigenvalues create an arc with the imaginary part decreasing, while the real part first increases then decreases. This arc is created quite slowly and follows the slowly evolving equilibria above. When the equilibria start evolving rapidly, the values of the eigenvalues rapidly shift with increasing real part. Specifically, the real part turns around and increases very rapidly to the Hopf bifurcation. The different CT patients have slightly different changes in their eigenvalues, particularly in what happens to the imaginary part. All cases of the CT patients have their eigenvalues with positive real part, which is to be expected. The frequency of the eigenvalues varies from $0.1641$ to $0.2698$, yielding periods in the range of 23.3 to 38.3 days, which are consistent with the simulations in Sect.~\ref{subsec:fitting}.

\section{Fitting of cyclic thrombocytopenia patient data}
\label{app:ABC}

In this appendix, we describe the statistical procedure used in Sect.~\ref{sec:ctp} to fit the parameters $\tau_{e}$, $\alpha_P$, $\alpha_T$, and $k_T$ of our model to 15 published platelet and TPO data sets of patients with CT \cite{bruin,cohen,Connor11,engstrom1966periodic,helleberg1995cyclic,kimura96,Kosugi1994809,morley1969platelet,rocha1991danazol,vonschulthess1986,skoog1957metabolic,wilkinson1966idiopathic,Yanabu1993155,zent}. Of these 15 data sets, only four contained both platelet and TPO data \cite{bruin,Connor11,kimura96,zent}. We chose to analyze data sets of untreated CT patients only, as treatments may have altered  platelet or TPO dynamics or both and is thus outside the scope of this model.

We first introduce some notation before describing the ABC-MCMC algorithm and the fitting procedure \cite{marjoram2003markov}. Let $\boldsymbol \allpar$ be the vector of all the parameters, that is,
$\boldsymbol \allpar=(\tau_{e}, \alpha_P, \alpha_T, k_T)^\top.$ Let $\data$ denote the observed data, including $\mathbf P_{data}$ and $\mathbf T_{data}$.	
Let $\data^{'}$ denote the model response, including $\mathbf P_{model}$ and $\mathbf T_{model}$.
Let  $\dist(\data^{'}, \data)$ denote the distance between $\data^{'}$ and  $\data$. Let $\thre$ be a prefixed threshold.
The gist of the fitting procedure for a data set is as follows.

\begin{enumerate}
	\item \label{item:initial} Choose an initial set of values of parameters for $\boldsymbol \allpar$.
	\item \label{item:abc_propose} Propose a move from the current value of $\boldsymbol \allpar$ to $\boldsymbol \allpar^{'}$ according to a transition kernel $\trans(\cdot|\allpar)$.
	\item  Simulate $\data^{'}$ using the model with parameters $\boldsymbol \allpar^{'}$.
	\item  If $\dist(\data^{'}, \data)\leq \thre$, go to step \ref{item:abc_mh}, and otherwise stay at $\boldsymbol \allpar$ and return to step \ref{item:abc_propose}.
	\item \label{item:abc_mh} Calculate
	\begin{equation*} \label{eq:abc_mh}
	\mh(\boldsymbol \allpar, \boldsymbol \allpar^{'}) = \min\left(1, \frac{\prior(\boldsymbol \allpar^{'}) \trans(\boldsymbol \allpar|\boldsymbol \allpar^{'})}{\prior(\boldsymbol \allpar) \trans(\boldsymbol \allpar^{'}|\boldsymbol \allpar)}\right),
	\end{equation*}
	update $\boldsymbol \allpar$ to $\boldsymbol \allpar^{'}$ with this probability, and store the value of $\boldsymbol \allpar^{'}$.
	\item Repeat steps 2-5 using $\boldsymbol \allpar'$ as the new initial set of values of parameters for a sufficient number of times, and finally pick $\boldsymbol \allpar'$ that minimizes $\dist(\data^{'}, \data)$ among all stored values of $\boldsymbol \allpar'$.
\end{enumerate}

To measure the distance between the simulated data  $\data^{'}$ and the observed data $\data$, we use the following sum of squared errors (SSE):
\begin{equation} \label{eq:Err_patient_fits}
\dist(\data^{'}, \data) = \frac{\|\frac{2}{3}\mathbf P_{model} - \mathbf P_{data}\|_{2}}{\|\mathbf P_{data}\|_{2}} + \frac{\|\mathbf T_{model}-\mathbf T_{data}\|_{2}}{\|\mathbf T_{data}\|_{2}},
\end{equation}
where the factor of 2/3 accounts for the fraction of platelets that circulate in blood in our model.

For the fits in this paper, in step \ref{item:initial} we chose the initial parameters so that the model generated oscillations, as discussed in Sect.~\ref{subsec:generating_oscillations}, and gave a rough approximation to the data set. These parameters defined the initial data $\data$, and the initial SSE was computed using Eq.~\eqref{eq:Err_patient_fits}. We fixed the threshold $\thre$ to 1.15 times the initial SSE and computed steps 2-5 for 250 successful iterations before choosing the vector of parameters $\boldsymbol \allpar$ minimizing $\dist(\data^{'}, \data)$.

In step \ref{item:abc_mh} of the ABC-MCMC implementation, we use a uniform distribution as the prior distribution for the parameters, which leads to $\prior(\boldsymbol \allpar^{'})/\prior(\boldsymbol \allpar)=1$. In addition, we choose a Gaussian distribution to be the transition kernel, which implies
$\trans(\boldsymbol \allpar|\boldsymbol \allpar^{'})/\trans(\boldsymbol \allpar^{'}|\boldsymbol \allpar)=1$.  As a result, $\mh(\boldsymbol \allpar, \boldsymbol \allpar^{'})$ is simplified to 1.

The stationary distribution of the MCMC chain is the posterior distribution of the parameters given the simulated data are close enough to the observed data ($\dist(\data^{'}, \data)\leq \thre$). In other words,	with a good initial choice of parameter values, this ABC-MCMC algorithm guarantees convergence to a good fit.
		
\section{Numerical Analysis}\label{app:numerics}

Numerical simulations of the system of equations \eqref{eq:de_tpo} and \eqref{eq:platelet_prod_rate} are necessary both to illustrate results for any parameter set, and also as part of the parameter fitting described in Appendix~\ref{app:ABC}. Accurate numerical solution of these equations is complicated by their structure as distributed delay differential equations (DDEs) with the terms $m_{e}(t,\tau_{e})$ and $M_{e}(t)$ both defined by integrals of the solution functions from time $t-\tau_{e}-\tau_{m}$ to $t$, with $M_{e}(t)$ requiring the computation of the integral of the product of the exponential of two integrals.

Traditional Runge-Kutta methods for ordinary differential equations (ODEs) only define a numerical approximation to the solution on a discrete set of time points. This is problematical when the solution is required at off mesh time values, which arises for example for the accurate evaluation of integrals, as is the case in our problem.
Continuous Runge-Kutta (CRK) methods were developed \cite{bellen2003,HNWI} to produce continuous output suitable for the numerical solution of both ODEs and DDEs. These are the methods currently most often used to
solve discrete DDEs, with the Matlab~\cite{matlab} software package containing built in
functions (\texttt{dde23}, \texttt{ddesd})
for the solution of discrete constant delay and state-dependent variable delay DDEs. However, these methods are not appropriate for problems with vanishing or distributed delays, because they become fully implicit. To see how this arises suppose the system \eqref{eq:de_tpo}--\eqref{eq:platelet_prod_rate} is already solved up to time $t_n$ and consider the computation of the next step. To compute the $j^{th}$ stage of the CRK method for the
next step requires the computation of the right-hand side of the system of equations at time $t_n+c_jh$, where $h=t_{n+1}-t_n$ is the step-size of the method and the $c_j$ are the abscissa of the CRK method.  But this requires us to evaluate $m_{e}(t_n+c_jh,\tau_{e})$ and $M_{e}(t_n+c_jh)$, for which we need integrals of $T(t)$ up to time $t_n+c_jh$, but until all the stages of the current step are computed we only have the solution of $T(t)$ available up to time $t_n$. This problem does not arise in the first order forward Euler method (which has just one stage with $c_1=0$), but all higher order CRK methods become fully implicit. Ad-hoc methods for approximating the missing integral result in a reduction in the order of the method, and so the only CRK methods that are appropriate for our problem are the first order forward Euler method or higher order implicit methods.
To obtain accurate solutions efficiently we do not use such methods.

Methods that remain explicit for distributed DDEs and DDEs with vanishing delays were first proposed by
Tavernini \cite{tavernini1971}, and have more recently been developed into a class of methods called Functional Continuous Runge-Kutta (FCRK) methods \cite{bellen2009,maset2005}. In Bellen~\cite{bellen2006} it was proposed to apply FCRK methods to distributed DDEs in biomathematics, but we are not aware of any implementation of FCRK methods for distributed DDEs before the current work.

To solve the system \eqref{eq:de_tpo}--\eqref{eq:platelet_prod_rate} we implemented the explicit two-stage second order Heun FCRK method proposed in \cite{cryertaver1972}. This has Butcher tableau
$$\def\arraystretch{1.25}
\begin{array}{c|c}
c\; & \; A(\alpha) \\
\hline
& b(\alpha)
\end{array}\quad=\quad
\begin{array}{l|cc}
0\; & 0 & 0 \\
1 & \alpha & 0 \\ \hline
& \;\alpha - \frac{1}{2} \alpha^2 & \;\frac{1}{2}\alpha^2
\end{array}$$
which defines the parameters $c_{i}$, $b_{i}(\alpha)$ and $a_{i j}(\alpha)$ in the explicit FCRK method
\begin{gather*}
K_{i}  = hG(t_{n} + c_{i}h, Y^i_{t_{n} + c_{i}h}) \\
Y^{i}(s)  = u_{n}(s) \text{ for } s \leq t_{n} \qquad
Y^{i}(t_{n} + \alpha h)  = u_{n}(t_{n}) + \sum\limits_{j=1}^{i-1} a_{i j}(\alpha) K_{j} \text{ for } \alpha \in (0, 1] \\
u_{n+1}(s)  = u_{n}(s) \text{ for } s \leq t_{n} \qquad
u_{n+1}(t_{n} + \alpha h)  = u_{n}(t_{n}) + \sum\limits_{i=1}^s b_{i}(\alpha)K_{i} \text{ for } \alpha \in (0, 1]
\end{gather*}
where the $K_{i}$ are said to be the stage variables.  In the method, $u_{n}$ is the numerical solution
with step-size $h$ generated by the n$^{th}$ step of the method defined up to time $t_{n}$ for
the delayed functional differential equation
$$u'(t) = G(t, u_{t}) \text{ for } t > t_{0}, 	\qquad u(t) = \phi(t) \text{ for } t \leq t_{0},$$
where $\phi$ is said to be the starting data, $t_{0}$ is the initial time, and
$u_{t}$ is the continuous function $u_{t}(\sigma) = u(t + \sigma)$ for $\sigma \in [-\tau, 0]$, where $\tau$ is the largest delay in the system (which is $\tau=\tau_e+\tau_m$ for our problem).

The fundamental difference between CRKs and FCRKs which enables the FCRK methods to remain explicit is that the FCRKs are endowed with a continuous approximation $Y^i(t)$ associated with each stage, whereas the CRK methods only have the single continuous approximation $u_{n+1}(t)$ defined once the step is computed.

To implement this FCRK method for the system \eqref{eq:de_tpo}--\eqref{eq:platelet_prod_rate}, the integrals need to be evaluated numerically to sufficient accuracy to maintain the convergence order of the method. Although the composite trapezoidal rule would be sufficient for second order accuracy we used fourth order composite methods, to allow for the possible later implementation of a fourth order FCRK method. It is necessary to evaluate the integrals on the same computational mesh as the underlying FCRK method. Since the numerical approximation $T_{n+1}(t)$
to $T(t)$ is smooth on each interval $[t_n,t_{n+1}]$, but not differentiable at the mesh points $t_n$, the functions $\eta_m(T_{n+1}(t))$ and $\eta_m(T_{n+1}(t))$ will also not be differentiable at the mesh points, and the convergence theory of the composite quadrature methods will break down unless the mesh points $t_n$ of the FCRK method are included as quadrature points for the integration. The exact solutions of delayed functional differential equations also have discontinuous derivatives at breaking points as outlined in \cite{bellen2003}, which requires certain time points to be included in the computational mesh. But because of the additional smoothing afforded by the integrals in
\eqref{eq:de_tpo}--\eqref{eq:platelet_prod_rate}
the only breaking point that needs to be included in the mesh to obtain second order convergence is the initial point $t_0$.

Taking account of these considerations we used Simpson's method to evaluate $M_e(t_n+c_jh)$ using values of
$m_e(t_n+c_jh,kh)$ and $m_e(t_n+c_jh,(k+1/2)h)$, where $c_j=0$ or $1$ for Heun's method, and $k\in\{0,1,...,N\}$ is an integer with the step-size $h$ of the FCRK method chosen so that $\tau_e=Nh$. The necessary values of $m_e(t,a)$ were obtained from \eqref{eq:platelet_prod_rate}, evaluating the integrals in this formula using Milne's method. To evaluate the nested integrals efficiently we store values of $\eta_m(T(t))$ and $\eta_e(T(t))$ at relevant points ($t_n+kh/4$ for $k=0,1,2,3$) and also store integrals which would otherwise be recomputed at multiple steps, including
$$\int_{t_j-\tau_m}^{t_j} \! \eta_m(T(s)) \, \mathrm{d}a,$$
which appears in $m_e(t_j+kh,kh)$ and hence is required to evaluate $M_e(t_{j+k})$ for each of $k=0,1,\ldots,N$.

The method is written to return a function handle that is created from the continuous approximation to the solution generated by the method.  This is returned to the user, which allows for the evaluation of $P(t)$ and $T(t)$ at all points in the computational interval (not just at mesh points).

The explicit fourth order FCRK method proposed in \cite{cryertaver1972}
could be similarly implemented, but requires $6$ stages to evaluate one step to fourth order, and so is not as advantageous as the 4-stage fourth order RK method for ODEs, and would require the computation of some complicated integrals over $6$ stages.

\end{appendices}
	
\bibliographystyle{spbasic}

\end{document}